\documentclass[transmag]{IEEEtran}
\usepackage{amsmath,amssymb,amsfonts,amsthm}
\usepackage{mathtools}
\usepackage{textcomp}

\let\labelindent\relax
\usepackage{amsmath,amssymb,graphicx,enumitem}
\usepackage{epstopdf}
\usepackage{multirow}
\usepackage{algorithm}
\usepackage{algpseudocode}
\usepackage{soul}

\usepackage[binary-units=true]{siunitx}

\usepackage{xr}
\makeatletter
\let\NAT@parse\undefined
\makeatother

\usepackage{hyperref}
\usepackage{cleveref}

\usepackage[hyphenbreaks]{breakurl}

\newtheorem{theorem}{Theorem}
\newtheorem{definition}[theorem]{Definition}

\newcommand{\X}{\boldsymbol{\bar{{\mathcal{X}}}}}

\newcommand{\A}{\boldsymbol{A}}

\newcommand{\x}{\boldsymbol{x}}

\newcommand{\redA}[0]{\boldsymbol{A_{\bar{\mathcal{X}},\bar{\mathcal{Y}}}}}

\crefname{lemma}{Lemma}{Lemmas}
\crefname{definition}{Defn.}{Defns.}

\newcommand{\floor}[1]{\left\lfloor #1 \right\rfloor}

\title{A Compressed Sensing Approach to Pooled RT-PCR Testing for COVID-19 Detection}

\author{Sabyasachi Ghosh\dag, Rishi Agarwal \dag, Mohammad Ali Rehan \dag, Shreya Pathak \dag, \\
Pratyush Agarwal \dag, Yash Gupta \dag, Sarthak Consul $\ast$, Nimay Gupta \dag, Ritika \dag, Ritesh Goenka \dag, \\
Ajit Rajwade \dag\thanks{AR acknowledges support from SERB Matrics grant MTR/2019/000691. AR and MG acknowledge support from IITB WRCB grant \#10013976, and DST-Rakshak grant \#10013980.}, Manoj Gopalkrishnan $\ast$
\thanks{\dag Dept. of Computer Science \& Engg., IIT Bombay, India; \{sghosh, ajitvr\}@cse.iitb.ac.in}
\thanks{$\ast$ Dept. of Electrical Engineering, IIT Bombay; manojg@ee.iitb.ac.in}
\thanks{This work has been accepted for publication at the IEEE Open Journal of Signal Processing.}
}

\begin{document}

\IEEEtitleabstractindextext{
\begin{abstract}
 We propose `Tapestry', a novel approach to pooled testing with application to COVID-19
 testing with quantitative Reverse Transcription Polymerase Chain Reaction (RT-PCR) that can result in shorter
 testing time and conservation of reagents and testing kits. Tapestry combines ideas
 from compressed sensing and combinatorial group testing with a novel noise model for
 RT-PCR used for generation of synthetic data.
 Unlike Boolean group testing algorithms, the input is a quantitative readout
 from each test and the output is a list of viral loads for each sample relative to the pool with the highest viral load.
 While other pooling techniques require a second confirmatory assay, Tapestry obtains individual sample-level results in a single round of testing, at clinically acceptable false positive or false negative rates. We also propose designs for pooling matrices that facilitate good prediction of the infected samples while remaining practically viable.
When testing $n$ samples out of which $k \ll n$ are infected, our method needs only $O(k \log n)$ tests when using random binary pooling matrices, with high probability.
However, we also use deterministic binary pooling matrices based on combinatorial design ideas of Kirkman Triple Systems to balance between good reconstruction properties and matrix sparsity for ease of pooling.
A lower bound on the number of tests with these matrices for satisfying a sufficient condition for guaranteed recovery is $k\sqrt{n}$.
In practice, we have observed the need for  fewer tests with such matrices than with random pooling matrices.
 This makes  Tapestry capable of very large savings at low prevalence rates, while simultaneously remaining viable even at prevalence rates as high as 9.5\%. Empirically we find that single-round Tapestry pooling improves over two-round Dorfman pooling by almost a factor of 2 in the number of tests required. We describe how to combine combinatorial group testing and compressed sensing algorithmic ideas together to create a new kind of algorithm that is very effective in deconvoluting pooled tests. We validate Tapestry in simulations and wet lab experiments with oligomers in quantitative RT-PCR assays. 
 An accompanying Android application Byom Smart Testing makes the Tapestry protocol straightforward to implement in
 testing centres, and is made available for free download.
 Lastly, we describe use-case scenarios for deployment. 
\end{abstract}

}

\maketitle

\bstctlcite{IEEEexample:BSTcontrol}

\section{Introduction}
\label{sec:intro}
The coronavirus disease of 2019 (COVID-19) crisis has led to widespread lockdowns in several countries, and has had a major negative impact on the economy. Early identification of infected individuals can enable quarantining of the individuals and thus control the spread of the disease. Such individuals may often be asymptomatic for many days. Widespread testing with the RT-PCR (reverse transcription polymerase chain reaction) method can help identify the infected individuals. However, widespread testing is not an available option in many countries due to constraints on resources such as testing time ($\sim 3-4$ hours per round), basic equipment, skilled manpower and reagents. 

The current low rate of COVID-19 infection in the world population \cite{Benatia_arxiv} means that most samples tested are not infected, so that most tests are wasted on uninfected samples. Group testing is a process of pooling together samples of $n$ different people into multiple pools, and testing the pools instead of each individual sample. A negative result on a pool implies that all samples participating in it were negative. This saves a huge amount of testing resources, especially with low infection rates. Group testing for medical applications has a long history dating back to the 1940s when it was proposed for testing of blood samples for syphilis~\cite{Dorfman1943}. Simple two-round group testing schemes  have already been applied in the field by several research labs \cite{israel_gt,germany_gt} for COVID-19 testing. Such two-round group testing schemes require pooling of samples and a second round of sample handling for all samples in positive pools. This second round of sample handling can increase the time to result and be laborious to perform since it requires the technician to wear PPE one more time, do another round of RNA extraction, and PCR. In situations where the result needs to be delivered fast, a second round of sample handling and testing must be avoided. In such situations, these schemes are less attractive. 

We present Tapestry, a novel combination of ideas from combinatorial group testing and compressed sensing (CS) \cite{CandesWakin2008} which uses the quantitative output of PCR tests to reconstruct the viral load of each sample in a single round. Tapestry has been validated with wet lab experiments with oligomers~\cite{Ghosh2020.04.23.20077727}. In this work, we elaborate on the results from the algorithmic perspective for the computer science and signal processing communities. Tapestry has a number of salient features which we enumerate below.
\begin{enumerate}[leftmargin=0.4cm]
\item Tapestry delivers results in a single round of testing, without the need for a second confirmatory round, at clinically acceptable false negative and false positive rates. 
The number $m$ of required tests is only $O(k \log n)$ for random binary pooling matrix constructions, as per compressed sensing theory for random binary matrices \cite{kueng2017robust}. In the targeted use cases where the number of infected samples $k \ll n$, we see that $m \ll n$. However, our deterministic pooling matrix constructions based on Kirkman Triple Systems \cite{Kirkman, ray1971solution} require fewer tests in practice (see Sec. \ref{subsub:optimal_sensing_matrices} for a discussion on why this may be the case). Consequently we obtain significant savings in testing time and resources such as number of tests, quantity of reagents, and manpower.
\item Tapestry reconstructs relative viral loads i.e., ratio of viral amounts in each sample to the highest viral amount across pools. 
It is believed that super-spreaders and people with severe symptoms have higher viral load~\cite{beldomenico2020superspreaders,Liu2020}, so this quantitative information might have epidemiological relevance.
\item Tapestry takes advantage of quantitative information in PCR tests. Hence it returns far fewer false positives than traditional binary group testing algorithms such as \textsc{Comp} (Combinatorial Orthogonal Matching Pursuit)\cite{Chan2011}, while maintaining clincally acceptable false negative rates.
Furthermore, it takes advantage of the fact that a negative pool has viral load exactly zero. Traditional CS algorithms do not take advantage of this information. Hence, Tapestry demonstrates better sensitivity and specificity than CS algorithms.
\item Because each sample is tested in three pools, Tapestry can detect some degree of noise in terms of cross-contamination of samples and pipetting errors. 
\item Tapestry allows PCR test measurements to be noisy. We develop a novel noise model to describe noise in PCR experiments. Our algorithms are tested on this noise model in simulation.
\item All tuning parameters for execution of the algorithms are inferred on the fly in a data driven fashion. 
\item Each sample contributes to exactly three pools, and each pool has the same number of samples. This simplifies the experimental design, conserves samples, keeps pipetting overhead to a minimum, and makes sure that dilution due to pool size is in a manageable regime.
\end{enumerate}
The organization of the paper is as follows. We first present a brief overview of the RT-PCR method in Sec. \ref{sec:rtpcr}. The precise mathematical definition of the computational problem being solved in this paper is then put forth in Sec. \ref{subsec:problem_stmt}. We describe traditional and CS-based group-testing algorithms for this problem in Sec. \ref{subsec:gt}, \ref{subsec:cs_gt} and \ref{subsec:combined}. The Tapestry method is described in Sec. \ref{subsec:combined}. The sensing matrix design problem, as well as theoretical guarantees using Kirkman Triple Systems or random binary matrices, are described in Sec. \ref{subsec:matrix_design}. Results on synthetic data are presented in Sec. \ref{sec:results}. This is followed by results on data from lab experiments performed with oligomers to mimic the clinical situation as closely as possible. In Sec. \ref{sec:prev_work}, we compare our work to recent related approaches. We conclude in Sec. \ref{sec:concl} with a glance through different scenarios where our work could be deployed. The supplemental material contains several additional experimental details as well as proofs of some theoretical results.

\section{RT-PCR Method}
\label{sec:rtpcr}
We present here a brief summary of the RT-PCR process, referring to \cite{rt_pcr_iaea} for more details. In the RT-PCR method for COVID-19 testing, a sample in the form of naso- or oro-pharyngeal swabs is collected from a patient. 
The sample is then dispersed into a liquid medium. The RNA molecules of the virus present in this liquid medium are converted into complementary DNA (cDNA) via a process called reverse transcription. DNA fragments called primers complementary to cDNA from the viral genome are then added. They attach themselves to specific sections of the cDNA from the viral genome if the virus is present in the sample. The cDNA of these specific viral genes then undergoes a process of exponential amplification in an RT-PCR machine. Here, cDNA is put through several cycles of alternate heating and cooling in the presence of Taq polymerase and appropriate reagents. This triggers the creation of many new identical copies of specific portions of the target DNA, roughly doubling in number with every cycle of heating and cooling. The reaction volume contains sequence-specific fluorescent markers which report on the total amount of amplified DNA of the appropriate sequence. The resulting fluorescence is measured, and the increase can be observed on a computer screen in real time. The time when the amount of fluorescence exceeds the threshold level is known as the threshold cycle $C_t$, and is a quantitative readout from the experiment. A smaller $C_t$ indicates greater number of copies of the virus. Usually $C_t$ takes values anywhere between $16$ to $32$ cycles in real experiments. PCR can detect even single molecules. A single molecule typically would have $C_t$ value of around $40$ cycles. 
A typical RT-PCR setup can test 96 samples in parallel. The test takes about 3-4 hours to execute.

\section{Testing Methods}

\subsection{Statement of the Computational Problem}
\label{subsec:problem_stmt}
Let $\boldsymbol{x}$ denote a vector of $n$ elements where $x_i$ is the viral load (i.e. viral amount) of the $i^\textrm{th}$ person. Throughout this paper we assume that only one sample per person is extracted. Hence $\boldsymbol{x}$ contains the viral loads corresponding to $n$ different people. Note that $x_i = 0$ implies that the $i^\textrm{th}$ person is not infected. Due to the low infection rate for COVID-19 as yet even in severely affected countries \cite{Benatia_arxiv}, $\boldsymbol{x}$ is considered to be a sparse vector with at the most $k \ll  n$ positive-valued elements. In group testing, small and equal volumes of the samples of a subset of these $n$ people are pooled together according to a sensing or pooling matrix $\boldsymbol{A} = (A_{ji})_{m\times n}$ whose entries are either $0$ or $1$. The viral loads of the pools will be given by:
\begin{equation}
z_j = \sum_{i=1}^n A_{ji} x_i = \boldsymbol{A^j x}, 1 \leq j \leq m, 1 \leq i \leq n, 
\end{equation}
where $A_{ji} = 1$ if a portion of the sample of the $i^\textrm{th}$ person is included in the $j^{\textrm{th}}$ pool, and $\boldsymbol{A^j}$ is the $j^{\textrm{th}}$ row of $\boldsymbol{A}$. In all, some $m < n$ pools are created and individually tested using RT-PCR. We now have the relationship
$\boldsymbol{z} = \boldsymbol{Ax}$, 
where $\boldsymbol{z}$ is the $m$-element vector of viral loads in the mixtures, and $\boldsymbol{A}$ denotes a $m \times n$ binary `pooling matrix' (also referred to as a `sensing matrix' in CS literature). Note that each positive RT-PCR test will yield a noisy version of $z_j$, which we refer to as $y_j$. The relation between the `clean' and noisy versions is given as follows (also see Eqn. \ref{eq:yj_zj}):
\begin{equation}\label{eq:noisy_viral_load}
y_j = z_j (1+q)^{e_j} = (1+q)^{e_j} \boldsymbol{A^j x} ,     
\end{equation}
where $e_j \sim \mathcal{N}(0,\sigma^2)$ and $q \in (0,1)$ is the fraction of viral cDNA that replicates in each cycle. The factor $(1+q)^{e_j}$ reflects the stochasticity in the growth of the numbers of DNA molecules during PCR. Here $\sigma$ is known and constant. 
Equivalently for positive tests, we have:
\begin{equation}
\log y_j = \log (\boldsymbol{A^j x}) + \log (1+q) e_j.  \label{eq:noisemodel}
\end{equation}
In case of negative tests, $y_j$ as well as $z_j$ are 0-valued, and no logarithms need be computed. In non-adaptive group testing, the core computational problem is to estimate $\boldsymbol{x}$ given $\boldsymbol{y}$ and $\boldsymbol{A}$ without requiring any further pooled measurements. It should be noted that though we have treated each element of $\boldsymbol{x}$ to be a fixed quantity, it is in reality a random variable of the form $x_i \sim \textrm{Poisson}(\lambda_i)$ where $\lambda_i \geq 0$. If matrix $\boldsymbol{A}$ contains only ones and zeros, this implies that $z_j \sim \textrm{Poisson}(\boldsymbol{A^j x})$ because the sum of Poisson random variables is also a Poisson random variable.

\subsubsection{Derivation of Noise Model}
\label{subsec:noisemodel}
For a positive pool $j$, the quantitative readout from RT-PCR is not its viral load but the 
observed cycle time $t_j$ when its fluorescence reaches a given threshold $F$ (see Sec. \ref{sec:rtpcr}).
In order to be able to apply CS techniques (see Sec. \ref{subsec:cs_gt}), we derive a relationship between the cycle time of
a sample and its viral load.
Because of exponential growth (see \cite{thermofisher}), the number of molecules of viral cDNA in pool $j$ at cycle time $t$, denoted by $v_j(t)$ is given by:
\begin{equation}
    v_j(t) = z_j(1+q)^{t}.
\end{equation}
Also, $t$ is a real number, with $\floor{t}$ indicating the number of PCR cycles that have passed, and $t - \floor{t}$ indicating the fraction of wall-clock time within the
current cycle. The fluorescence of the pool, $f_j(t)$, is directly proportional to the number of virus molecules $v_j(t)$. That is,
\begin{equation}
\label{eqn:fluorescence}
    f_j(t) = Kv_j(t) = Kz_j(1+q)^t,
\end{equation}
where $K$ is a constant of proportionality.
Suppose the fluorescence of pool $j$ should reach the threshold value $F$ at cycle time $\tau_j$, according to Eqn. \ref{eqn:fluorescence}.
Due to the stochastic
nature of the reaction, as well as measurement error in the PCR machine, the threshold
cycle output by the machine will not reflect this true cycle time. We model this 
discrepancy as Gaussian noise.
Hence, the true cycle time $\tau_j$ and the observed cycle time $t_j$ are related as $\tau_j = t_j + e_j$,
where $e_j \sim \mathcal{N}(0,\sigma^2)$ as before. Now, since $f_j(\tau_j) = F$, using Eqn. \ref{eqn:fluorescence}, we have
\begin{equation}
\label{eq:cycle_threshold}
    F = Kz_j(1+q)^{\tau_j} = K y_j (1+q)^{t_j}.
\end{equation}
The latter equality is since we use the noisy cycle threshold $t_j$ to compute viral load, where $y_j$ is defined to be the noisy viral load of pool $j$. Hence we find
\begin{equation}
    y_j = z_j(1+q)^{\tau_j - t_j} = z_j(1+q)^{e_j} = (1+q)^{e_j} \boldsymbol{A^j x},
    \label{eq:yj_zj}
\end{equation}
obtaining the relationship from Eqn. \ref{eq:noisy_viral_load}. 

Constants $F$ and $K$ are unknown. Hence it is not possible to directly obtain $y_j$
from $t_j$ without additional machine-specific calibration. However, we can find the ratio between the noisy viral loads of two
pools using Eqn. \ref{eq:cycle_threshold}.
Let $y_{min}$ be the noisy viral load of the pool with the minimum observed threshold cycle ($t_{min}$) among all pools. Then we define
relative viral loads as: 
\begin{align}
    \label{eq:relative_viral_loads}
    &\widetilde y_j = \frac{y_j}{y_{min}} = (1+q)^{t_{min} - t_j}, \widetilde z_j = \frac{z_j}{y_{min}}, \widetilde {\boldsymbol{x}} = \frac{\boldsymbol{x}}{y_{min}}
\end{align}
where $\widetilde z_j$ is the relative viral load of a pool, $\widetilde y_j$ is its noisy version, and
$\widetilde {\boldsymbol{x}}$ is the vector of relative viral loads of each sample. We note that due to Eqn. \ref{eq:yj_zj}, the following relation holds:
\begin{equation}
\widetilde{y_j} = \widetilde{z_j} (1+q)^{e_j} = (1+q)^{e_j} \boldsymbol{A^j \widetilde{x}} ,     
\end{equation}
Hence we can apply CS techniques from Sec. \ref{subsec:cs_gt}  to determine the relative magnitudes of viral loads without
knowing $F$ and $K$. We provide more comments about the settings of various noise model parameters for our experiments, in Sec. \ref{sec:results}, particularly in Sec. \ref{subsub:parameters}.

\subsection{Combinatorial Group-Testing}
\label{subsec:gt}
Combinatorial Orthogonal Matching Pursuit (\textsc{Comp}) is a Boolean nonadaptive group testing method \cite[Sec. 2.3]{aldridge2014group}. Here one uses the simple idea that if a mixture $\widetilde{y}_j$ tests negative then any sample $\widetilde{x}_i$ for which $A_{ji} = 1$ must be negative. Note that pools which test negative are regarded as noiseless observations, as argued in Sec. \ref{subsec:noisemodel}. The other samples are all considered to be positive. This algorithm guarantees that there are no `false negatives'. However it can produce a very large number of `false positives'. For example, a sample $\widetilde{x}_k$ will be falsely reported to be positive if every mixture $\widetilde{y}_j$ it is part of, also contains at least one other genuinely positive sample. The \textsc{Comp} algorithm is largely insensitive to noise. Moreover a small variant of it can also produce a list of `high confidence positives', after identifying the (sure) negatives. This happens when a positive mixture $\widetilde{y}_j$ contains only one sample $\widetilde{x}_i$, not counting the other samples which were declared sure negatives in the earlier step. Such a step of identifying `high confidence positives' is included in the so-called \textbf{Definite Defectives} (\textsc{Dd}) Algorithm \cite[Sec. 2.4]{aldridge2014group}. However \textsc{Dd} labels all remaining items to be negative, potentially leading to a large number of false-negatives. The performance guarantees for \textsc{Comp} have been analyzed in \cite{Chan2011} and show that \textsc{Comp} requires $ek(1+\delta)\log n$ tests for an error probability less than $n^{-\delta}$ (see Sec. \ref{subsub:optimal_sensing_matrices}). This analysis has been extended to include the case of noisy test results as well \cite{Chan2011}. However \textsc{Comp} can result in a large number of false positives if not enough tests are used, and it also does not predict viral loads. 

\subsection{Compressed Sensing for Pooled Testing}
\label{subsec:cs_gt}
Group testing is intimately related to the field of compressed sensing (CS) \cite{Gilbert2008}, which has emerged as a significant sub-area of signal and image processing \cite{CandesWakin2008}, with many applications in biomedical engineering \cite{Zhao2019,Zhang2012,Liu2015}. In CS, an image or a signal $\boldsymbol{x}$ with $n$ elements, is directly acquired in compressed format via $m$ linear measurements of the form $\boldsymbol{y} = \boldsymbol{Ax} + \boldsymbol{\eta}$. Here, the measurement vector $\boldsymbol{y}$ has $m$ elements, and $\boldsymbol{A}$ is a matrix of size $m \times n$, and $\boldsymbol{\eta}$ is a vector of noise values. If $\boldsymbol{x}$ is a sparse vector with $k \ll n$ non-zero entries, and $\boldsymbol{A}$ obeys the so-called restricted isometry property (RIP), then \emph{exact} recovery of $\boldsymbol{x}$ from $\boldsymbol{y}, \boldsymbol{A}$ is possible \cite{Candes2008} if $\boldsymbol{\eta} = \boldsymbol{0}$. In the case of measurement noise, the recovery of $\boldsymbol{x}$ produces a solution that is provably close to the original $\boldsymbol{x}$. A typical recovery problem \textsc{P0} consists of optimizing the following cost function:
\begin{equation}
    \textrm{min} \|\boldsymbol{x}\|_0 \textrm{ s.t. } \|\boldsymbol{y}-\boldsymbol{Ax}\|_2 \leq \varepsilon,
\label{eq:P0}
\end{equation}
where $\varepsilon$ is an upper bound (possibly a high probability upper bound) on $\|\boldsymbol{\eta}\|_2$, and $\|\boldsymbol{x}\|_0$ is the number of non-zero elements in $\boldsymbol{x}$. In the absence of noise, a unique and exact solution to this problem is possible with as few as $2k$ measurements in $\boldsymbol{y}$ if $\boldsymbol{x}$ has $k$ non-zero elements \cite{Candes2008}. Unfortunately, this optimization problem \textsc{P0} is NP-Hard and the algorithm requires brute-force subset enumeration. Instead, the following problem \textsc{P1} (often termed `Basis Pursuit Denoising' or \textsc{Bpdn}) is solved in practice:
\begin{equation}
    \textrm{min} \|\boldsymbol{x}\|_1 \textrm{ s.t. } \|\boldsymbol{y}-\boldsymbol{Ax}\|_2 \leq \varepsilon.
\label{eq:P1}
\end{equation}
\textsc{P1} is a convex optimization problem which yields the same solution as the earlier problem (with similar conditions on $\boldsymbol{x},\boldsymbol{A}$) at significantly lower computational cost, albeit with $O(k \log n)$ measurements (i.e. typically greater than $2k$) \cite{CandesWakin2008,Candes2008}.

The order $k$ restricted isometry constant (RIC) of a matrix $\boldsymbol{A}$ is defined as the smallest constant $\delta_k$, for which the following relationship holds for all $k$-sparse vectors $\boldsymbol{x}$ (i.e. all vectors with at the most $k$ non-zero entries): $(1-\delta_k)\|\boldsymbol{x}\|^2_2 \leq \|\boldsymbol{Ax}\|^2_2 \leq (1+\delta_k) \|\boldsymbol{x}\|^2_2$.
The matrix $\boldsymbol{A}$ is said to obey the order $k$ restricted isometry property (RIP) if $\delta_k$ is close to 0. This property essentially implies that no $k$-sparse vector (other than the zero vector) can lie in the null-space of $\boldsymbol{A}$. Unique recovery of $k$-sparse signals requires that no $2k$-sparse vector lies in the nullspace of $\boldsymbol{A}$ \cite{Candes2008}. A matrix $\boldsymbol{A}$ which obeys RIP of order $2k$ satisfies this property. It has been proved that matrices with entries randomly and independently drawn from distributions such as Rademacher or Gaussian, obey the RIP of order $k$ with high probability \cite{Baraniuk2008}, provided they have at least $O(k \log n)$ rows. There also exist \textbf{deterministic binary sensing matrix designs} (e.g. \cite{Devore2007}) which require $O(\textrm{max}(k^2,\sqrt{n}))$ measurements. However it has been shown recently \cite{lotfi2020compressed} that the constant factors in the deterministic case are \emph{significantly smaller} than those in the former random case when $n < 10^5$, making the deterministic designs more practical for typically encountered problem sizes. The solution to the optimization problems \textsc{P0} and \textsc{P1} in Eqns. \ref{eq:P0} and \ref{eq:P1} respectively, are provably robust to noise \cite{CandesWakin2008}, and the recovery error decreases with decrease in noise magnitude. The error bounds for \textsc{P0} in Eqn. \ref{eq:P0} are of the form, for solution $\boldsymbol{\hat{x}}$ \cite{davenport_duarte_eldar_kutyniok_2012}:
\begin{equation}
 \dfrac{\varepsilon}{\sqrt{1+\delta_{2k}}} \leq  \|\boldsymbol{x}-\boldsymbol{\hat{x}}\|_2 \leq \dfrac{\varepsilon}{\sqrt{1-\delta_{2k}}},
\end{equation}
whereas those for \textsc{P1} in Eqn. \ref{eq:P1} have the form \cite{davenport_duarte_eldar_kutyniok_2012}:
\begin{equation}
 \|\boldsymbol{x}-\boldsymbol{\hat{x}}\|_2 \leq \varepsilon \zeta(\delta_{2k}).
\end{equation}
Here $\zeta(\delta_{2k})$ is a monotonically increasing function of $\delta_{2k} \in (0,1)$ and has a small value in practice. 

The Restricted Isometry Property as defined above is also known as RIP-2, because it uses the $\ell_2$-norm. Many other sufficient conditions for recovery of $k$-sparse vectors exist.
We define the following which we use later in Sec. \ref{subsec:matrix_design} and supplemental Sec. \ref{sup-sec:comp_ric} to prove theoretical guarantees of our method.
\begin{definition}
\label{def:rip-1}
\textbf{RIP-1: } \cite[Defn. 8]{Berinde2008} A $m\times n$ matrix $\A$ is said to obey RIP-1 of order $k$
if $\exists$ $ \delta_k \in (0, 1) $ such that 
for all $k$-sparse vectors $\x \in \mathbb{R}^n$,
$$\|\boldsymbol{x}\|_1 \leq \|\boldsymbol{Ax}\|_1 \leq (1+\delta_k)\|\boldsymbol{x}\|_1$$.
\end{definition}

\begin{definition}
\label{def:rnsp}
\textbf{RNSP: } \cite[Eqn. 12]{lotfi2020compressed} A $m\times n$ matrix $\A$ is said to obey the Robust Nullspace Property (RNSP) of order $k$ if $\exists$ $\rho < 1$ and $\tau > 0 $ such that for all $\x \in \mathbb{R}^n$ it holds that
$$||\x_S||_2 \leq {\rho}||\boldsymbol{x}_{\bar{S}}||_1 + \tau ||\A \x||_2 \text{   } $$
for all $S \subset \{1\dots n\}$ with $|S| \leq k$.
\end{definition}

\begin{definition}
\label{def:l2_rnsp}
\textbf{$\ell_2$-RNSP: } \cite[Defn. 1]{kueng2017robust} A $m\times n$ matrix $\A$ is said to obey the $\ell_2$-robust Nullspace Property ($\ell_2$-RNSP) of order $k$
if $\exists$ $\rho \in (0, 1)$ and $\tau > 0 $ such that for all $\x \in \mathbb{R}^n$ it holds that
$$||\x_S||_2 \leq \frac{\rho}{\sqrt{k}}||\boldsymbol{x}_{\bar{S}}||_1 + \tau ||\A \x||_2 \text{   } $$
for all $S \subset \{1\dots n\}$ with $|S| \leq k$.
\end{definition}
Over the years, a variety of different techniques for compressive recovery have been proposed. We use some of these for our experiments in Sec. \ref{subsec:combined}. These algorithms use different forms of sparsity and incorporate different types of constraints on the solution. 

\subsection{CS and Traditional GT Combined} \label{sec:cs_gt_combined}
\begin{algorithm}
\caption{Tapestry Method}
\label{alg:tapestry}
\begin{algorithmic}[1]
    \Statex Input: $n$ samples, $m\times n$ pooling matrix $\boldsymbol{A}$ 
    \State Perform pooling according to pooling matrix $\boldsymbol{A}$ and create $m$ pooled samples 
    \State Run RT-PCR test on these $m$ pooled samples and receive $m \times 1$ vector of cycle threshold values $\boldsymbol{t}$ 
    \State Compute $m\times 1$ vector of relative viral loads $\widetilde{ \boldsymbol{y}}$ from $\boldsymbol t$
    \State Use \textsc{Comp} to filter out negative tests and sure negative samples. Compute submatrix 
    $\boldsymbol{A_{\bar{\mathcal{X}},\bar{\mathcal{Y}}}}$, $\boldsymbol{\widetilde{y}_{\bar{\mathcal{Y}}}}$ and list $\mathcal{HCP}$ of `high-confidence positives' along with their viral loads (see Sec. \ref{subsec:gt}). 
    \State Use a CS decoder to recover relative viral loads
    $\boldsymbol{\widetilde{x}_{\bar{\mathcal{X}}}}$ from 
    $\boldsymbol{\widetilde{y}_{\bar{\mathcal{Y}}}},
    \boldsymbol{A_{\bar{\mathcal{X}},\bar{\mathcal{Y}}}}$
    \State Compute $n\times 1$ relative viral load vector $\widetilde{\x}$ by setting its entries from $\widetilde{\x}_{\X}$, and setting remaining entries to $0$.
    \State \Return $\boldsymbol{\widetilde{x}}$, $\mathcal{HCP}$.  
\end{algorithmic}
\end{algorithm}
\label{subsec:combined}
The complete pipeline of the Tapestry method is presented in Algorithm \ref{alg:tapestry}. First, a wet lab technician performs pooling of $n$ samples into $m$ pools according to a $m \times n$ pooling matrix $\boldsymbol A$. Then they run the RT-PCR test on these $m$ pools (in parallel). The output of the RT-PCR tests -- the threshold cycle ($C_t$) values of each pool -- is processed to find the relative viral load vector $\boldsymbol{\widetilde y}$ of the $m$ pools (as shown in Eqn. \ref{eq:relative_viral_loads}). This is given as input to the Tapestry decoding algorithm, which outputs a sparse relative viral load vector $\boldsymbol{\widetilde x}$.

The Tapestry decoding algorithm, our approach toward group-testing for COVID-19, involves a two-stage procedure\footnote{The two-stage procedure is purely algorithmic. It does not require two consecutive rounds of testing in a lab.}. In the first stage, we apply the \textsc{Comp} algorithm described in Sec. \ref{subsec:gt}, to identify the sure negatives (if any) in $\boldsymbol{\widetilde{x}}$ to form a set $\mathcal{X}$. Let $\mathcal{Y}$ be the set of zero-valued measurements in $\boldsymbol{\widetilde{y}}$ (i.e. negative tests). Please refer to Sec. \ref{subsec:noisemodel} for the definition of $\boldsymbol{\widetilde{x}}, \boldsymbol{\widetilde{y}}$. Moreover, we define $\bar{\mathcal{X}}, \bar{\mathcal{Y}}$ as the complement-sets of $\mathcal{X}, \mathcal{Y}$ respectively. Also, let $\boldsymbol{y_{\bar{\mathcal{Y}}}}$ be the vector of $m-|\mathcal{Y}|$ measurements which yielded a positive result. Let $\boldsymbol{x_{\bar{\mathcal{X}}}}$ be the vector of $n-|\mathcal{X}|$ samples, which does not include the $|\mathcal{X}|$ surely negative samples. Let $\boldsymbol{A_{\bar{\mathcal{X}},\bar{\mathcal{Y}}}}$ be the submatrix of $\boldsymbol{A}$, having size $(m-|\mathcal{Y}|) \times (n-|\mathcal{X}|)$, which excludes rows corresponding to zero-valued measurements in $\boldsymbol{y}$ and columns corresponding to negative elements in $\boldsymbol{x}$. In the second stage, we apply a CS algorithm to recover $\boldsymbol{\widetilde{x}_{\bar{\mathcal{X}}}}$ from $\boldsymbol{\widetilde{y}_{\bar{\mathcal{Y}}}}, \boldsymbol{A_{\bar{\mathcal{X}},\bar{\mathcal{Y}}}}$. \emph{To avoid symbol clutter, we henceforth just stick to the notation $\boldsymbol{y}, \boldsymbol{x}, \boldsymbol{A}, m, n$, even though they respectively refer to $\boldsymbol{\widetilde{y}_{\bar{\mathcal{Y}}}}, \boldsymbol{\widetilde{x}_{\bar{\mathcal{X}}}}, \boldsymbol{A_{\bar{\mathcal{X}},\bar{\mathcal{Y}}}},m-|\mathcal{Y}|,n-|\mathcal{X}|$}.

Note that the CS stage following \textsc{Comp} is very important for the following reasons:
\begin{enumerate}[leftmargin=0.4cm]
    \item \textsc{Comp} typically produces a large number of false positives. The CS algorithms help reduce the number of false positives as we shall see in later sections.
    \item \textsc{Comp} does not estimate viral loads, unlike CS algorithms.
    \item In fact, unlike CS algorithms, \textsc{Comp} treats the measurements in $\boldsymbol{y}$ as also being binary, thus discarding a lot of useful information.
    \item \textsc{Comp} preserves the RIP-1, RIP-2, RNSP, and $\ell_2$-RNSP of the pooling matrix, i.e. if $\boldsymbol{A}$ obeys any of RIP-1, RIP-2, RNSP or $\ell_2$-RNSP of order $k$, then $\boldsymbol{A_{\bar{\mathcal X},\bar{\mathcal Y}}}$ also obeys the same property of the same order $k$ with the same parameters. We formalize and prove these claims in the supplemental section \ref*{sup-sec:comp_ric}.
\end{enumerate}

However, the \textsc{Comp} algorithm prior to applying the CS algorithm is also very important for the following reasons:
\begin{enumerate}[leftmargin=0.4cm]
    \item Viral load in negative pools is exactly $0$. \textsc{Comp} identifies the sure negatives in $\boldsymbol{x}$ from the negative measurements in $\boldsymbol{y}$. Traditional CS algorithms do not take advantage of this information, since they assume all tests to be noisy (Eqns. \ref{eq:P0} and \ref{eq:P1}). It is instead easier to discard the obvious negatives before applying the CS step.
    \item  Since \textsc{Comp} identifies the sure negatives, therefore, it effectively reduces the size of the problem to be solved by the CS step from $(m,n)$ to $(m-|\mathcal{Y}|,n-|\mathcal{X}|)$.
\item In a few cases, a (positive) pool in $\bar{\mathcal{Y}}$ may contain only one contributing sample in $\bar{\mathcal{X}}$, after negatives have been eliminated by \textsc{Comp}. Such a sample is called a `high-confidence positive', and we denote the list of high-confidence positives as $\mathcal{HCP}$. 
In rare cases, the CS decoding algorithms we employed (see further in this section) did not recognize such a positive.
However, such samples will still be returned by our algorithm as positives, in the set $\mathcal{HCP}$ (see last step of Alg. \ref{alg:tapestry}, and `definite defectives' in Sec. \ref{subsec:gt}).
\end{enumerate}
For CS recovery, we employ one of the following algorithms after \textsc{Comp}: the non-negative LASSO (\textsc{Nnlasso}), non-negative orthogonal matching pursuit (\textsc{Nnomp}), Sparse Bayesian Learning (\textsc{Sbl}), and non-negative absolute deviation regression (\textsc{Nnlad}). For problems of small size, we also apply a brute force (\textsc{Bf}) search algorithm to solve a problem similar to \textsc{P0} from Eqn. \ref{eq:P0} combinatorially.

\subsubsection{The Non-negative LASSO (\textsc{Nnlasso})}
\label{subsubsec:nnlasso}
The LASSO (least absolute shrinkage and selection operator) is a penalized version of the constrained problem \textsc{P1} in Eqn. \ref{eq:P1}, and seeks to minimize the following cost function:
    \begin{equation}
        J_{lasso}(\boldsymbol{x};\boldsymbol{y},\boldsymbol{A}) := \|\boldsymbol{y}-\boldsymbol{Ax}\|^2_2 + \lambda \|\boldsymbol{x}\|_1.
    \end{equation}
    Here $\lambda$ is a regularization parameter which imposes sparsity in $\boldsymbol{x}$. The LASSO has rigorous theoretical guarantees \cite{THW2015} (chapter 11) for recovery of $\boldsymbol{x}$ as well as recovery of the support of $\boldsymbol{x}$ (i.e. recovery of the set of non-zero indices of $\boldsymbol{x}$). Given the non-negative nature of $\boldsymbol{x}$, we implement a variant of LASSO with a non-negativity constraint, leading to the following optimization problem:
    \begin{equation}
        J_{nnlasso}(\boldsymbol{x};\boldsymbol{y},\boldsymbol{A}) := \|\boldsymbol{y}-\boldsymbol{Ax}\|^2_2 + \lambda \|\boldsymbol{x}\|_1 \textrm{  s.t.  } \boldsymbol{x} \geq \boldsymbol{0}.
    \end{equation}
    
    \textbf{Selection of $\lambda$}: There are criteria defined in \cite{THW2015} for selection of $\lambda$ under iid Gaussian noise, so as to guarantee statistical consistency. However, in practice, \textbf{cross-validation} (CV) can be used for optimal choice of $\lambda$ in a purely data-driven fashion from the available measurements. The details of this are provided in the supplemental section \ref*{sup-sec:crossval}.

\subsubsection{Non-negative Orthogonal Matching Pursuit (\textsc{Nnomp})}
Orthogonal Matching Pursuit (OMP) \cite{Pati1993} is a greedy approximation algorithm to solve the optimization problem in Eqn. \ref{eq:P0}. Rigorous theoretical guarantees for OMP have been established in \cite{Cai2011}. OMP proceeds by maintaining a set $\mathcal{H}$ of `selected coefficients' in $\boldsymbol{x}$ corresponding to columns of $\boldsymbol{A}$. In each round a column of $\boldsymbol{A}$ is picked greedily, based on the criterion of maximum absolute correlation with a residual vector $\boldsymbol{r} := \boldsymbol{y} - \sum_{k \in \mathcal{H}}\boldsymbol{A_k}\hat{x}_k$. Each time a column is picked, \emph{all} the coefficients extracted so far (i.e. in set $\mathcal{H}$) are updated. This is done by computing the orthogonal projection of $\boldsymbol{y}$ onto the subspace spanned by the columns in $\mathcal{H}$. The OMP algorithm can be quite expensive computationally. Moreover, in order to maintain non-negativity of $\boldsymbol{x}$, the orthogonal projection step would require the solution of a non-negative least squares problem, further adding to computational costs. However, a fast implementation of a non-negative version of OMP (\textsc{Nnomp}) has been developed in \cite{Yaghoobi2015}, which is the implementation we adopt here.
For the choice of $\varepsilon$ in Eqn. \ref{eq:P0}, we can use CV as described in Sec. \ref{subsubsec:nnlasso}. 

\subsubsection{Sparse Bayesian Learning (\textsc{Sbl})}
Sparse Bayesian Learning (\textsc{Sbl}) \cite{Tipping2001,Wipf2004} is a non-convex optimization algorithm based on Expectation-Maximization (EM) that has empirically shown superior reconstruction performance  to most other CS algorithms with manageable computation cost~\cite{Marques2019}. In \textsc{Sbl}, we consider the case of Gaussian noise in $\boldsymbol{y}$ and a Gaussian prior on elements of $\boldsymbol{x}$, leading to:
\begin{eqnarray}
p(\boldsymbol{y}|\boldsymbol{x}) = \dfrac{\exp(-\|\boldsymbol{y}-\boldsymbol{Ax}\|^2_2/(2\sigma^2))}{(2\pi\sigma^2)^{n/2}} \\
p(x_i;\varphi_i) = \dfrac{\exp(-x^2_i/(2\varphi_i))}{\sqrt{2\pi\varphi_i}}; \varphi_i \geq 0. 
\end{eqnarray}
Since both $\boldsymbol{x}$ and $\boldsymbol{\varphi}$ (the vector of the 
$\{\varphi_i\}_{i=1}^n$ values) are unknown, the optimization for these
quantities can be performed using an EM algorithm. In the following, we shall
denote $\boldsymbol{\Phi} := \textrm{diag}(\boldsymbol{\varphi})$.
Moreover, we shall use the notation $\boldsymbol{\Phi}^{(l)}$ for the
estimate of $\boldsymbol{\Phi}$ in the $l^{\textrm{th}}$ iteration. The E-step of the EM algorithm here involves computing $Q(\boldsymbol{\Phi}|\boldsymbol{\Phi}^{(l)}) := E_{\boldsymbol{x}|\boldsymbol{y};\boldsymbol{\Phi}^{(l)}} \log p(\boldsymbol{y},\boldsymbol{x};\boldsymbol{\Phi})$. It is to be noted that the posterior distribution $p(\boldsymbol{x}|\boldsymbol{y};\boldsymbol{\Phi}^{(l)})$ has the form $\mathcal{N}(\boldsymbol{\mu},\boldsymbol{\Sigma})$ where $\boldsymbol{\mu} := \boldsymbol{\Sigma} \boldsymbol{A}^T \boldsymbol{y} / \sigma^2$ and $\boldsymbol{\Sigma} := (\boldsymbol{A}^T \boldsymbol{A}/\sigma^2 + (\boldsymbol{\Phi}^{(l)})^{-1})^{-1}$. The M-step involves maximization of $Q(\boldsymbol{\Phi}|\boldsymbol{\Phi}^{(l)})$,
leading to the update $\boldsymbol{\Phi}^{(l+1)} = \textrm{diag}(\mu^2_i + \Sigma_{ii})$. The E-step and M-step are executed alternately until convergence. Convergence to a fixed-point is guaranteed, though the fixed point may or may not be a local minimum. However, all local minima are guaranteed to produce sparse solutions for $\boldsymbol{x}$ (even in the presence of noise) because most of the $\varphi_i$ values shrink towards 0. The \textsc{Sbl} procedure can also be modified to dynamically update the noise variance $\sigma^2$ (as followed in this paper), if it is unknown. All these results can be found in \cite{Wipf2004}. Unlike \textsc{Nnlasso} or \textsc{Nnomp}, the \textsc{Sbl} algorithm from \cite{Wipf2004} expressly requires Gaussian noise. However we use it as is in this paper for the simplicity it affords. Unlike \textsc{Nnomp} or \textsc{Nnlasso}, there is no explicit non-negativity constraint imposed in the basic \textsc{Sbl} algorithm. In our implementation, the non-negativity is simply imposed at the end of the optimization by setting to 0 any negative-valued elements in $\boldsymbol{\mu}$, though more principled, albeit more computationally heavy, approaches such as \cite{Nalci2018} can be adopted. 

\subsubsection{Non-negative Absolute Deviation Regression (\textsc{Nnlad})}
\label{subsubsec:nnlad}
The Non-Negative Absolute Deviation Regression (\textsc{Nnlad}) \cite{petersen2020efficient} and Non-negative Least squares (\textsc{Nnls}) \cite{kueng2017robust} seek to respectively minimize
\begin{eqnarray}
    J_{nnlad}(\boldsymbol x;\boldsymbol y,A) := \|\boldsymbol{y} - \boldsymbol A \boldsymbol{x} \|_1 \textrm{ s.t. } \boldsymbol{x} \geq \boldsymbol{0}, \\
    J_{nnls}(\boldsymbol x;\boldsymbol y,A) := \|\boldsymbol{y} - \boldsymbol A \boldsymbol{x} \|_2 \textrm{ s.t. } \boldsymbol{x} \geq \boldsymbol{0}.
\end{eqnarray}
It has been shown in \cite{petersen2020efficient} that \textsc{Nnlad} is sparsity promoting for certain conditions on the sensing matrix $\boldsymbol{A}$, and that its minimizer $\boldsymbol{x}^*$ obeys bounds of the form $||\boldsymbol{x} - \boldsymbol{x}^*||_1 \le C ||\boldsymbol{\eta}||_1$, where $C$ is a constant independent of $\boldsymbol{x}, \boldsymbol{x}^*, \boldsymbol{\eta}, \boldsymbol{y}$. A salient feature of \textsc{Nnlad}/\textsc{Nnls} is that they do not require any parameter tuning. This property makes them useful for matrices of smaller size where cross-validation may be unreliable. 
\subsection{Generalized Binary Search Techniques}
There exist \textit{adaptive group testing} techniques which can determine $k$ infected samples in $O(k\log n)$ tests via repeated binary search. These techniques are impractical in our setting due to their sequential nature and large pool sizes. We provide details of these techniques in the supplemental section \ref*{sup-sec:binary_search}. We also compare with a two-stage approach called Dorfman's method \cite{Dorfman1943} in Sec. \ref{subsubsec:dorfman}.

\subsection{Sensing Matrix Design}
\label{subsec:matrix_design}
\subsubsection{Physical Requirements of the Sensing Matrix}
The sensing matrix $\boldsymbol{A}$ must obey some properties specific to this application such as being non-negative. For ease and speed of pipetting, it is desirable that the entries of $\boldsymbol{A}$ be (1) binary (where $A_{ji} = 0$ indicates that sample $i$ did not contribute to pool $j$, and $A_{ji} = 1$ indicates that a fixed volume of sample $i$ was pipetted into pool $j$), and (2) sparse. Sparsity  ensures that not too many samples contribute to a pool, and that a single sample does not contribute to too many pools. The former is important because typically the volume of sample that is added in a PCR reaction is fixed. Increasing pool size means each sample contributes a smaller fraction of that volume. This leads to dilution which manifests as a shift of the $C_t$ value towards larger numbers. If care is not taken in this regard, this can affect the power of PCR to discriminate between positive and negative samples. The latter is important because contribution of one sample to a large number of pools could lead to depletion of sample.  

\subsubsection{RIP-1 of Expander Graph Adjacency Matrices}
The Restricted Isometry Property (RIP-2) of sensing matrices is a sufficient condition for good CS recovery as described in Sec. \ref{subsec:cs_gt}. However the matrices which obey the aforementioned physical constraints are not guaranteed to obey RIP-2.
Instead, we consider sensing matrices which are adjacency matrices of \textbf{expander graphs}. A left-regular bipartite graph $G((\mathcal{V_I},\mathcal{V_O}), \mathcal{E} \subseteq \mathcal{V_I} \times \mathcal{V_O})$ with degree of each vertex in $\mathcal{V_I}$ being $d$, is said to be a $(k,\epsilon)$-\textbf{unbalanced expander graph} for some integer $k > 0$ and some real-valued $\epsilon \in (0,1)$, if for every subset $\mathcal{S} \subseteq \mathcal{V_I}$ with $|\mathcal{S}| \leq k$, we have $|N(\mathcal{S})| \geq (1-\epsilon)d |\mathcal{S}|$. 
Here $N(\mathcal{S})$ denotes the union set of neighbors of all nodes in $\mathcal{S}$.
Intuitively a bipartite graph is an expander if every `not too large' subset has a `large' boundary.
It can be proved that a randomly generated left-regular bipartite graph with $|\mathcal{V_O}| \geq O(k \log n)$, $n = |\mathcal{V_I}|$ is an expander, with high probability \cite{Lotfi2018,Raginsky2011}.
Moreover, it has been shown in \cite[Thm. 1]{Berinde2008} that the scaled adjacency matrix $\boldsymbol{A}/d$ of a $(k,\epsilon)$-unbalanced expander graph obeys RIP-1 (\cref{def:rip-1}) of order $k$. Here columns of $\A$ correspond to vertices in $\mathcal {V_I}$, and rows correspond to vertices in $\mathcal{V_O}$.
That is, for any $k$-sparse vector $\boldsymbol{x}$, the following relationship holds: $\|\boldsymbol{x}\|_1 \leq \|\boldsymbol{Ax}\|_1/d \leq (1+C\epsilon)\|\boldsymbol{x}\|_1$ for some absolute constant $C > 1$.
This property again implies that the null-space of  $\boldsymbol{A}$ cannot contain vectors that are `too sparse' (apart from the zero-vector). This summarizes the motivation behind the use of expanders in compressive recovery of sparse vectors, and also in group testing \cite{Berinde2008}. 

\subsubsection{Matrices derived from Kirkman Triple Systems}
\label{subsubsec:kirkman}
Although randomly generated left-regular bipartite graphs are expanders, we would need to verify whether a \emph{particular} such graph is a good expander, which may take prohibitively long in practice \cite{Lotfi2018}. In the application at hand, this can prove to be a critical limitation since matrices of various sizes may have to be served, depending on the number of samples arriving in that batch at the testing centre, and the number of tests available to be performed. Hence, we have chosen to employ deterministic procedures to design such matrices, based on objects from combinatorial design theory known as \textbf{Kirkman triples} (see \cite{Kirkman, ray1971solution}). 

\begin{figure}[t]
    \centering
    \includegraphics[width=8.5cm]{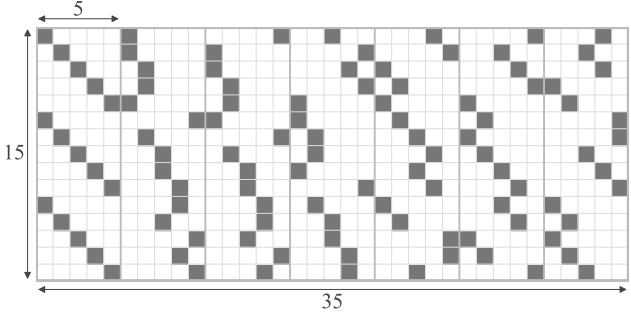}
    \caption{A full Kirkman matrix with $m=15$ rows and $n={m \choose 2}/3 = 35$ columns. Each cell denotes an entry of the matrix, with white cells denoting the location of a $0$ entry and the greyed out cells indicating the location of a $1$ entry. Each column has exactly $3$ entries with value $1$. Each row has $7$ entries with value $1$. There are $(m-1)/2 = 7$ groups of columns, each consisting of $m/3 = 5$ columns. Each row in a column group has exactly one $1$ entry. Matrices of size $15\times 20$, $15\times 25$, $15\times 30$ or $15\times 35$ may be served by choosing the first $4$, $5$, $6$, or $7$ column groups, while keeping the number of $1$ entries in each row equal.} \label{fig:kirkman_15_35}
\end{figure}

We first recall Kirkman Triple Systems (an example of which is illustrated in Fig. \ref{fig:kirkman_15_35}) which are Steiner Triple Systems with an extra property. Steiner Triple Systems consist of $n = {m \choose 2}/3$ column vectors with $m$ elements each, with each entry being either $0$ or $1$ such that each column has exactly three $1$s, every pair of rows has dot product equal to $1$ and every pair of columns has dot product at most $1$ \cite{Steiner}.
This means that each column of a Steiner Triple System corresponds to a triplet of rows (i.e. contains exactly three 1s), and every pair of rows occurs together in exactly one such triplet (i.e. for every pair of rows indexed by $i,j$, there exists exactly one column index $k$ for which $A_{ik} = A_{jk} = 1$).
If the columns of a Steiner Triple System can be arranged such that the sum of columns from $i$ to $i+m/3 - 1$ equals $\boldsymbol{1} \in\mathbb{R}^m$ for every $i \equiv 1$ modulo $m/3$ then the Steiner Triple System is said to be \textbf{resolvable}, and is known as a \textbf{Kirkman Triple System} \cite{Kirkman}.
That is, the set of columns of a Kirkman Triple System can be partitioned into $(m-1)/2$ disjoint groups, each consisting of $m/3$ columns, such that each row has exactly one $1$ entry in a given such group of columns.
Because of this property, we may choose any $l$ such groups of columns of a Kirkman Triple System to form a $m\times n$ matrix, $n > m$, with $n = lm/3$, and $3 < l \leq (m-1)/2$, while keeping the number of $1$ entries in each row the same. From here on, we refer to such matrices as \textbf{Kirkman matrices}. If $l = (m-1)/2$, then we refer to it as a \textbf{full} Kirkman matrix, else it is referred to as a \textbf{partial} Kirkman matrix.
Note that in a partial Kirkman matrix, the dot product of any two rows may be at most $1$, whereas in a full Kirkman matrix, it must be equal to $1$.

Notice that $m = 6t+3$ for some $t \in Z_{\geq 0}$ for a Kirkman Triple System to exist, since $m-1$ must be divisible by $2$, and $m$ must be divisible by $3$. This, and the existence of Kirkman Triple Systems for all $t \in Z_{\geq0}$ have been proven in \cite{ray1971solution}. Explicit constructions of Kirkman Triple Systems for $m \leq 99$ exist \cite{Kirkman}.
Generalizations of Kirkman Triple Systems under the name of the Social Golfer Problem is an active area of research (see \cite{SocialGolfer, SocialGolferMathPuzzle}).
The Social Golfer Problem asks if it is possible for $g\times p$ golfers to play in $g$ groups of $p$ players each for $w$ weeks, such that no two golfers play in the same group more than once \cite[Sec. 1.1]{triska2008solution}. Kirkman Triple Systems with $m$ rows and ${m\choose 2}/3$ columns are a solution to the Social Golfer Problem for the case when $p = 3$, $g = m/3$ and $w=(m-1)/2$.
Full or partial Kirkman matrices may be constructed via greedy search techniques used for solving the Social Golfer Problem (such as in \cite{dotu2005scheduling}).
Previously, Kirkman matrices have been proposed for use as Low-Density Parity Check codes in \cite{johnson2001construction}, due to high girth\footnote{The girth of a graph is equal to the length of the shortest cycle in it.} of Kirkman matrix bipartite graphs and the ability to serve only part of the matrix while keeping the row weights\footnote{defined as the number of $1$ entries in a row } equal. Matrices derived from Steiner Triple Systems have previously been used for pooled testing for transcription regulatory network mapping in \cite{vermeirssen2007matrix}. Further, matrices derived from Steiner Systems \cite{SteinerSystem}, a generalization of Steiner Triple Systems, have been proposed for optimizing $2$-stage binary group testing in \cite{tonchev2008steiner}.

\subsubsection{RIP-1 and Expansion Properties of Kirkman Matrices}
\label{subsubsec:kirkman_rip_1}
We show that Kirkman matrix bipartite graphs are $(k, \epsilon)$-unbalanced expanders, with $\epsilon = (k-1)/2d$, where $d$ is the left-degree of the graph and is $3$ for Kirkman matrices.
Given a set $S$ of column vertices such that $|S| \leq k$, we note that the size of the union set of neighbours of $S$, $|N(S)|$, is at least $|S|d - pr$, where $p = {|S| \choose 2}$ is the number of (unordered) pairs of columns in $S$, and $r$ is the maximum number of row vertices in common between any two column vertices. For a Kirkman matrix, since any two columns have dot product at most $1$, hence $r = 1$. Therefore, $|N(S)| \geq d|S|(1 - (|S| -1)/2d)$. Since $|S| \leq k$, therefore $|N(S)| \geq d|S|(1-(k-1)/2d)$. This implies that Kirkman matrix biparite graphs are $(k, \epsilon)$-unbalanced expanders, with $\epsilon = (k-1)/2d$. If we put in the requirement that $d = 3$ for Kirkman matrices and $\epsilon < 1$, we find that $k < 7$. 
Hence it follows from \cite[Thm. 1]{Berinde2008} that the scaled Kirkman matrix has RIP-1 of order $k$ for $k < 7$ and $\epsilon = (k-1)/6$. This suggests exact recovery for upto $3$ infected samples using CS. However, in practice, we observe that using our method we are able to recover much higher number of positives, at the cost of an acceptable number of false positives and rare false negatives (Sec. \ref{sec:results}).

\subsubsection{Optimality of Girth 6 Matrices}
\label{subsub:girth_6}
A Steiner Triple System bipartite graph does not have a cycle of length $4$.
If it did, then there would exist two rows $a$ and $b$, and two columns $u$ and $v$ of the Steiner Triple System matrix $\A$ such that $A_{au} = A_{bu} = 1$ and $A_{av} = A_{bv} = 1$.
This would violate the property that dot product of any two rows of the Steiner Triple System must be equal to $1$.
Furthermore, \cite[Lemma 1]{johnson2001construction} show that Steiner Triple System bipartite graphs have girth equal to $6$. Since Kirkman Triple Systems are resolvable Steiner Triple Systems (see definitions earlier in this section), their bipartite graphs also have girth equal to $6$. For a bipartite graph constructed from a partial Kirkman matrix, the girth is at least $6$, since dropping some column vertices will not introduce new cycles in the graph. Furthermore, it is shown in \cite[Thm. 10]{lotfi2020compressed} that adjacency matrices of left-regular graphs with girth at least 6 satisfy RNSP (\cref{def:rnsp}) of order $k$ (for suitable $k$).
Consequently, they may be used for CS decoding \cite[Thm. 5]{lotfi2020compressed}.
They also give lower bounds on the number of rows $m$ of left-regular bipartite graph matrices whose column weight\footnote{defined as the number of 1 entries in a column} is more than $2$, for them to have high girth and consequently satisfy RNSP of order $k$, given $k$ and $n$ \cite[Eqn. 32, 33]{lotfi2020compressed}.
Given $k$ and $n$, these lower bounds are minimized for graphs of girth 6 and 8, and the bounds are, respectively, $m\geq k\sqrt{n}$ and $m \geq k^{3/2}\sqrt{n}$ (\cite[Eqn. 37]{lotfi2020compressed}).
However, with the additional requirement that $m < n$ for CS, it is found that girth 6 matrices can recover $k < \sqrt n$ defects, while girth 8 matrices can only recover $k < \sqrt[3]{n}$ defects. Hence, matrices whose bipartite graphs have girth equal to 6 are optimal in this sense. Full Kirkman matrix bipartite graphs are left-regular and have girth 6, as argued earlier, and hence they satisfy RNSP, may be used for compressive sensing, and are optimal in the sense of being able to handle most number of defects while minimizing the number of measurements. {We note that since we employ Kirkman triples, each column has only three 1s. The theoretical guarantees for such matrices hold for signals with $\ell_0$ norm less than or equal to 2. However, we have obtained acceptable false positive and false negative rates in practice for much larger sparsity levels, as will be seen in Sec. \ref{sec:results}.}

\subsubsection{Disjunctness Property of Kirkman Matrices}
\label{subsub:disjunctness}

In order for a matrix to be suitable for our method, it should not only be good for CS decoding algorithms, but also for \textsc{Comp}. Kirkman matrices are $2$-disjunct, and can recover up to $2$ defects exactly using \textsc{Comp}. 
In a $k$-disjunct matrix, there does not exist any column such that its support is a subset of the union of the support of $k$ other columns \cite{aldridge2014group}. Matrices which are $k$-disjunct have exact support recovery guarantee for $k$-sparse vectors, using \textsc{Comp} (see \cite{aldridge2014group}).
Disjunctness follows from the following properties of Kirkman matrices -- that two columns in a Kirkman matrix have at most one row in common with an entry of $1$, and that each column has exactly three $1$ entries.
Consider $R_a$, $R_b$, and $R_c$, the sets of rows for which the three columns $a$, $b$ and $c$ respectively have a $1$ entry. Note that $|R_a| = |R_b| = |R_c| = 3$, and $|R_p \cap R_q| \leq 1$ for $p, q \in \{a, b, c\}, p\neq q$. If $R_c \subseteq R_{a} \cup R_b$, then either $|R_c \cap R_a| > 1$ or $|R_c \cap R_b| > 1$, which presents a contradiction. 

Empirically we find that even for $k > 2$, \textsc{Comp} reports only a small fraction of the total number of samples as positives when using Kirkman matrices (Table \ref{tab:comp}). In Sec. \ref*{sup-sec:kirkman_f} (Proposition 6) of the supplemental material, we prove that \emph{if a fraction $f \in (0, 1)$ of the tests come out to be positive, then \textsc{Comp} reports strictly less than fraction $f^2$ of the samples as positive for a full Kirkman matrix}. This provides intuition behind why Kirkman matrices may be well-suited for our combined \textsc{Comp} + CS method, since most samples are already eliminated by \textsc{Comp}.
On the other hand, CS decoding (without the earlier \textsc{Comp} step) on the full Kirkman matrix does not perform as well, as shown in the supplemental section \ref*{sup-sec:cs_algos_only}.

\subsubsection{Advantages of using Kirkman Matrices}
\label{subsub:advantages}
As we have seen in earlier sections, Kirkman matrices are suitable for use in compressed sensing due to their expansion, RIP-1 and high girth properties, as well as for binary group testing due to disjunctness.
Furthermore, the dot product between two columns of a Kirkman matrix being at most $1$ ensures that no two samples participate in more than one test together. 
This has favourable consequences in terms of placing an upper bound on the \textbf{mutual coherence} of the matrix, defined as:
\begin{equation}
\mu(\boldsymbol{A}) := \textrm{max}_{i \neq j} \dfrac{|\boldsymbol{A_i}^t \boldsymbol{A_j}|}{\|\boldsymbol{A_i}\|_2 \|\boldsymbol{A_j}\|_2},
\label{eq:mu_mc}
\end{equation}
where $\boldsymbol{A_i}$ refers to the $i^{\textrm{th}}$ column of $\boldsymbol{A}$. Matrices with lower $\mu(\boldsymbol{A})$ values have lower values of worst case upper bounds on the reconstruction error \cite{Studer2014}. These bounds are looser than those based on the RIC that we saw in previous sections. However, unlike the RIC, the mutual coherence is efficiently computable.

A practical benefit of Kirkman triples that is not shared by Steiner triples is that the former can be served for number of samples far less than $n={m \choose 2}/3$ while keeping pools balanced (i.e. ensuring that each pool is created from the same number of samples).
In fact, we can choose $n$ to be any integer multiple of $m/3$, and ensure that every pool gets the same number of samples, as discussed in section \ref{subsubsec:kirkman}.
Notice that the expansion, RIP-1, high girth and disjunctness properties hold for full as well as partial Kirkman matrices, as proven in previous sections.
This allows us to characterize the properties of the full Kirkman matrix, and use that analysis to predict how it will behave in the clinical situation where the pooling matrix to be served may require very specific values of $m, n$ depending on the prevalence rate.

\textbf{Column weight:} Kirkman matrices have column weight equal to $3$ - that is, each sample goes to $3$ pools. It is possible to construct matrices with higher number of pools per sample (such as those derived from the Social Golfer Problem \cite{SocialGolfer}, which will retain several benefits of the Kirman matrices: (1) They would have the ability to serve only part of the matrix; (2) They would retain the the expander and RIP-1 properties, following a proof similar to the one in  Sec. \ref{subsubsec:kirkman_rip_1}; (3) They would not have any $4$-cycles in the corresponding bipartite graph, following a similar argument as in Sec. \ref{subsub:girth_6}; and (4) They would possess the disjunctness property following a proof similar to the one in Sec. \ref{subsub:disjunctness}). Nevertheless, the time and effort needed for pooling increases with more pools per sample. Further, higher pools per sample will come at the cost of a larger number of tests (if pool size is kept constant), or larger pool size (if number of tests is kept constant). Higher number of tests is undesirable for obvious reasons, while larger pool size may lead to dilution of the sample within a pool, leading to individual RT-PCR tests failing.

\subsubsection{Optimal Binary Sensing Matrices with Random Construction}
\label{subsub:optimal_sensing_matrices}

While Kirkman matrices which satisfy RNSP of order $k$ must have at least $k\sqrt{n}$ measurements, we can get much better bounds in theory if we use random constructions.
From \cite[Prop. 10]{kueng2017robust} we see that with high probability, $0/1$ Bernoulli($p$) matrices need only $O(k\log n)$ measurements in order to satisfy $\ell_2$-RNSP (\cref{def:l2_rnsp}) of order $k$, with $p\in (0, 1)$ being the probability with which each entry of the matrix is independently $1$.

In the supplemental section \ref*{sup-sec:comp_ric}, we prove that $\ell_2$-RNSP is preserved by \textsc{Comp}. That is, the reduced matrix $\redA$ obeys $\ell_2$-RNSP of order $k$ with the same parameters as the original matrix $\A$. Hence our method only needs $O(k\log n)$ measurements for robust recovery of $k$-sparse vectors with such random matrix constructions. Bernoulli($p$) matrices are also good for \textsc{Comp} -- \cite[Thm. 4]{Chan2011} shows that Bernoulli($p$) matrices with $p=1/k$ need only $O(k \log n)$ measurements for exact support recovery of $k$-sparse vectors with \textsc{Comp} with vanishingly small probability of error.

In practice, we observe that Kirkman matrices perform better than Bernoulli($p$) matrices using our method in the regime of our problem size. This gap between theory and practice may be arising due to the following reasons:
(1) The $k\sqrt{n}$ lower bound for Kirkman triples is for a sufficient but not necessary condition for sparse recovery;
(2) The $O(k \log n)$ may be ignoring a very large constant factor which affects the performance of moderately-sized problems such as the ones reported in this paper; and
(3) The theoretical bounds are for exact recovery with vanishingly small error, whereas we allow some false positives and rare false negatives in our experiments.
Similar comparisons between binary and Gaussian random matrices have been recently put forth in \cite{lotfi2020compressed}. Moreover, the average column weight of Bernoulli($p$) matrices is $pm$, where $m$ is the number of measurements. This is typically much higher than column weight $3$ of Kirkman matrices and hence undesirable (see Sec. \ref{subsub:advantages}). In the supplemental section \ref*{sup-sec:sensing_matrix}, we compare the performance of Kirkman matrices with Bernoulli($0.1$) and Bernoulli($0.5$) matrices.

\subsubsection{Mutual Coherence optimized Sensing Matrices}
\label{subsub:mutual_coherence}
As mentioned earlier, the mutual coherence from Eqn. \ref{eq:mu_mc} is efficient to compute and optimize over. Hence, there is a large body of literature on designing CS matrices by minimizing $\mu(\boldsymbol{A})$ w.r.t. $\boldsymbol{A}$, for example \cite{Abdoghasemi2010}. We followed such a procedure for designing sensing matrices for some of our experimental results in Sec. \ref{subsec:real_results}. For this, we follow simulated annealing to update the entries of $\boldsymbol{A}$, starting with an initial condition where $\boldsymbol{A}$ is a random binary matrix. For synthetic experiments, we compared such matrices with Bernoulli($p$) random matrices, adjacency matrices of biregular random sparse graphs (i.e. matrices in which each column has the same weight, and each row has the same weight - which may be different than the column weight), and Kirkman matrices. We found that matrices of Kirkman triples perform very well empirically in the regime of sizes we are interested in, besides facilitating easy pipetting, and hence the results are reported using only Kirkman matrices.

\section{Experimental Results}
\label{sec:results}
In this section, we show a suite of experimental results on synthetic data as well as on real data. 

\subsection{Results on Synthetic Data}
\subsubsection{Choice of Sensing Matrix}
Recall from section \ref{sec:rtpcr} that a typical RT-PCR setup can test 96 samples in parallel.
Three of these tests are used as control by the RT-PCR technician in order to have confidence that the RT-PCR process has worked.
Hence, in order to optimize the available test bandwidth of the RT-PCR setup, the number of tests we perform in parallel should be $\leq 93$, and as close to $93$ as possible. Since in Kirkman matrices, the number of rows must be $6t+3$ for some $t \in \mathbb{Z}_{\geq 0}$, hence we choose $93$.
With this choice, the number of samples tested $n$ has to be a multiple of $93/3 = 31$, hence we chose $n = 961$.
This matrix is not a full Kirkman matrix -- a full matrix with $93$ rows will have $1426$ columns. However, we keep the number of columns of the matrix under $1000$ due to challenges in pooling large number of samples.
Furthermore, $n=961$, $m = 93$ satisfies more than $10$x factor improvement in testing while detecting $1\%$ infected samples with reasonable sensitivity and specificity and is in a regime of interest for widespread screening or repeated testing.

We also present results with a $45\times 105$ partial Kirkman matrix in the supplemental Sec. \ref{sup-sec:45_105_results}. This matrix gives $2.3$x improvement in testing while detecting $9.5\%$ infected samples with reasonable sensitivity and specificity. Further, two such batches of $105$ tests in $45$ pools may be run in parallel in a single RT-PCR setup.

\label{subsec:synthetic_results}
\subsubsection{Signal/Measurement Generation} For the case of synthetic data, we generated $k$-sparse signal vectors $\boldsymbol{x}$ of dimension $n = 961$, for each $k$ in $\{5, 8, 10, 12, 15, 17, 20\}$. We choose a wide range of $k$ in order to demonstrate that not only do our algorithms have high sensitivity and specificity for large values of $k$, they also keep performing reasonably, well beyond the typical operating regime. The support of each signal vector $\boldsymbol x$ -- given $k$ -- was chosen by sampling a $k$-sparse binary vector uniformly at random from the set of all $k$-sparse binary vectors. The magnitudes of the non-zero elements of $\boldsymbol{x}$ were picked uniformly at random from the range $[1,32768]$. This high dynamic range in the value of $\boldsymbol{x}$ was chosen to reflect a variance in the typical threshold cycle values ($C_t$) of real PCR experiments, which can be between $16$ and $32$. From Eqn. \ref{eq:cycle_threshold}, we can infer that viral loads vary roughly as $2^{-C_t}$ (setting $ q = 1$), up to constant multiplicative terms. 
In all cases, $m=93$ noisy measurements in $\boldsymbol{y}$ were simulated following the noise model in Eqn. \ref{eq:noisemodel} with $\sigma = 0.1$ and $q = 0.95$.
A $93 \times 961$ Kirkman sensing matrix was used for generating the measurements.
The Poisson nature of the elements of $\boldsymbol{x}$ in Eqn. \ref{eq:noisemodel} was ignored. This approximation was based on the principle that if $X \sim \textrm{Poisson}(\lambda)$, then $\textrm{Std. Dev.}(X)/E(X) = \sqrt{\lambda}/\lambda = 1/\sqrt{\lambda}$ which becomes smaller and smaller as $\lambda$ increases. 
The recovery algorithms were tested on $Q = 1000$ randomly generated signals for each value of $k$. 

\subsubsection{Algorithms tested} The following algorithms were compared:
\begin{enumerate}[leftmargin=0.4cm]
    \item \textsc{Comp} (see Table \ref{tab:comp})
    \item \textsc{Comp} followed by \textsc{Nnlasso} (see Table \ref{tab:comp_nnlasso})
    \item \textsc{Comp} followed by \textsc{Sbl} (see Table \ref{tab:comp_sbl})
    \item \textsc{Comp} followed by \textsc{Nnomp} (see Table \ref{tab:comp_nnomp})
    \item \textsc{Comp} followed by \textsc{Nnlad} (see Table \ref{tab:comp_nnlad})
    \item \textsc{Comp} followed by \textsc{Nnls} (see Table \ref*{tab:comp_nnls} in the Supplementary)
\end{enumerate}
For each algorithm any positives missed during the CS stage but caught by \textsc{Dd} were declared as positives, as mentioned in Sec. \ref{sec:cs_gt_combined}. For small sample sizes we also tested \textsc{Comp}-\textsc{Bf}, i.e. \textsc{Comp} followed by brute-force search for samples in $\boldsymbol{x}$ with non-zero values. Details of this algorithm and experimental results with it are presented in the supplemental section \ref*{sup-sec:comp_bf}.
\subsubsection{Comparison Criteria}\label{subsec:comparison_criteria} In the following, $\boldsymbol{\hat{x}}$ denotes the estimate of $\boldsymbol{x}$. Most numerical algorithms do not produce vectors that are exactly sparse and have many entries with very tiny magnitude, due to issues such as choice of convergence criterion. Since in this application, support recovery is of paramount importance to identify which samples in $\boldsymbol{x}$ were infected, we employed the following post-processing step: All entries in $\boldsymbol{\hat{x}}$ whose magnitude fell below a threshold $\tau := 0.2 \times x_{min}$ were set to zero, yielding a vector $\boldsymbol{\bar{x}}$. Here $x_{min}$ refers to the least possible value of the viral load, and this can be obtained offline from practical experiments on individual samples. In these synthetic experiments, we simply set $x_{min} := 1$. We observed that varying the value of $\tau$ over a fairly wide range had negligible impact on the results, as can be observed from Sec. \ref*{sup-sec:tau} of the supplemental material. For \textsc{Sbl}, we set $\tau$ to $0$ and also set also negative entries in the estimate to 0. For \textsc{Nnomp}, such thresholding was inherently not needed. 
The various algorithms were compared with respect to the following criteria:
\begin{enumerate}[leftmargin=0.4cm]
    \item RMSE := $\|\boldsymbol{x}-\boldsymbol{\bar{x}}\|_2/\|\boldsymbol{x}\|_2$
    \item Number of false positives (FP) := $|\{i:x_i=0, \hat{x}_i > 0\}|$
    \item Number of false negatives (FN) := $|\{i:x_i > 0, \hat{x}_i = 0\}|$
    \item Sensitivity (also called Recall or True Positive rate) := $\#$correctly detected positives/$\#$actual positives
    \item Specificity (also called True Negative Rate) := $\#$correctly detected negatives/$\#$actual negatives.
\end{enumerate}

\subsubsection{Main Results}

\begin{table*}
\centering
\caption{Performance of \textsc{Comp} and \textsc{Dd} (on synthetic data) for $93 \times 961$ Kirkman triple matrix. For each criterion and each $k$ value, mean and standard deviation values are reported, across 1000 signals.}
\label{tab:comp}
\begin{tabular}{|c|c|c|c|c|c|c|}
\hline
$k$ & \textbf{RMSE} & \textbf{\#FN} & \textbf{\#FP} & \textbf{Sens.} & \textbf{Spec.} & {$\mathbf{\#\mathcal{HCP}}$} \\ \hline
5          & 1.000 $\pm$ 0.000   & 0.0 $\pm$ 0.0       & 1.6 $\pm$ 1.2       & 1.0000 $\pm$ 0.0000  & 0.9983 $\pm$ 0.0013  & $4.7$\\ \hline
8          & 1.000 $\pm$ 0.000   & 0.0 $\pm$ 0.0       & 7.9 $\pm$ 3.0       & 1.0000 $\pm$ 0.0000  & 0.9918 $\pm$ 0.0031  & $4.3$\\ \hline
10         & 1.000 $\pm$ 0.000   & 0.0 $\pm$ 0.0       & 15.3 $\pm$ 4.5      & 1.0000 $\pm$ 0.0000  & 0.9839 $\pm$ 0.0048  & $2.5$\\ \hline
12         & 1.000 $\pm$ 0.000   & 0.0 $\pm$ 0.0       & 25.3 $\pm$ 6.7      & 1.0000 $\pm$ 0.0000  & 0.9733 $\pm$ 0.0070  & $1.1$\\ \hline
15         & 1.000 $\pm$ 0.000   & 0.0 $\pm$ 0.0       & 46.1 $\pm$ 10.3     & 1.0000 $\pm$ 0.0000  & 0.9512 $\pm$ 0.0109  & $0.2$\\ \hline
17         & 1.000 $\pm$ 0.000   & 0.0 $\pm$ 0.0       & 62.3 $\pm$ 13.7     & 1.0000 $\pm$ 0.0000  & 0.9340 $\pm$ 0.0146  & $0.1$\\ \hline
20         & 1.000 $\pm$ 0.000   & 0.0 $\pm$ 0.0       & 91.5 $\pm$ 18.1     & 1.0000 $\pm$ 0.0000  & 0.9028 $\pm$ 0.0192  & $0.0$\\ \hline
\end{tabular}

\end{table*}

\begin{table*}
\centering
\caption{Performance of Comp followed by Nnlasso (on synthetic data) for $93 \times 961$ Kirkman triple matrix. For each criterion and each $k$ value, mean and standard deviation values are reported, across 1000 signals.}
\label{tab:comp_nnlasso}
\begin{tabular}{|c|c|c|c|c|c|}
\hline
\textbf{k}    & \textbf{RMSE}       & \textbf{\#FN}   & \textbf{\#FP}     & \textbf{Sens.}        & \textbf{Spec.}        \\ \hline
\textbf{$5$}  & $ 0.047 \pm 0.020 $ & $ 0.0 \pm 0.1 $ & $ 0.8 \pm 0.9 $   & $ 0.9990 \pm 0.0141 $ & $ 0.9991 \pm 0.0009 $ \\ \hline
\textbf{$8$}  & $ 0.069 \pm 0.028 $ & $ 0.1 \pm 0.2 $ & $ 4.0 \pm 2.1 $   & $ 0.9925 \pm 0.0307 $ & $ 0.9958 \pm 0.0022 $ \\ \hline
\textbf{$10$} & $ 0.100 \pm 0.049 $ & $ 0.2 \pm 0.5 $ & $ 7.9 \pm 3.4 $   & $ 0.9780 \pm 0.0454 $ & $ 0.9917 \pm 0.0035 $ \\ \hline
\textbf{$12$} & $ 0.149 \pm 0.092 $ & $ 0.6 \pm 0.7 $ & $ 12.9 \pm 5.0 $  & $ 0.9538 \pm 0.0591 $ & $ 0.9864 \pm 0.0052 $ \\ \hline
\textbf{$15$} & $ 0.295 \pm 0.166 $ & $ 1.0 \pm 1.1 $ & $ 28.5 \pm 16.5 $ & $ 0.9316 \pm 0.0722 $ & $ 0.9699 \pm 0.0175 $ \\ \hline
\textbf{$17$} & $ 0.404 \pm 0.184 $ & $ 1.1 \pm 1.4 $ & $ 46.3 \pm 23.5 $ & $ 0.9355 \pm 0.0817 $ & $ 0.9509 \pm 0.0249 $ \\ \hline
\textbf{$20$} & $ 0.563 \pm 0.172 $ & $ 1.1 \pm 1.8 $ & $ 78.4 \pm 26.7 $ & $ 0.9452 \pm 0.0923 $ & $ 0.9167 \pm 0.0284 $ \\ \hline
\end{tabular}
\end{table*}

\begin{table*}
\centering
\caption{Performance of Comp followed by Sbl (on synthetic data) for $93 \times 961$ Kirkman triple matrix. For each criterion and each $k$ value, mean and standard deviation values are reported, across 1000 signals.}
\label{tab:comp_sbl}
\begin{tabular}{|c|c|c|c|c|c|}
\hline
$k$ & \textbf{RMSE} & \textbf{\#FN} & \textbf{\#FP} & \textbf{Sens.} & \textbf{Spec.}  \\ \hline
5          & 0.043 $\pm$ 0.017   & 0.0 $\pm$ 0.0       & 0.9 $\pm$ 0.9       & 0.9998 $\pm$ 0.0063  & 0.9991 $\pm$ 0.0010  \\ \hline
8          & 0.058 $\pm$ 0.021   & 0.0 $\pm$ 0.2       & 4.3 $\pm$ 2.1       & 0.9958 $\pm$ 0.0227  & 0.9955 $\pm$ 0.0023  \\ \hline
10         & 0.071 $\pm$ 0.025   & 0.1 $\pm$ 0.2       & 8.2 $\pm$ 3.1       & 0.9937 $\pm$ 0.0247  & 0.9913 $\pm$ 0.0033  \\ \hline
12         & 0.094 $\pm$ 0.035   & 0.1 $\pm$ 0.4       & 13.6 $\pm$ 4.4      & 0.9886 $\pm$ 0.0310  & 0.9856 $\pm$ 0.0046  \\ \hline
15         & 0.123 $\pm$ 0.108   & 0.3 $\pm$ 0.6       & 25.1 $\pm$ 6.9      & 0.9804 $\pm$ 0.0396  & 0.9735 $\pm$ 0.0073  \\ \hline
17         & 0.165 $\pm$ 0.179   & 0.5 $\pm$ 0.8       & 35.1 $\pm$ 9.9      & 0.9713 $\pm$ 0.0491  & 0.9628 $\pm$ 0.0105  \\ \hline
20         & 0.318 $\pm$ 0.305   & 1.3 $\pm$ 1.6       & 54.5 $\pm$ 13.2     & 0.9349 $\pm$ 0.0803  & 0.9420 $\pm$ 0.0140  \\ \hline
\end{tabular}
\end{table*}


\begin{table*}
\centering
\caption{Performance of Comp followed by Nnomp (on synthetic data) for $93 \times 961$ Kirkman triple matrix. For each criterion and each $k$ value, mean and standard deviation values are reported, across 1000 signals.}
\label{tab:comp_nnomp}
\begin{tabular}{|c|c|c|c|c|c|}
\hline
$k$ & \textbf{RMSE} & \textbf{\#FN} & \textbf{\#FP} & \textbf{Sens.} & \textbf{Spec.} \\ \hline
5          & 0.043 $\pm$ 0.019   & 0.0 $\pm$ 0.1       & 0.3 $\pm$ 0.6       & 0.9982 $\pm$ 0.0209  & 0.9997 $\pm$ 0.0006  \\ \hline
8          & 0.060 $\pm$ 0.025   & 0.1 $\pm$ 0.4       & 1.8 $\pm$ 2.0       & 0.9831 $\pm$ 0.0472  & 0.9981 $\pm$ 0.0021  \\ \hline
10         & 0.077 $\pm$ 0.035   & 0.3 $\pm$ 0.5       & 3.7 $\pm$ 3.2       & 0.9739 $\pm$ 0.0541  & 0.9961 $\pm$ 0.0034  \\ \hline
12         & 0.115 $\pm$ 0.067   & 0.5 $\pm$ 0.7       & 7.8 $\pm$ 4.9       & 0.9565 $\pm$ 0.0560  & 0.9918 $\pm$ 0.0051  \\ \hline
15         & 0.242 $\pm$ 0.190   & 1.5 $\pm$ 1.4       & 15.6 $\pm$ 6.0      & 0.9013 $\pm$ 0.0951  & 0.9835 $\pm$ 0.0064  \\ \hline
17         & 0.361 $\pm$ 0.243   & 2.8 $\pm$ 2.2       & 20.8 $\pm$ 5.6      & 0.8329 $\pm$ 0.1268  & 0.9780 $\pm$ 0.0059  \\ \hline
20         & 0.589 $\pm$ 0.282   & 6.1 $\pm$ 3.0       & 27.0 $\pm$ 5.2      & 0.6941 $\pm$ 0.1520  & 0.9713 $\pm$ 0.0055  \\ \hline
\end{tabular}
\end{table*}

\begin{table*}
\centering
\caption{Performance of Comp followed by Nnlad (on synthetic data) for $93 \times 961$ Kirkman triple matrix. For each criterion and each $k$ value, mean and standard deviation values are reported, across 1000 signals.}
\label{tab:comp_nnlad}

\begin{tabular}{|c|c|c|c|c|c|}
\hline
\textbf{k}    & \textbf{RMSE}       & \textbf{\#FN}   & \textbf{\#FP}     & \textbf{Sens.}        & \textbf{Spec.}        \\ \hline
\textbf{$5$}  & $ 0.050 \pm 0.021 $ & $ 0.0 \pm 0.0 $ & $ 1.0 \pm 1.0 $   & $ 0.9996 \pm 0.0089 $ & $ 0.9990 \pm 0.0010 $ \\ \hline
\textbf{$8$}  & $ 0.077 \pm 0.034 $ & $ 0.0 \pm 0.2 $ & $ 4.9 \pm 2.3 $   & $ 0.9939 \pm 0.0270 $ & $ 0.9949 \pm 0.0024 $ \\ \hline
\textbf{$10$} & $ 0.107 \pm 0.050 $ & $ 0.2 \pm 0.4 $ & $ 9.3 \pm 3.0 $   & $ 0.9809 \pm 0.0441 $ & $ 0.9903 \pm 0.0032 $ \\ \hline
\textbf{$12$} & $ 0.167 \pm 0.095 $ & $ 0.5 \pm 0.7 $ & $ 14.4 \pm 5.1 $  & $ 0.9574 \pm 0.0586 $ & $ 0.9848 \pm 0.0054 $ \\ \hline
\textbf{$15$} & $ 0.296 \pm 0.160 $ & $ 0.9 \pm 1.0 $ & $ 29.7 \pm 16.7 $ & $ 0.9393 \pm 0.0687 $ & $ 0.9686 \pm 0.0176 $ \\ \hline
\textbf{$17$} & $ 0.401 \pm 0.171 $ & $ 0.8 \pm 1.2 $ & $ 50.2 \pm 23.9 $ & $ 0.9549 \pm 0.0717 $ & $ 0.9468 \pm 0.0253 $ \\ \hline
\textbf{$20$} & $ 0.530 \pm 0.166 $ & $ 0.2 \pm 0.9 $ & $ 87.9 \pm 25.1 $ & $ 0.9884 \pm 0.0427 $ & $ 0.9066 \pm 0.0267 $ \\ \hline
\end{tabular}

\end{table*}

It should be noted that all algorithms were evaluated on 1000 randomly generated sparse signals, given the same sensing matrix. The average value as well as standard deviation of all quality measures (over the 1000 signals) are reported in the Tables \ref{tab:comp}, \ref{tab:comp_nnlasso}, \ref{tab:comp_sbl}, \ref{tab:comp_nnomp}, \ref{tab:comp_nnlad}, \ref{tab:comp_nnls}. A comparison of Table \ref{tab:comp} to Tables \ref{tab:comp_nnlasso}, \ref{tab:comp_sbl}, \ref{tab:comp_nnomp}, \ref{tab:comp_nnlad}, \ref{tab:comp_nnls} indicates that \textsc{Comp} followed by \textsc{Nnlasso}/\textsc{Sbl}/\textsc{Nnomp}/\textsc{Nnlad}/\textsc{Nnls} significantly reduces the false positives at the cost of a rare false negative. The RMSE is also significantly improved, since \textsc{Comp} does not estimate viral loads. 
At the same time, \textsc{Comp} significantly reduces the size of the problem for the CS stage. For example, for the $93 \times 961$ Kirkman matrix, when number of infected samples $k$ is $12$, the average size of the matrix after \textsc{Comp} filtering is $\sim 30 \times 37$.
From Table \ref{tab:comp} we see that Definite Defectives classifies many positives as high-confidence positives, for $k$ upto $8$.
We note that the experimental results reported in these tables are quite encouraging, since these experiments are challenging due to small $m$ and fairly large $k,n$, albeit with testing on synthetic data. We noticed that running the CS algorithms without the \textsc{Comp} step did not perform as well, results for which are presented in the supplemental section \ref*{sup-sec:cs_algos_only}.
We observed that the advantages of our combined group testing and compressed sensing approach holds regardless of the sensing matrix size. For comparison, results of running our algorithms using a $45 \times 105$ Kirkman matrix instead of the $93 \times 961$ Kirkman matrix are presented in supplemental section \ref*{sup-sec:45_105_results}. 

\subsubsection{Parameter Selection}
\label{subsub:parameters}
As mention earlier, the regularization parameters in various estimators such as \textsc{Comp}-\textsc{Nnlasso}, \textsc{Comp}-\textsc{Nnlad}, \textsc{Comp}-\textsc{Nnomp}, etc. are estimated via cross-validation. For these estimators, we therefore do not require knowledge of the $\sigma$ parameter in the noise model from Eqn. \ref{eq:noisemodel}. The $q$ parameter in the noise model is set to $0.95$ in all our experiments.
It is a reasonable choice as the molecule count is known to roughly double in each cycle of RT-PCR \cite{thermofisher}.
Moreover, variation of $q$ in the range from 0.7 to 1 showed negligible variation in the results of our wet-lab experiments as can be seen in Sec. \ref*{sup-sec:viral_load_q} and Table \ref*{tab:viral_loads_q} of the supplemental material. 
Also note that we only report viral loads relative to $y_{\min}$ (see Eqn. \ref{eq:relative_viral_loads}) - we do not attempt to estimate $y_{\min}$. These relative viral loads are interpretable by the RT-PCR technicians since they know $t_{\min}$, the minimum $C_t$ (threshold cycle) value observed in that experiment. Note as well that since $y_{\min}$ is the viral load of the pool with the minimum $C_t$ value -- it corresponds to the pool with the maximum viral load in that experiment.

\subsubsection{Comparison with Dorfman Pooling}
\label{subsubsec:dorfman}

\begin{table}[h]
\centering
\caption{Expected number of tests needed by optimal Dorfman Testing for number of samples ($n$) 105 and 961 for various $k$. Note that our proposed methods based on CS require much fewer tests (45 and 93) typically, and do not require two rounds of testing.}
\label{tab:dorfman}
\begin{tabular}{|c|c|c|c|c|c|}
\hline
\multicolumn{3}{|c|}{$N = 105$}              & \multicolumn{3}{c|}{$N = 961$}               \\ \hline
$k$ & \textbf{\# Tests} & \textbf{Pool Size} & $k$ & \textbf{\# Tests} & \textbf{Pool Size} \\ \hline
5          & 43.7       & 5                  & 5          & 136.5      & 14                 \\ \hline
8          & 55.3       & 4                  & 8          & 172.2      & 11                 \\ \hline
10         & 61.3       & 4                  & 10         & 192.2      & 11                 \\ \hline
12         & 67.0       & 4                  & 12         & 209.6      & 9                  \\ \hline
15         & 73.9       & 3                  & 15         & 233.7      & 9                  \\ \hline
17         & 78.2       & 3                  & 17         & 248.7      & 8                  \\ \hline
20         & 84.3       & 3                  & 20         & 269.4      & 7                  \\ \hline
\end{tabular}
\end{table}

We also performed a comparison of our algorithms with the popular two-stage Dorfman pooling method (an adaptive method), with regard to the number of tests required. In the first stage of the Dorfman pooling technique, the $n$ samples are divided into $n/g$  pools, each of size $g$. Each of these $n/g$ pools are tested, and a negative result leads to all members of that pool being considered negative (i.e. non-infected). However, the pools that are tested positive are passed onto a second stage, where all members of those pools are individually tested.
The optimal pool size $g^*$ will minimize the expected number of tests taken by this process (given that the membership in each pool is decided randomly).
A formula for the expected number of tests taken by Dorfman testing is derived in \cite{Dorfman1943}.
The derivation in \cite{Dorfman1943} assumes the following: (1) Any given sample may be positive with probability $p$, independently of the other samples; (2) The number of samples $n$ is divisible by the pool size $g$.
We modify the formula from \cite{Dorfman1943} for the case that $n$ is not divisible by $g$ (supplemental section \ref*{sup-sec:optimal_dorfman}), and find $g^*$ by choosing the value of $g$ which minimizes this number.
We set $p=k/n$, so that out of $n$ samples, the number of infected samples is $k$ in expectation.
Table \ref{tab:dorfman} shows the expected number of tests computed from the formula in supplemental section \ref*{sup-sec:optimal_dorfman}, assuming that the expected number of infected samples $k$ (and thus the optimal pool size $g^*$) is known in advance. 
We also empirically verified the expected number of tests by performing $1000$ Monte Carlo simulations of Dorfman testing with the optimal pool size $g^*$ for each case, and did not observe much deviation from the numbers reported in Table \ref{tab:dorfman}.
Comparisons of Tables \ref{tab:comp}, \ref{tab:comp_nnlasso}, \ref{tab:comp_sbl}, \ref{tab:comp_nnomp} with the two-stage Dorfman pooling method in \ref{tab:dorfman} show that our methods require much fewer tests, albeit with a slight increase in number of false negatives. Moreover, all our methods are single-stage methods and therefore require less time for testing, unlike the Dorfman method which requires two stages of testing. 

\subsubsection{Estimation of number of infected samples}
\label{subsec:estimating_k}

\begin{table}
\centering
\caption{True sparsity $k$ versus estimated sparsity $k_{est}$ (on synthetic data) for $93 \times 961$ Kirkman matrix. Mean and standard deviation of estimated sparsity is computed over 1000 signals for each $k$.}
\label{tab:sparsity_estimate}
\begin{tabular}{|c|c|c|}
\hline
$k$ & $k_{est}$ \\ \hline
5          & 5.01 $\pm$ 0.33              \\ \hline
10         & 10.07 $\pm$ 0.69              \\ \hline
15         & 15.13 $\pm$ 1.17              \\ \hline
20         & 20.26 $\pm$ 1.63              \\ \hline
25         & 25.54 $\pm$ 2.03              \\ \hline
30         & 30.53 $\pm$ 2.61              \\ \hline
\end{tabular}
\end{table}
\begin{table}
\centering
\caption{Comparison of mean number of false negative and false positives for COMP, COMP-SBL and COMP-SBL with graceful failure mode for high values of ${k}$ for the $93\times 961$ Kirkman matrix. The algorithm goes into graceful failure mode when estimated sparsity is greater than or equal to $20$}
\label{tab:graceful_failure}
\begin{tabular}{|c|c|c|c|c|c|c|}
\hline
\multicolumn{1}{|l|}{} & \multicolumn{2}{c|}{\textbf{COMP}} & \multicolumn{2}{c|}{\textbf{COMP-SBL}} & \multicolumn{2}{c|}{\textbf{COMP-SBL-graceful}} \\ \hline
$k$             & \textbf{\#FN}    & \textbf{\#FP}   & \textbf{\#FN}      & \textbf{\#FP}     & \textbf{\#FN}          & \textbf{\#FP}          \\ \hline
15                     & 0                & 45.3            & 0.3                & 24.9              & 0.3                    & 24.8                   \\ \hline
20                     & 0                & 92.7            & 1.3                & 55.4              & 0.4                    & 80.2                   \\ \hline
25                     & 0                & 151.2           & 4                  & 97.5              & 0                      & 151.2                  \\ \hline
30                     & 0                & 212.1           & 6.9                & 140.6             & 0                      & 212.1                  \\ \hline
\end{tabular}
\end{table}
The number of CS measurements for successful recovery depends on the number of non-zero elements ($\ell_0$ norm) of the underlying signal. For example, this varies as $O(k \log n)$ for randomized sensing matrices \cite{CandesWakin2008} or as $O(\textrm{max}(k^2,\sqrt{n})$ for deterministic designs \cite{Devore2007}.
There is a lower bound of $k\sqrt{n}$ measurements for certain types of expander matrices to satisfy a sufficient (but not necessary) condition for recovery \cite{lotfi2020compressed}.
However, in practice $k$ is always unknown, which leads to the question as to how many measurements are needed as a minimum for a \emph{particular} problem instance. To address this, we adopt the technique from \cite{Bioglio2015} to estimate $k$ on the fly from the compressive measurements. This technique does not require signal recovery for estimating $k$. The relative error in the estimate of $k$ is shown to be $O(\sqrt{\log m/m})$ \cite{Ravazzi2018}, which diminishes as $m$ increases (irrespective of the true $k$). Table \ref{tab:sparsity_estimate} shows the accuracy of our sparsity estimate on synthetic data.

The advantage of this estimate of $k$ is that it can drive the \textsc{Comp}-\textsc{Bf} algorithm, as well as act as an indicator of whether there exist any false negatives.
We can use this knowledge to enable a \textit{graceful failure mode}.
In this mode, if our estimate of $k$ is larger than what the CS algorithms can handle, we return only the output of the \textsc{Comp} stage.
Hence in such rare cases, it minimizes the number of false negatives, at the cost of many false positives.
In these cases a second stage of individual testing must be done on the samples which were declared positive.
Table \ref{tab:graceful_failure} shows the effect of using graceful failure mode with \textsc{Comp} followed by \textsc{Sbl} for large values of $k$.
In these experiments, output of \textsc{Comp} is returned if the estimated sparsity, $k_{est}$, is greater than or equal to $20$.
We see that \textsc{Comp}-\textsc{Sbl} with graceful failure mode matches the behaviour of \textsc{Comp}-\textsc{Sbl} at sparsity value lower than $20$, and that of \textsc{Comp} at sparsity value greater than $20$.
At sparsity equal to $20$, it compromises between the high false positives of \textsc{Comp}, and the high false negatives of \textsc{Comp}-\textsc{Sbl}.
This is because of the variability in $k_{est}$, which can occasionally be less than $20$ even if $k$ is equal to $20$.

\subsection{Results on Real Data}
\label{subsec:real_results}
We acquired real data in the form of test results on pooled samples from two labs: one at the National Center of Biological Sciences (NCBS) in India, and the other at the Wyss Institute at the Harvard Medical School, USA. In both cases, viral RNA was artificially injected into $k$ of the $n$ samples where $k \ll n$. From these $n$ samples, a total of $m$ mixtures were created.
For the datasets obtained from NCBS that we experimented with, we had $m = 16$, $n = 40$, $k \in \{1,2,3,4\}$. For the data from the Wyss Institute, we had $m = 24$, $n = 60$, $k = 2$ and $m = 30$, $n = 120$, $k = 2$. The results for all these datasets are presented in Table \ref{tab:experimental_data}. The $16\times 40$ and $24\times 60$ pooling matrices were obtained by performing a simulated annealing procedure to minimize the mutual coherence (see Sec. \ref{subsub:mutual_coherence}), starting with a random sparse binary matrix as initial condition. The $30\times 120$ pooling matrix was a Kirkman matrix.
We used $q=0.95$ in all cases to obtain relative viral loads from $C_t$ values, using Eqn. \ref{eq:relative_viral_loads}. While $q$ may be estimated from raw RT-PCR data (Sec. \ref*{sup-sec:viral_load_q}, supplemental material), we found $q=0.95$ to be a reasonable choice, and did not observe any variation  in the number of reported positives when this parameter was changed between 0.7 to 1. For \textsc{Nnlasso}, \textsc{Nnls} and \textsc{Nnlad}, we use $\tau = 0.2 \times \widetilde{y}_{\max}$ as the threshold below which an estimated relative viral load is set to $0$, since value of $x_{\min}$ may not always be available for real experiments.
Here $\widetilde{y}_{\max}$ is the relative viral load of the pool with the largest $C_t$ value, and consequently the smallest viral amount.
We see that the CS algorithms reduce the false positives, albeit with an introduction of occasional false negatives for higher values of $k$. We also refer the reader to our work in \cite{Ghosh2020.04.23.20077727} for a more in-depth description of results on real experimental data.
\begin{table*}[]
\centering
\caption{Results of lab experiments with each algorithm}
\label{tab:experimental_data}
\begin{tabular}{|c|c|c|c|c|}
\hline
\textbf{Dataset}                             & \textbf{Algorithm} & \textbf{\# true pos} & \textbf{\# false neg} & \textbf{\#false pos} \\ \hline
\multirow{4}{*}{\textbf{Harvard $24 \times 60, k = 2$}}      & \textsc{Comp}               & 2                    & 0                     & 1                    \\ \cline{2-5} 
                                             & \textsc{Comp}-\textsc{Sbl}           & 2                    & 0                     & 1                    \\ \cline{2-5} 
                                             & \textsc{Comp}-\textsc{Nnomp}         & 2                    & 0                     & 0                    \\ \cline{2-5} 
                                             & \textsc{Comp}-\textsc{Nnlasso}        & 2                    & 0                     & 1                    \\ \cline{2-5}
                                             & \textsc{Comp}-\textsc{Nnlad}        & 2                    & 0                     & 1                    \\ \cline{2-5}
                                             & \textsc{Comp}-\textsc{Nnls}        & 2                    & 0                     & 1                    \\ \hline
\hline
\multirow{4}{*}{\textbf{Harvard $30 \times 120, k = 2$}}      & \textsc{Comp}               & 2                    & 0                     & 1                    \\ \cline{2-5} 
                                             & \textsc{Comp}-\textsc{Sbl}           & 2                    & 0                     & 1                    \\ \cline{2-5} 
                                             & \textsc{Comp}-\textsc{Nnomp}         & 2                    & 0                     & 1                    \\ \cline{2-5} 
                                             & \textsc{Comp}-\textsc{Nnlasso}        & 2                    & 0                     & 1                    \\ \cline{2-5}
                                             & \textsc{Comp}-\textsc{Nnlad}        & 2                    & 0                     & 1                    \\ \cline{2-5}
                                             & \textsc{Comp}-\textsc{Nnls}        & 2                    & 0                     & 1                    \\ \hline
\hline
\multirow{4}{*}{\textbf{NCBS-0 $16 \times 40, k = 0$}} & \textsc{Comp}               & 0                    & 0                     & 0                    \\ \cline{2-5} 
                                             & \textsc{Comp}-\textsc{Sbl}           & 0                    & 0                     & 0                    \\ \cline{2-5} 
                                             & \textsc{Comp}-\textsc{Nnomp}         & 0                    & 0                     & 0                    \\ \cline{2-5} 
                                             & \textsc{Comp}-\textsc{Nnlasso}        & 0                    & 0                     & 0                    \\ \cline{2-5}
                                             & \textsc{Comp}-\textsc{Nnlad}        & 0                    & 0                     & 0                    \\ \cline{2-5}
                                             & \textsc{Comp}-\textsc{Nnls}        & 0                    & 0                     & 0                    \\ \hline
\hline
\multirow{4}{*}{\textbf{NCBS-1 $16 \times 40, k = 1$}} & \textsc{Comp}               & 1                    & 0                     & 0                    \\ \cline{2-5} 
                                             & \textsc{Comp}-\textsc{Sbl}           & 1                    & 0                     & 0                    \\ \cline{2-5} 
                                             & \textsc{Comp}-\textsc{Nnomp}         & 1                    & 0                     & 0                    \\ \cline{2-5} 
                                             & \textsc{Comp}-\textsc{Nnlasso}        & 1                    & 0                     & 0                    \\ \cline{2-5}
                                             & \textsc{Comp}-\textsc{Nnlad}        & 1                    & 0                     & 0                    \\ \cline{2-5}
                                             & \textsc{Comp}-\textsc{Nnls}        & 1                    & 0                     & 0                    \\ \hline
\hline
\multirow{4}{*}{\textbf{NCBS-2 $16 \times 40, k = 2$}} & \textsc{Comp}               & 2                    & 0                     & 0                    \\ \cline{2-5} 
                                             & \textsc{Comp}-\textsc{Sbl}           & 2                    & 0                     & 0                    \\ \cline{2-5} 
                                             & \textsc{Comp}-\textsc{Nnomp}         & 2                    & 0                     & 0                    \\ \cline{2-5} 
                                             & \textsc{Comp}-\textsc{Nnlasso}        & 2                    & 0                     & 0                    \\ \cline{2-5}
                                             & \textsc{Comp}-\textsc{Nnlad}        & 2                    & 0                     & 0                    \\ \cline{2-5}
                                             & \textsc{Comp}-\textsc{Nnls}        & 2                    & 0                     & 0                    \\ \hline
\hline
\multirow{4}{*}{\textbf{NCBS-3 $16 \times 40, k = 3$}} & \textsc{Comp}               & 3                    & 0                     & 1                    \\ \cline{2-5} 
                                             & \textsc{Comp}-\textsc{Sbl} & 2 & 1                      & 1                    \\ \cline{2-5} 
                                             & \textsc{Comp}-\textsc{Nnomp}         & 2                    & 1                     & 0                    \\ \cline{2-5} 
                                             & \textsc{Comp}-\textsc{Nnlasso}        & 2                    & 1                     & 1             
\\ \cline{2-5} 
                                             & \textsc{Comp}-\textsc{Nnlad}        & 3                    & 0                     & 1                    \\ \cline{2-5} 
                                             & \textsc{Comp}-\textsc{Nnls}        & 2                    & 1                     & 1                    \\ \cline{2-5} 
                                             & \textsc{Comp}-\textsc{Bf} & 2 & 1                      & 1                                              
                                             \\ \hline
\hline
\multirow{4}{*}{\textbf{NCBS-4 $16 \times 40, k = 4$}} & \textsc{Comp}               & 4                    & 0                     & 3                    \\ \cline{2-5} 
                                             & \textsc{Comp}-\textsc{Sbl}           & 3                    & 1                     & 2                    \\ \cline{2-5} 
                                             & \textsc{Comp}-\textsc{Nnomp}         & 2                    & 2                     & 2                    \\ \cline{2-5} 
                                             & \textsc{Comp}-\textsc{Nnlasso}        & 3                    & 1                     & 2                    \\ \cline{2-5} 
                                             & \textsc{Comp}-\textsc{Nnlad}        & 2                    & 2                     & 2                    \\ \cline{2-5} 
                                             & \textsc{Comp}-\textsc{Nnls}        & 3                    & 1                     & 2                    \\ \cline{2-5} 
                                             & \textsc{Comp}-\textsc{Bf} & 2 & 2                      & 2 \\
                                             
                                             \hline
\end{tabular}
\end{table*}

\subsection{Discussion}
\label{subsec:discussion}

Each algorithm we ran presented a different set of tradeoffs between sensitivity and specificity. While \textsc{Comp} provides us with sensitivity equal to $1$, it suffers many false positives, especially for higher $k$. For other algorithms, in general both the sensitivity and the specificity decrease as $k$ is increased. \textsc{Comp}-\textsc{Nnomp} (Table \ref{tab:comp_nnomp}) has the highest specificity, but it comes at the cost of sensitivity.
\textsc{Comp}-\textsc{Sbl} (Table \ref{tab:comp_sbl}) has the best sensitivity for most values of $k$ amongst the CS algorithms. \textsc{Comp}-\textsc{Nnlasso} (Table \ref{tab:comp_nnlasso}) has better specificity than \textsc{Comp}-\textsc{Sbl} for small values of $k$, but loses out for $k \geq 15$. \textsc{Comp}-\textsc{Nnlad} and \textsc{Comp}-\textsc{Nnls} (Tables \ref{tab:comp_nnlad} and \ref{tab:comp_nnls}) start behaving like \textsc{Comp} for higher values of $k$, effectively bounding the number of false negatives. However, their number of false positives is almost as much as those with \textsc{Comp}.

Ideally, we want both high sensitivity and high specificity while catching a large number of infected samples. Hence, we look at $k^*$, which is the maximum number of infected samples $k$ for which the sensitivity and specificity of the algorithm are greater than or equal to some threshold values. For the $45 \times 105$ Kirkman matrix, we chose the sensitivity threshold as $0.99$ and the specificity threshold as $0.95$. For the $93 \times 961$ Kirkman matrix, we chose both thresholds to be $0.99$, since a specificity threshold of $0.95$ gives too many false positives for $961$ samples. We observed that \textsc{Comp}-\textsc{Sbl} has $k^* = 10$ for both matrices, which is the highest amongst all algorithms tested. 
Typically we do not know the number of infections, but a prevalence rate of infection. The number of infected samples out of a given set of $n$ samples may be treated as a Binomial random variable with probability of success equal to the prevalence rate. Under this assumption, using \textsc{Comp}-\textsc{Sbl} with the $93 \times 961$ Kirkman matrix, we observed that the maximum prevalence rate for which sensitivity and specificity are both above $0.99$ is $1\%$.
Similarly, using \textsc{Comp}-\textsc{Sbl} with the $45\times 105$ Kirkman matrix, we observed that the maximum prevalence rate for which sensitivity is above $0.99$ and specificity is above $0.95$ is $9.5\%$. Thus, Tapestry is viable at prevalence rates as high as $9.5\%$, while reducing testing cost by a factor of $2.3$. On the other hand, if the prevalence rate is only $1\%$ or less, it can reduce testing cost by a factor of $10.3$.

\textbf{Comments about sensitivity and specificity:} We observe that the sensitivity and specificity of our method on synthetic data is \emph{within the recommendations of the U.S. Food and Drugs Administration (FDA)}, as provided in this document \cite{fda_pooled_testing_eua}.
The document provides recommendations for percent positive agreement (PPA) and percent negative agreement (PNA) of a COVID-19 test with a gold standard test (such as RT-PCR done on individual samples).
PPA and PNA are used instead of sensitivity and specificity when ground-truth positives are not known. Since for synthetic data we know the ground truth positives, we compare their PPA and PNA recommendations with the sensitivity and specificity observed by us. We use \textsc{Comp}-\textsc{Sbl} for comparison, since we consider it to be our best method.

For `Testing patients suspected of COVID-19 by their healthcare provider' (point G.4.a, page 7 of \cite{fda_pooled_testing_eua}), the document considers positive and negative agreement of $\geq 95\%$ as acceptable clinical performance (page 9, row 2 of table in \cite{fda_pooled_testing_eua}). The sensitivity and specificity of our method on the $93\times961$ Kirkman matrix is within this range for $k \leq 17$ infected samples (Table \ref{tab:comp_sbl}). For the $45\times 105$ matrix, it is within this range for $k \leq 10$ infected samples (Table \ref{tab:comp_sbl_45_105}).

For `Screening individuals without symptoms or other reasons to suspect COVID-19 with a previously unauthorized test' (point G.4.c, page 10 of \cite{fda_pooled_testing_eua}), the document considers positive agreement of $\geq 95\%$ and negative agreement of $\geq 98\%$ as acceptable (along with the lower bounds of two-sided $95\%$ confidence interval to be $> 76\%$ and $> 95\%$ respectively). Similarly, for `Adding population screening of individuals without symptoms or other reasons to suspect COVID-19 to an authorized test' (point G.4.d, page 12 of \cite{fda_pooled_testing_eua}) the document has the same criterion as for point G.4.c.
Our sensitivity and specificity are within the ranges specified for the $93\times 961$ Kirkman matrix for $k \leq 12$ (Table \ref{tab:comp_sbl}).
While we do not report confidence intervals (as suggested for point G.4.c and G.4.d of \cite{fda_pooled_testing_eua}), the standard deviation of sensitivity and specificity reported by us are fairly low, and we believe the performance of our method is within the recommendations of \cite{fda_pooled_testing_eua}. Since our numbers are on synthetic data - these numbers may vary upon full clinical validation, especially considering that there may be more sources of error in a real test. Nonetheless, we find these numbers to be encouraging. 

Further, we note that while our method incurs an occasional false negative, the viral loads of these false negative values are fairly small.
This means that super-spreaders (who are believed to have high viral load \cite{beldomenico2020superspreaders}) will almost always be caught by our method. In the supplemental material, we discuss this in more detail in Sec. \ref*{sup-sec:viral_loads_false_neg}, and provide a table of mean and standard deviations of viral loads of false negatives (Table \ref*{tab:viral_loads_false_negatives}) for all our methods on synthetic data.

Tapestry can detect certain errors caused by incorrect pipetting, pool contamination, or failed RT-PCR amplification of some pools. This is done by performing a consistency check after the \textsc{Comp} stage. If there is a pool which is positive, but all of the samples involved in it have been declared negative by \textsc{Comp}, this is indicative of error. In case of error, we list all samples categorized by the number of tests that they are positive in. However, the \textsc{Comp} consistency check will not catch all errors. Alternately, the noisy \textsc{Comp} \cite{Chan2011} algorithm may be used to correct for errors in the \textsc{Comp} stage. A full exposition on detection and correction of errors is left as future work.

Although Tapestry can work with a variety of sensing matrix designs, we found Kirkman matrices to be most suitable for our purposes. This is due to lower sparsity and smaller pool sizes presented by Kirkman matrices. Our algorithms also exhibit a more stable behaviour over a wide range of the number of infected samples $k$ when using Kirkman matrices. We compare some alternative matrix designs in section \ref{sup-sec:sensing_matrix}.

\section{Relation to Previous Work}
\label{sec:prev_work}
We review some recent work which apply CS or combinatorial group testing for COVID-19 testing. The works in \cite{Yi_arxiv, Shental_sciencemag, petersen2020practical} adopt a nonadaptive CS based approach.
The works in \cite{Baron_Arxiv, seong2020group, taufer2020rapid}
use combinatorial group testing. 
Compared to these methods, our work is different in the following ways (also see \cite{Ghosh2020.04.23.20077727}):
\begin{enumerate}[leftmargin=0.4cm]
    \item \textit{Real/Synthetic data}: Our work as well as that in \cite{Shental_sciencemag} have tested results on real data, while the rest present only numerical or theoretical results.
    \item \textit{Quantitative Noise model}: Our work uses the physically-derived noise model in Eqn. \ref{eq:noisemodel} (as opposed to only Gaussian noise).
    This noise model is not considered in \cite{Yi_arxiv}. 
    The work in \cite{petersen2020practical} considers unknown noise.
    Combinatorial group testing methods \cite{Baron_Arxiv, seong2020group, taufer2020rapid} do not make use of quantitative information. The work in \cite{Shental_sciencemag} uses only binary test information, even though the decoding algorithm is based on CS.
    
    \item \textit{Algorithms}: The work in \cite{Yi_arxiv} adopts the \textsc{Bpdn} technique (i.e \textsc{P1} from Eqn. \ref{eq:P1}) as well as the brute-force search method for reconstruction. The work in \cite{Shental_sciencemag,Nida2016} uses the \textsc{Lasso}, albeit with a ternary representation for the viral loads. The work in  \cite{petersen2020practical} uses \textsc{Nnlad}. We use the \textsc{Lasso} with a non-negative constraint, the brute-force method, \textsc{Nnlad}, as well as other techniques such as \textsc{Sbl} and \textsc{Nnomp}, all in combination with \textsc{Comp}. The work in \cite{Yi_arxiv} assumes knowledge of the (Gaussian) noise variance for selection of $\varepsilon$ in the estimator in Eqn. \ref{eq:P1}, whereas we use cross-validation for all our estimators. The technique in \cite{Shental_sciencemag} uses a slightly different form of cross-validation for selection of the regularization parameter in LASSO. 
    Amongst combinatorial algorithms, \cite{taufer2020rapid} uses \textsc{Comp}, while \cite{Baron_Arxiv} and \cite{seong2020group} use message passing.
    
    \item \textit{Sensing matrix design}: The work in \cite{Yi_arxiv} uses randomly generated expander graphs, whereas we use Kirkman matrices.  The work in \cite{Shental_sciencemag} uses randomly generated sparse Bernoulli matrices or Reed-Solomon codes, while \cite{seong2020group} uses Low-Density Parity Check (LDPC) codes \cite{mackay1999good}. The work in \cite{petersen2020practical} uses Euler square matrices \cite{naidu2016deterministic}, and the work in \cite{taufer2020rapid} uses the Shifted Transversal Design \cite{thierry2006new}. Both are deterministic disjunct matrices like Kirkman matrices. Each sample in our matrix participates in 3 pools as opposed to 5 pools as used in \cite{seong2020group}, 6 pools as used in \cite{Shental_sciencemag} and \cite{taufer2020rapid}, and 8 pools as used in \cite{petersen2020practical}, which is advantageous from the point of view of pipetting time.
    
    \item \textit{Sparsity estimation:} Our work uses an explicit sparsity estimator and does not rely on any assumption regarding the prevalence rate. 
    
    \item \textit{Numerical comparisons:} We found that \textsc{Comp}-\textsc{Nnlad} works better than the \textsc{Nnlad} method used in \cite{petersen2020practical} on our matrices (see Tables \ref{tab:comp_nnlad} and \ref*{tab:only_nnlad}) . We also found that \textsc{Comp}-\textsc{Nnlasso} and \textsc{Comp}-\textsc{Sbl} have better sensitivity and specificity than \textsc{Comp}-\textsc{Nnlad} (see Tables \ref{tab:comp_nnlasso}, \ref{tab:comp_sbl}, and \ref{tab:comp_nnlad}). The method in \cite{Shental_sciencemag} can correctly identify up to 5/384 (1.3\%) of samples with 48 tests, with an average number of false positives that was less than 2.75, and an average number of false negatives that was less than 0.33. On synthetic simulations with their $48 \times 384$ Reed-Solomon code based matrix (released by the authors) for a total of 100 $\boldsymbol{x}$ vectors with $\ell_0$ norm of 5 using \textsc{Comp}-\textsc{Nnlasso}, we obtained 1.51 false positives and 0.02 false negatives on an average with a standard deviation of 1.439 and 0.14 respectively. Using \textsc{Comp}-\textsc{Sbl} instead of \textsc{Comp}-\textsc{Nnlasso} with all other settings remaining the same, we obtained 1.4 false positives and 0.0 false negatives on an average with a standard deviation of 1.6 and 0.1 respectively. As such, a direct numerical comparison between our work and that in \cite{Shental_sciencemag} is not possible, due to lack of available real data, however these numbers yield some indicator of performance. 
    \item \textit{Number of Tests:} We use $93$ tests for $961$ samples while achieving more than $0.99$ sensitivity and specificity for $k=10$ infections using \textsc{Comp}-\textsc{Sbl}. In a similar setting, \cite{seong2020group} use $108$ tests for $Q=1000$ samples under prevalence rate $0.01$ for exact 2-stage recovery. The work in \cite{taufer2020rapid} uses $186$ tests for $961$ samples under the same prevalence rate, albeit for sensitivity equal to $1$ and very high specificity. Matrix sizes studied in other work are very different than ours. The work in \cite{heidarzadeh2020two} builds on top of our Tapestry scheme to reduce the number of tests, but it is a two-stage adaptive technique and hence will require much more testing time. 
    
\end{enumerate}

\section{Conclusion}
\label{sec:concl}
We have presented a non-adaptive, single-round technique for prediction of infected samples as well as the viral loads, from an array of $n$ samples, using a compressed sensing approach. We have empirically shown on synthetic data as well as on some real lab acquisitions that our technique can correctly predict the positive samples with a very small number of false positives and false negatives. Moreover, we have presented techniques for appropriate design of the mixing matrix. Our single-round testing technique can be deployed in many different scenarios such as the following:
\begin{enumerate}[leftmargin=0.4cm]
    \item Testing of 105 symptomatic individuals in 45 tests.
    \item Testing of 195 asymptomatic individuals in 45 tests assuming a low rate of infection. A good use case for this is airport security personnel, delivery personnel, or hospital staff.
    \item Testing of 399 individuals in 63 tests. This can be used to test students coming back to campuses, or police force, or asymptomatic people in housing blocks and localities currently under quarantine.
    \item Testing of 961 people in 93 tests, assuming low infection rate. This might be suitable for airports and other places where samples can be collected and tested immediately, and it might be possible to obtain liquid handling robots.
\end{enumerate}
\textbf{Outputs:} We have designed an Android app named Byom Smart Testing to make our Tapestry protocol easy to deploy in the future. The app can be accessed at \cite{ByomApp}. We are also sharing our code and some amount of data at \cite{CodeGithub}. More information is also available at our website \cite{Website}. \\
\textbf{Future work:} Future work will involve extensive testing on real COVID-19 data, and extensive implementation of a variety of algorithms for sensing matrix design as well as signal recovery, keeping in mind the accurate statistical noise model and accounting for occasional pipetting errors. 

\section*{Acknowledgement}
AR acknowledges support from SERB Matrics grant MTR/2019/000691. AR and MG acknowledge support from IITB WRCB grant \#10013976, and DST-Rakshak grant \#10013980. The authors thank the two anonymous reviewers as well as the Associate Editor for careful review of the previous version of this paper and helpful suggestions which have greatly improved this paper.
\nocite{CGT2000}
\nocite{Zhang2014}
\nocite{Li2018}

\bibliographystyle{IEEEtran}
\bibliography{refs}

\end{document}


\maketitle
\section{Notations Used}
We refer the reader to the main paper for the meaning of various symbols and notations employed in this document. 

\section{Generalized Binary Search Techniques}
\label{sup-sec:binary_search}
In the class of `adaptive group testing' techniques, the $n$ samples are distributed into two or more groups, each of smaller size, and the smaller groups are then individually tested. In one particular adaptive method called generalized binary splitting (GBS) \cite{CGT2000}, this procedure is repeated (in a binary search fashion) until a single infected sample is identified. This requires $O(\log n)$ sequential tests, where each test requires mixing up to $n/2$ samples. This sample is then discarded, and the entire procedure is performed on the remaining $n-1$ samples. Such a procedure does not introduce any false negatives, and does not require prior knowledge of the number of infected samples $k$. It requires a total of only $O(k \log n)$ tests, if $k$ is the number of infected samples. However such a multi-stage method is impractical to be deployed due to its sequential nature, since each RT-PCR stage requires nearly 3-4 hours. Moreover, each mixture that is tested contains contributions from as many as $O(n)$ samples, which can lead to significant dilution or may be difficult to implement in the lab. Hence in this work, we do not pursue this particular approach. Such an approach may be very useful if each individual test had a quick turn-around time. 

\section{Details of Cross-Validation for Compressed Sensing}
\label{sup-sec:crossval}
For this, the measurements in $\boldsymbol{y}$ are divided into two randomly chosen disjoint sets: one for reconstruction ($\mathcal{R}$) and the other for validation ($\mathcal{V}$). A decoding algorithm such as \textsc{Nnlasso} is executed independently on multiple values of the regularization parameter $\lambda$ from a candidate set $\Lambda$. (Other decoding algorithms will have their own parameters. For example, \textsc{Nnomp} or \textsc{BPDN} will use the parameter $\varepsilon$, i.e. the bound on the noise magnitude.) For each $\lambda$ value, an estimate $\boldsymbol{\hat{x}_{\lambda}}$ is produced using measurements only from $\mathcal{R}$, and the CV error $v_e(\lambda) := \sum_{i \in \mathcal{V}} (y_i - \boldsymbol{A^i \hat{x}_{\lambda}})^2$ is computed. The value of $\lambda$ which yields the least value of $v_e(\lambda)$ is chosen, and a final estimate of $\boldsymbol{x}$ is obtained by executing the algorithm again, but now using all measurements from $\mathcal{R} \cup \mathcal{V}$. If $\mathcal{V}$ is large enough, then $v_e(\lambda)$ is shown to be a good estimate of the actual error $\|\boldsymbol{x}-\boldsymbol{\hat{x}_{\lambda}}\|^2$, as has been shown for Gaussian noise \cite{Zhang2014}. Nonetheless, it should be noted that CV is a method of choice for parameter selection in CS even under a variety of other noise models such as Poisson \cite{Li2018}, etc, and we have experimentally observed that it works well even in the case of our noise model in Eqn. \ref*{eq:noisemodel} of the main paper.

\section{Brute-force search method}
\label{sup-sec:comp_bf}
We refer to \textsc{Comp}-\textsc{Bf} as a method where we apply \textsc{Comp} followed by a brute-force search to minimize the cost function in Eqn. \ref{eq:J_BF} below. The brute-force search is computationally feasible only when $C(n,k)$ is `reasonable' in value (note that the effective $n$ is often reduced after application of \textsc{Comp}), and so we employ it only for small-sized matrices. The method essentially enumerates all possible supports of $\boldsymbol{x}$ which have size $k$. For each such candidate support set $\mathcal{Z}$, the following cost function is minimized using the \texttt{fmincon} routine of MATLAB which implements an interior-point optimizer
\footnote{\url{https://in.mathworks.com/help/optim/ug/choosing-the-algorithm.html\#bsbwxm7}}:
\begin{equation}
    J(\boldsymbol{x_{\mathcal{Z}}}) := \|\log \boldsymbol{y} - \log \boldsymbol{A_{\mathcal{Z}} x_{\mathcal{Z}}}\|_2 \textrm{ such that } \boldsymbol{x_{\mathcal{Z}}} \geq \boldsymbol{0}.
    \label{eq:J_BF}
\end{equation}
Results with the \textsc{Comp}-\textsc{Bf} method are shown in Table \ref{tab:comp_bf}. The special advantage of the brute-force method is that it requires only $m=2k$ pools, which is less than $O(k \log n)$. However, such a method requires prior knowledge of $k$, or an estimate thereof. We employ a method to estimate $k$ directly from $\boldsymbol{y}, \boldsymbol{A}$. This is described in Sec. \ref*{subsec:estimating_k} of the main paper. The results in Table \ref{tab:comp_bf} assume that the exact $k$ was known, or that the estimator predicted the exact $k$. However, we observed that the estimator from Sec. \ref*{subsec:estimating_k} can sometimes over-estimate $k$. Hence, we also present results with \textsc{Comp}-\textsc{Bf} where the brute-force search assumed that the sparsity was (over-estimated to be) $k+1$ instead of $k$. These are shown in Table \ref{tab:comp_bf2}. A comparison of Tables \ref{tab:comp_bf} and \ref{tab:comp_bf2} shows that RMSE deteriorates if $k$ is incorrectly estimated. However there is no adverse effect on the number of false negatives, and only a small adverse effect on the number of false positives. 
\begin{table}
\centering
\caption{Performance of \textsc{Comp} followed by brute-force search (\textsc{Comp}-\textsc{Bf}) to minimize the function in Eqn. \ref{eq:J_BF} for a $16 \times 40$ matrix optimized on mutual coherence, with different values of $k$, assuming that the true value of $k$ was known. Results for both methods are reported on synthetic data.}
\label{tab:comp_bf}
\begin{tabular}{|c|c|c|c|c|c|c|}
\hline
Method & $k$  & RMSE & \#false neg. & \#false pos. & sens. & spec. \\ \hline
\textsc{Comp} & 2 & 1.00 & 0.00 & 0.50 & 1.00 & 0.99 \\ \hline
\textsc{Comp}-\textsc{Bf} & 2 & 0.03 & 0.00 & 0.00 & 1.00 & 1.00 \\ \hline
\textsc{Comp} & 3 & 1.00 & 0.00 & 1.85 & 1.00 & 0.95\\ \hline
\textsc{Comp}-\textsc{Bf} & 3 & 0.05 & 0.00 & 0.00 & 1.00 & 1.00\\ \hline
\textsc{Comp} & 4 & 1.00 & 0.00 & 3.35 & 1.00 & 0.91\\ \hline
\textsc{Comp}-\textsc{Bf} & 4 & 0.05 & 0.00 & 0.00 & 1.00 & 1.00\\ \hline
\end{tabular}
\end{table}

\begin{table}
\centering
\caption{Performance of \textsc{Comp} followed by brute-force search (\textsc{Comp}-\textsc{Bf}) to minimize the cost function in Eqn. \ref{eq:J_BF} for $16 \times 40$ matrix optimized on mutual coherence, with different values of $k$, assuming that the sparsity value was estimated to be $k+1$. Results for both methods are reported on synthetic data.}
\label{tab:comp_bf2}
\begin{tabular}{|c|c|c|c|c|c|c|}
\hline
Method & $k$  & RMSE & \#false neg. & \#false pos. & sens. & spec. \\ \hline
\textsc{Comp} & 2 & 1.00 & 0.00 & 0.75 & 1.00 & 0.98\\\hline
\textsc{Comp}-\textsc{Bf} & 2 & 0.47 & 0.00 & 0.55 & 1.00 & 0.99\\\hline
\textsc{Comp} & 3 & 1.00 & 0.00 & 1.70 & 1.00 & 0.95\\\hline
\textsc{Comp}-\textsc{Bf} & 3 & 0.24 & 0.00 & 0.80 & 1.00 & 0.98\\\hline
\textsc{Comp} & 4 & 1.00 & 0.00 & 3.00 & 1.00 & 0.92\\\hline
\textsc{Comp}-\textsc{Bf} & 4 & 0.15 & 0.00 & 0.90 & 1.00 & 0.97\\\hline
\end{tabular}
\end{table}

\section{RIP and RNSP Preservation after \textsc{Comp}}
\label{sup-sec:comp_ric}
We prove that the reduced matrix $\redA$ obtained after \textsc{Comp} (Sec. \ref*{sec:cs_gt_combined} of the main paper) preserves RIP-1, RIP-2 and RNSP of the full matrix $\A$, of same order and with the same parameters. Recall definition of $\Xn$, $\Yn$, $\X$ and $\Y$ from section \ref*{sec:cs_gt_combined} of the main paper.

First, notice that in all the rows of the matrix $\A$ for which the measurement vector $\y$ was 0, the entries in $\A$ corresponding to indices from $\X$ contained only 0. That is, 
\begin{align}
\A_{ij} = 0 \text{ }\forall i \in \Yn, j \in \X. \label{comp-zeros}
\end{align}

This is because \textsc{Comp} eliminates all samples for which $\A_{ij} > 0$ and $y_i = 0$ -- hence if for any sample $j$, $A_{ij} > 0$ and $y_i = 0$, then $j \in \Xn$.
Second, we observe the following:

\begin{lemma}
\label{p-norm-equiv}
 Let $\xbar$ denote a $|\X|$ dimensional vector. Consider a $n$-dimensional vector $\x$ such that $\x_{\X} = \xbar$ and $\x_{\Xn} = \boldsymbol{0}$. Then, $\|\A\x\|_p = \|\A_{\X, \Y}\xbar\|_p$ for any $p \in \mathbb{Z}_{+}$.
\end{lemma}
\begin{proof}

Let $\A^i$ denote the $1\times n$ $i^{\text{th}}$ row vector of the matrix $\A$.
\begin{align*}
\|\A\x\|_p &= \Big(\sum_i |\A^i\x|^p \Big)^{1/p}\\
&= \Big(\sum_{i \in \Y} |\A^i\x|^p + \sum_{i \in \Yn} |\A^i\x|^p \Big)^{1/p}\\
&= \Big(\sum_{i \in \Y} |\A^i\x|^p + \sum_{i \in \Yn} \Big|\sum_{j \in \X} \A_{ij}\x_j + \sum_{j \in \Xn} \A_{ij}\x_j\Big|^p \Big)^{1/p}. \numberthis \label{three-terms}
\end{align*}
Using ~\cref{comp-zeros}, the second term in \cref{three-terms} is $0$. The third term is $0$ as well, since $\x_j = 0$ $\forall j \in \Xn$. Hence,
\begin{align*}
\|\A\x\|_p &= \Big(\sum_{i \in \Y} |\A^i\x|^p\Big)^{1/p}\\
&= \Big(\sum_{i \in \Y} \Big|\sum_{j \in \X} \A_{ij}\x_j +\sum_{j \in \Xn} \A_{ij}\x_j  \Big|^p\Big)^{1/p}. \numberthis \label{two-terms}
\end{align*}
Again, since $\x_j = 0$ $\forall j \in \Xn$, the second term in \cref{two-terms} is $0$. Hence, 
\begin{align*}
\|\A\x\|_p &= \Big(\sum_{i \in \Y} \Big|\sum_{j \in \X} \A_{ij}\x_j \Big|^p\Big)^{1/p}. \numberthis \label{one-term}
\end{align*}
Notice that the RHS in \cref{one-term} is simply $\|\A_{\X, \Y}\xbar\|_p$ by definition of $\redA$ and $\xbar$. Hence,
$\|\A\x\|_p = \|\A_{\X, \Y}\xbar\|_p$.
\end{proof}
This lemma is critical to proving the following RIP-preservation theorems:
\begin{theorem}
Let $d$ be a constant. If the scaled matrix $\A/d$ satisfies RIP-1 of order $k$ with $k$-order RIC $\delta_k \in (0,1)$, and $k\le|\X|$, then the scaled \textsc{Comp}-reduced matrix $\A_{\X, \Y}/d$ also satisfies RIP-1 of order $k$ with RIC $\delta_k$. That is, if 
$$\|\x\|_1\le \frac{1}{d}\|\A\x\|_1 \le (1+\delta_k)\|\x\|_1,$$
for all $n$-dimensional $k$-sparse vectors $\x$
then,
$$\|\xbar\|_1\le \frac{1}{d}\|\A_{\X, \Y}\xbar\|_1 \le (1+\delta_k)\|\xbar\|_1$$
for all $|\X|$-dimensional $k$-sparse vectors $\xbar$.
\end{theorem}
\begin{proof}
Let $\xbar$ be a $|\X|$ dimensional $k$-sparse vector. Consider a $n$-dimensional vector $\x$ such that $\x_{\X} = \xbar$ and $\x_{\Xn} = \boldsymbol{0}$. $\x$ is also a $k$-sparse vector.
By definition of RIP-1,
$$\|\x\|_1\le \frac{1}{d}\|\A\x\|_1 \le (1+\delta_k)\|\x\|_1.$$
Using ~\cref{p-norm-equiv}, $\|\A\x\|_1 = \|\A_{\X, \Y}\xbar\|_1$. Hence,
$$\|\x\|_1\le \frac{1}{d}\|\A_{\X, \Y}\xbar\|_1 \le (1+\delta_k)\|\x\|_1.$$
Also note that $\|\x\|_1 = \|\xbar\|_1$. Hence,
$$\|\xbar\|_1\le \frac{1}{d}\|\A_{\X, \Y}\xbar\|_1 \le (1+\delta_k)\|\xbar\|_1$$.
\end{proof}

\begin{theorem}
Let $d$ be a constant. If the scaled matrix $\A/d$ satisfies RIP-2 of order $k$ with RIC $\delta_k \in (0,1)$ and $k\le|\X|$, then the scaled \textsc{Comp}-reduced matrix $\A_{\X, \Y}/d$ also satisfies RIP-2 of order $k$ with RIC $\delta_k$. That is, if
$$(1-\delta_k)\|\x\|_2^2\le \frac{1}{d^2}\|\A\x\|_2^2 \le (1+\delta_k)\|\x\|_2^2.$$
for all $n$-dimensional $k$-sparse vectors $\x$
then,
$$(1-\delta_k)\|\xbar\|_2^2\le \frac{1}{d^2}\|\A_{\X, \Y}\xbar\|_2^2 \le (1+\delta_k)\|\xbar\|_2^2$$
for all $|\X|$-dimensional $k$-sparse vectors $\xbar$.
\end{theorem}
\begin{proof}
Let $\xbar$ be a $|\X|$ dimensional $k$-sparse vector.
Consider a $n$-dimensional vector $\x$ such that $\x_{\X} = \xbar$ and $\x_{\Xn} = \boldsymbol{0}$.
Hence $\x$ is also a $k$-sparse vector.
By definition of RIP-2,
$$(1-\delta_k)\|\x\|_2^2\le \frac{1}{d^2}\|\A\x\|_2^2 \le (1+\delta_k)\|\x\|_2^2.$$
Using ~\cref{p-norm-equiv}, $\|\A\x\|_2 = \|\A_{\X, \Y}\xbar\|_2$. Hence,
$$(1-\delta_k)\|\x\|_2^2\le \frac{1}{d^2}\|\A_{\X, \Y}\xbar\|_2^2 \le (1+\delta_k)\|\x\|_2^2.$$
Also note that $\|\x\|_2 = \|\xbar\|_2$.
Hence,
$$(1-\delta_k)\|\xbar\|_2^2\le \frac{1}{d^2}\|\A_{\X, \Y}\xbar\|_2^2 \le (1+\delta_k)\|\xbar\|_2^2$$.
\end{proof}

\begin{theorem}\label{thm:l2-rnsp}
If the matrix $\A$ satisfies $\ell_2$-RNSP of order $k$ with parameters $\rho$ and $\tau$, then the \textsc{Comp}-reduced matrix $\A_{\X, \Y}$ also satisfies $\ell_2$-RNSP of order $k$ with the same parameters $\rho$ and $\tau$. That is, if
$$\|\x_S\|_2 \leq \frac{\rho}{\sqrt{k}}\|\boldsymbol{x}_{\bar{S}}\|_1 + \tau \|\A \x\|_2 \text{   } \forall \x \in \mathbb{R}^n$$
holds for all $S \subset \{1\dots n\}$ with $|S| \leq k$
then,
$$\|\xbar_B\|_2 \leq \frac{\rho}{\sqrt{k}}\|\xbar_{\bar{B}}\|_1
+ \tau \|\redA \xbar\|_2 \text{   } \forall \xbar \in \mathbb{R}^{|\X|}$$
holds for all $B \subset \{1\dots |\X|\}$ with $|B| \leq k$.
\end{theorem}

\begin{proof}
Let $\xbar$ be a $|\X|$-dimensional vector.
Consider a $n$-dimensional vector $\x$ such that $\x_{\X} = \xbar$ and $\x_{\Xn} = \boldsymbol{0}$.
For any set $B \subset \{1,\dots,|\X|\}$, consider the set $S \subset \X$, such that $\x_{S} = \xbar_{B}$, and the set $\bar S = \Xn \cup (\X - S)$.
By definition of $\ell_2$-RNSP,
$$\|\x_S\|_2 \leq \frac{\rho}{\sqrt{k}}\|\boldsymbol{x}_{\bar{S}}\|_1 + \tau \|\A \x\|_2 \text{   }$$
Using ~\cref{p-norm-equiv}, $\|\A\x\|_2 = \|\A_{\X, \Y}\xbar\|_2$. Also note that $\|\x_S\|_2 = \|\xbar_B\|_2$, and $\|\x_{\bar S}\|_1 = \|\xbar_{\bar B}\|_1$. Hence,
$$\|\xbar_B\|_2 \leq \frac{\rho}{\sqrt{k}}\|\xbar_{\bar{B}}\|_1
+ \tau \|\redA \xbar\|_2.$$

\end{proof}

\begin{theorem}
If the matrix $\A$ satisfies RNSP of order $k$ with parameters $\rho$ and $\tau$, then the \textsc{Comp}-reduced matrix $\A_{\X, \Y}$ also satisfies RNSP of order $k$ with the same parameters $\rho$ and $\tau$. That is, if
$$\|\x_S\|_2 \leq {\rho}\|\boldsymbol{x}_{\bar{S}}\|_1 + \tau \|\A \x\|_2 \text{   } \forall \x \in \mathbb{R}^n$$
holds for all $S \subset \{1\dots n\}$ with $|S| \leq k$
then,
$$\|\xbar_B\|_2 \leq {\rho}\|\xbar_{\bar{B}}\|_1
+ \tau \|\redA \xbar\|_2 \text{   } \forall \xbar \in \mathbb{R}^{|\X|}$$
holds for all $B \subset \{1\dots |\X|\}$ with $|B| \leq k$.
\end{theorem}

\begin{proof}
The proof follows the same argument as for \cref{thm:l2-rnsp}, except the coefficient $\frac{\rho}{\sqrt{k}}$ is replaced by $\rho$.

\end{proof}

\section{Sensing Matrix Comparison}
\label{sup-sec:sensing_matrix}

\begin{table*}
\centering
\caption{Performance of COMP followed by SBL (on synthetic data) for $93 \times 961$ matrix optimized for low mutual coherence. For each criterion, mean and standard deviation values are reported, across 1000 signals.}
\label{tab:sensing_mats_mutual_coherence}
\begin{tabular}{|c|c|c|c|c|c|}
\hline
$k$ & \textbf{RMSE} & \textbf{\#FN} & \textbf{\#FP} & \textbf{Sens.} & \textbf{Spec.} \\ \hline
5          & 0.031 $\pm$ 0.012   & 0.0 $\pm$ 0.0       & 0.1 $\pm$ 0.3       & 1.0000 $\pm$ 0.0000  & 0.9999 $\pm$ 0.0003  \\ \hline
8          & 0.041 $\pm$ 0.013   & 0.0 $\pm$ 0.1       & 1.8 $\pm$ 1.5       & 0.9996 $\pm$ 0.0068  & 0.9982 $\pm$ 0.0015  \\ \hline
10         & 0.049 $\pm$ 0.013   & 0.0 $\pm$ 0.2       & 5.7 $\pm$ 3.3       & 0.9973 $\pm$ 0.0168  & 0.9940 $\pm$ 0.0034  \\ \hline
12         & 0.062 $\pm$ 0.018   & 0.1 $\pm$ 0.3       & 13.0 $\pm$ 6.3      & 0.9920 $\pm$ 0.0262  & 0.9863 $\pm$ 0.0066  \\ \hline
15         & 0.098 $\pm$ 0.030   & 0.2 $\pm$ 0.5       & 32.0 $\pm$ 12.8     & 0.9843 $\pm$ 0.0320  & 0.9662 $\pm$ 0.0135  \\ \hline
17         & 0.120 $\pm$ 0.029   & 0.4 $\pm$ 0.6       & 49.8 $\pm$ 18.4     & 0.9777 $\pm$ 0.0356  & 0.9473 $\pm$ 0.0195  \\ \hline
20         & 0.126 $\pm$ 0.029   & 0.6 $\pm$ 0.7       & 86.2 $\pm$ 28.6     & 0.9718 $\pm$ 0.0368  & 0.9084 $\pm$ 0.0304  \\ \hline
\end{tabular}
\end{table*}

\begin{table*}
\centering
\caption{Performance of COMP followed by SBL (on synthetic data) for $93 \times 961$ Bernoulli($0.5$) matrix. For each criterion, mean and standard deviation values are reported, across 100 signals.}
\label{tab:sensing_mats_bernoulli_0.5}
\begin{tabular}{|c|c|c|c|c|c|}
\hline
\textbf{k} & \textbf{RMSE} & \textbf{\#FN} & \textbf{\#FP} & \textbf{Sens.} & \textbf{Spec.} \\ \hline
5  & 0.123 $\pm$ 0.051 & 0.1 $\pm$ 0.3 & 127.9 $\pm$ 127.8 & 0.9840 $\pm$ 0.0543 & 0.8662 $\pm$ 0.1337 \\ \hline
8  & 0.147 $\pm$ 0.027 & 0.3 $\pm$ 0.5 & 405.9 $\pm$ 132.3 & 0.9675 $\pm$ 0.0652 & 0.5741 $\pm$ 0.1388 \\ \hline
10 & 0.167 $\pm$ 0.026 & 0.4 $\pm$ 0.6 & 466.4 $\pm$ 65.7  & 0.9580 $\pm$ 0.0603 & 0.5096 $\pm$ 0.0690 \\ \hline
12 & 0.190 $\pm$ 0.033 & 0.5 $\pm$ 0.6 & 475.1 $\pm$ 49.5  & 0.9558 $\pm$ 0.0533 & 0.4994 $\pm$ 0.0521 \\ \hline
15 & 0.254 $\pm$ 0.050 & 1.0 $\pm$ 0.9 & 479.5 $\pm$ 37.0  & 0.9353 $\pm$ 0.0607 & 0.4932 $\pm$ 0.0391 \\ \hline
17 & 0.288 $\pm$ 0.064 & 1.3 $\pm$ 0.9 & 484.6 $\pm$ 37.1  & 0.9241 $\pm$ 0.0554 & 0.4867 $\pm$ 0.0393 \\ \hline
20 & 0.365 $\pm$ 0.096 & 1.9 $\pm$ 1.3 & 488.7 $\pm$ 44.9  & 0.9040 $\pm$ 0.0651 & 0.4806 $\pm$ 0.0477 \\ \hline
\end{tabular}
\end{table*}

\begin{table*}
\centering
\caption{Performance of COMP followed by SBL (on synthetic data) for $93 \times 961$ Bernoulli($0.1$) matrix. For each criterion, mean and standard deviation values are reported, across 1000 signals.}
\label{tab:sensing_mats_bernoulli_0.1}
\begin{tabular}{|c|c|c|c|c|c|}
\hline
\textbf{k} & \textbf{RMSE} & \textbf{\#FN} & \textbf{\#FP} & \textbf{Sens.} & \textbf{Spec.} \\ \hline
5  & 0.035 $\pm$ 0.015 & 0.0 $\pm$ 0.0 & 1.5 $\pm$ 1.6    & 0.9998 $\pm$ 0.0063 & 0.9985 $\pm$ 0.0017 \\ \hline
8  & 0.052 $\pm$ 0.018 & 0.0 $\pm$ 0.1 & 8.0 $\pm$ 5.6    & 0.9974 $\pm$ 0.0179 & 0.9916 $\pm$ 0.0058 \\ \hline
10 & 0.073 $\pm$ 0.028 & 0.1 $\pm$ 0.2 & 17.9 $\pm$ 9.7   & 0.9940 $\pm$ 0.0242 & 0.9811 $\pm$ 0.0102 \\ \hline
12 & 0.106 $\pm$ 0.039 & 0.2 $\pm$ 0.4 & 33.7 $\pm$ 18.3  & 0.9863 $\pm$ 0.0334 & 0.9645 $\pm$ 0.0193 \\ \hline
15 & 0.137 $\pm$ 0.034 & 0.4 $\pm$ 0.6 & 68.8 $\pm$ 30.9  & 0.9757 $\pm$ 0.0386 & 0.9273 $\pm$ 0.0327 \\ \hline
17 & 0.138 $\pm$ 0.031 & 0.5 $\pm$ 0.6 & 98.6 $\pm$ 38.5  & 0.9720 $\pm$ 0.0382 & 0.8955 $\pm$ 0.0408 \\ \hline
20 & 0.141 $\pm$ 0.031 & 0.6 $\pm$ 0.8 & 155.8 $\pm$ 51.7 & 0.9679 $\pm$ 0.0394 & 0.8345 $\pm$ 0.0549 \\ \hline
\end{tabular}
\end{table*}

We generated sensing matrices optimized to have low mutual coherence, as described in Sec. \ref*{subsec:matrix_design}. We have observed that these matrices, besides originating from randomly generated matrices, have a much higher number of non-zero elements and higher pool sizes compared to Kirkman or STS matrices. This leads to difficulty in pooling, increased pooling time, wastage of sample, and sample dilution. A well-tuned $93 \times 961$ matrix that we generated using this method had $6517$ non-zero elements, with its smallest pool being of size $66$. In comparison, the Kirkman triple matrix of the same dimensions had $2883$ non-zero elements and a pool size equal to $31$. Table \ref{tab:sensing_mats_mutual_coherence} shows the performance of \textsc{Comp} followed by \textsc{Sbl} on this matrix designed to minimize coherence, using synthetic data. For smaller values of $k$ (upto $12$), it has lower false negatives and false positives than the same algorithm on the $93 \times 961$ Kirkman matrix (Table \ref*{tab:comp_sbl}). However, the number of false positives increases significantly for higher number of infections.

We also considered Bernoulli($p$) random matrices, whose entries are $1$ with probability $p$, and $0$ otherwise. Here $p$ is the Bernoulli parameter and lies in the range $(0, 1)$. It is shown in \cite{kueng2017robust} that Bernoulli($p$) matrices are good for compressed sensing. That is, they satisfy a robust nullspace property of order $k$ if the number of rows $m = O(k\log n/k)$, with very high probability. On the other hand, Bernoulli($1/k$) sensing matrices with $O(k \log n)$ rows incur low probability of error when \textsc{Comp} is used to determine the support of $k$-sparse vectors \cite{Chan2011}.
Table \ref{tab:sensing_mats_bernoulli_0.5} shows the performance of \textsc{Comp} followed by \textsc{Sbl} on a $93\times 961$ Bernoulli($0.5$) matrix. The \textsc{Comp} step does not help that much in this case, leading to a large number of false positives in the subsequent \textsc{Sbl} step. Table \ref{tab:sensing_mats_bernoulli_0.1} shows the performance of \textsc{Comp} followed by \textsc{Sbl} on a $93\times 961 $ Bernoulli($0.1$) matrix. Results on this matrix are better than the Bernoulli($0.5$) matrix. However, neither of these matrices come close to the performance exhibited by Kirkman triple matrices (e.g. Table \ref{tab:comp_sbl}).

\section{Non-negative Least Squares (\textsc{Nnls})}
\label{sup-sec:nnls}
We present the results of running \textsc{Comp} followed by Non-negative least squares (\textsc{Nnls}, section \ref{subsubsec:nnlad}) on synthetic data for the $93 \times 961$ Kirkman matrix in Table \ref{tab:comp_nnls}.

\begin{table*}
\centering
\caption{Performance of COMP followed by NNLS (on synthetic data) for $93 \times 961$ Kirkman triple matrix. For each criterion and each $k$ value, mean and standard deviation values are reported, across 1000 signals.}
\label{tab:comp_nnls}
\begin{tabular}{|c|c|c|c|c|c|}
\hline
\textbf{k}    & \textbf{RMSE}       & \textbf{\#FN}   & \textbf{\#FP}     & \textbf{Sens.}        & \textbf{Spec.}        \\ \hline
\textbf{$5$}  & $ 0.046 \pm 0.018 $ & $ 0.0 \pm 0.1 $ & $ 0.8 \pm 0.9 $   & $ 0.9992 \pm 0.0126 $ & $ 0.9992 \pm 0.0009 $ \\ \hline
\textbf{$8$}  & $ 0.070 \pm 0.031 $ & $ 0.1 \pm 0.3 $ & $ 4.0 \pm 2.2 $   & $ 0.9912 \pm 0.0324 $ & $ 0.9958 \pm 0.0023 $ \\ \hline
\textbf{$10$} & $ 0.097 \pm 0.044 $ & $ 0.2 \pm 0.5 $ & $ 7.9 \pm 3.1 $   & $ 0.9777 \pm 0.0468 $ & $ 0.9917 \pm 0.0032 $ \\ \hline
\textbf{$12$} & $ 0.154 \pm 0.093 $ & $ 0.5 \pm 0.7 $ & $ 13.1 \pm 5.5 $  & $ 0.9566 \pm 0.0586 $ & $ 0.9862 \pm 0.0058 $ \\ \hline
\textbf{$15$} & $ 0.288 \pm 0.164 $ & $ 0.9 \pm 1.0 $ & $ 29.1 \pm 17.0 $ & $ 0.9397 \pm 0.0685 $ & $ 0.9692 \pm 0.0180 $ \\ \hline
\textbf{$17$} & $ 0.397 \pm 0.174 $ & $ 0.8 \pm 1.2 $ & $ 50.0 \pm 24.1 $ & $ 0.9547 \pm 0.0715 $ & $ 0.9470 \pm 0.0256 $ \\ \hline
\textbf{$20$} & $ 0.528 \pm 0.167 $ & $ 0.2 \pm 0.9 $ & $ 87.8 \pm 25.3 $ & $ 0.9883 \pm 0.0428 $ & $ 0.9067 \pm 0.0269 $ \\ \hline
\end{tabular}
\end{table*}

\section{Synthetic data results on $45 \times 105$ Kirkman triple matrix}
\label{sup-sec:45_105_results}
We present performance of \textsc{Comp} followed by CS algorithms on synthetic data for the Kirkman triple matrix of size $45 \times 105$ in Tables \ref{tab:comp_45_105},
\ref{tab:comp_nnlasso_45_105}, \ref{tab:comp_sbl_45_105}, \ref{tab:comp_nnomp_45_105}, \ref{tab:comp_nnlad_45_105}, and \ref{tab:comp_nnls_45_105}.
We followed the same methodology as in Sec. \ref*{subsec:synthetic_results} of the main paper for these experiments. Using \textsc{Comp} without the CS step gives acceptable performance up to $k=8$. For larger values of $k$, significant improvements are seen by performing the additional CS step after \textsc{Comp}. \textsc{Comp}-\textsc{Sbl} is the best algorithm, achieving high sensitivity and specificity for a wide range of $k$. It is followed closely by \textsc{Comp}-\textsc{Nnlad}, \textsc{Comp}-\textsc{Nnlasso}, and \textsc{Comp}-\textsc{Nnls}, all of which give higher false negatives than \textsc{Comp}-\textsc{Sbl} for high values of $k$. \textsc{Comp}-\textsc{Nnomp} provides the highest specificity, albeit at the cost of lower sensitivity than the other algorithms, especially for high values of $k$. We note that \textsc{Comp}-\textsc{Sbl} and \textsc{Comp}-\textsc{Nnlad} achieve greater than $0.99$ sensitivity and $0.95$ specificity for $k$ upto $10$. \textsc{Comp}-\textsc{Sbl}'s performance degrades gracefully for higher values of $k$.

\section{Synthetic data results using CS algorithms only}

\label{sup-sec:cs_algos_only}
We provide results of running CS algorithms on synthetic data for the $93 \times 961$ and $45\times 105$ Kirkman triple matrices, without doing the \textsc{Comp} preprocessing step, in Tables 
\ref{tab:only_nnlasso}, \ref{tab:only_sbl}, \ref{tab:only_nnomp}, \ref{tab:only_nnlad}, \ref{tab:only_nnls},
\ref{tab:only_nnlasso_45_105}, \ref{tab:only_sbl_45_105}, \ref{tab:only_nnomp_45_105}, \ref{tab:only_nnlad_45_105} and \ref{tab:only_nnls_45_105}.
We find that running the CS algorithms without the \textsc{Comp} step increases false negatives or false positives by a large factor for each algorithm.
We observed that this method was computationally more expensive due to the algorithms running on the full matrix. We used the same methodology as in Sec. \ref*{subsec:synthetic_results}. 
Results for all algorithms are over $1000$ randomly generated signals for each $k$, except for \textsc{Nnlasso} on $93\times 961$ Kirkman triple matrix, for which only $100$ signals were used.

\section{Viral Loads of False Negatives}
\label{sup-sec:viral_loads_false_neg}

\begin{table*}
\centering
\caption{Viral load of false negative samples for the COMP followed by CS algorithms using a $93\times 961$ Kirkman matrix on synthetic data, normalized by the maximum possible viral load. Mean and standard deviation reported over $1000$ randomly generated signals.}
\label{tab:viral_loads_false_negatives}
\begin{tabular}{|c|c|c|c|c|c|}
\hline
$k$  & \textbf{COMP-SBL}      & \textbf{COMP-NNOMP}    & \textbf{COMP-NNLASSO}  & \textbf{COMP-NNLAD}    & \textbf{COMP-NNLS}     \\ \hline
\textbf{5}  & $0.013 \pm 0.007$ & $0.028 \pm 0.031$ & $0.021 \pm 0.011$ & $0.012 \pm 0.014$ & $0.011 \pm 0.008$ \\ \hline
\textbf{8}  & $0.015 \pm 0.017$ & $0.023 \pm 0.024$ & $0.024 \pm 0.022$ & $0.029 \pm 0.032$ & $0.029 \pm 0.030$ \\ \hline
\textbf{10} & $0.023 \pm 0.026$ & $0.032 \pm 0.052$ & $0.035 \pm 0.040$ & $0.034 \pm 0.036$ & $0.030 \pm 0.031$ \\ \hline
\textbf{12} & $0.047 \pm 0.080$ & $0.066 \pm 0.099$ & $0.057 \pm 0.066$ & $0.067 \pm 0.075$ & $0.059 \pm 0.067$ \\ \hline
\textbf{15} & $0.093 \pm 0.134$ & $0.151 \pm 0.181$ & $0.098 \pm 0.107$ & $0.084 \pm 0.083$ & $0.081 \pm 0.082$ \\ \hline
\textbf{17} & $0.192 \pm 0.227$ & $0.224 \pm 0.217$ & $0.110 \pm 0.113$ & $0.085 \pm 0.083$ & $0.086 \pm 0.088$ \\ \hline
\textbf{20} & $0.306 \pm 0.250$ & $0.299 \pm 0.240$ & $0.192 \pm 0.164$ & $0.104 \pm 0.120$ & $0.105 \pm 0.122$ \\ \hline
\end{tabular}

\end{table*}

As seen in Tables \ref{tab:comp} through \ref{tab:comp_nnlad} in the main paper, our method gives far fewer false positives when compared to \textsc{Comp}, at the cost of a rare false negative.
Furthermore, due to the numerical nature of our method, there is asymmetry in our mode of failure, and we fail less on samples with high viral load.
That is, even the rare false negatives that we fail to detect have very small viral loads.
Table \ref{tab:viral_loads_false_negatives} shows the mean and standard deviation of viral loads of false negative samples for obtained by our method on synthetic data for a range of values of $k$.
We see that for $k$ upto $10$, these false negative viral load values are very small.
For example, if we use \textsc{Comp} followed by \textsc{Sbl}, the average viral load value of false negative samples is $0.023 \pm 0.026$, around $40$ times lower than the maximum viral load ($1.0$). This means that our method almost never misses a sample with high viral load. This matters particularly because COVID-19 super-spreaders are believed to have a high viral load \cite{beldomenico2020superspreaders}.

\section{Dependence of relative viral loads on parameter $q$}
\label{sup-sec:viral_load_q}
\begin{table*}
\centering
\caption{Effect of parameter $q$ on the reported relative relative viral loads of real data, using COMP-SBL. Viral loads reported are relative to the viral load content of the pool with the smallest threshold cycle ($C_t$) value in that particular experiment. Sample IDs are the indices of the estimated viral load vector $\widetilde{\x}$ which have value more than $0$ in that experiment, for any value of $q$. In some cases, there may be more sample ID entries for an experiment than the number of ground truth positive samples $k$ in that experiment, due to false positives. Reported viral load of remaining samples is $0$.
}
\label{tab:viral_loads_q}
\begin{tabular}{llllllllll}
\hline
\multicolumn{1}{|l|}{\multirow{4}{*}{\textbf{Harvard $24 \times 60$, $k=2$}}}  & \multicolumn{2}{c|}{\textbf{q $\rightarrow$}}                                                               & \multicolumn{1}{l|}{\textbf{0.7}} & \multicolumn{1}{l|}{\textbf{0.75}} & \multicolumn{1}{l|}{\textbf{0.8}} & \multicolumn{1}{l|}{\textbf{0.85}} & \multicolumn{1}{l|}{\textbf{0.9}} & \multicolumn{1}{l|}{\textbf{0.95}} & \multicolumn{1}{l|}{\textbf{1}} \\ \cline{2-10} 
\multicolumn{1}{|l|}{}                                                         & \multicolumn{1}{l|}{\multirow{3}{*}{\textbf{Sample IDs}}} & \multicolumn{1}{l|}{\textbf{10}}  & \multicolumn{1}{l|}{0.13}         & \multicolumn{1}{l|}{0.11}          & \multicolumn{1}{l|}{0.09}         & \multicolumn{1}{l|}{0.07}          & \multicolumn{1}{l|}{0.06}         & \multicolumn{1}{l|}{0.04}          & \multicolumn{1}{l|}{0.03}       \\ \cline{3-10} 
\multicolumn{1}{|l|}{}                                                         & \multicolumn{1}{l|}{}                                     & \multicolumn{1}{l|}{\textbf{28}}  & \multicolumn{1}{l|}{0.86}         & \multicolumn{1}{l|}{0.85}          & \multicolumn{1}{l|}{0.84}         & \multicolumn{1}{l|}{0.84}          & \multicolumn{1}{l|}{0.83}         & \multicolumn{1}{l|}{0.82}          & \multicolumn{1}{l|}{0.82}       \\ \cline{3-10} 
\multicolumn{1}{|l|}{}                                                         & \multicolumn{1}{l|}{}                                     & \multicolumn{1}{l|}{\textbf{54}}  & \multicolumn{1}{l|}{0.08}         & \multicolumn{1}{l|}{0.09}          & \multicolumn{1}{l|}{0.09}         & \multicolumn{1}{l|}{0.10}          & \multicolumn{1}{l|}{0.10}         & \multicolumn{1}{l|}{0.11}          & \multicolumn{1}{l|}{0.11}       \\ \hline
                                                                               &                                                           &                                   &                                   &                                    &                                   &                                    &                                   &                                    &                                 \\ \hline
\multicolumn{1}{|l|}{\multirow{3}{*}{\textbf{Harvard $30 \times 120$, $k=2$}}} & \multicolumn{2}{c|}{\textbf{q $\rightarrow$}}                                                               & \multicolumn{1}{l|}{\textbf{0.7}} & \multicolumn{1}{l|}{\textbf{0.75}} & \multicolumn{1}{l|}{\textbf{0.8}} & \multicolumn{1}{l|}{\textbf{0.85}} & \multicolumn{1}{l|}{\textbf{0.9}} & \multicolumn{1}{l|}{\textbf{0.95}} & \multicolumn{1}{l|}{\textbf{1}} \\ \cline{2-10} 
\multicolumn{1}{|l|}{}                                                         & \multicolumn{1}{l|}{\multirow{2}{*}{\textbf{Sample IDs}}} & \multicolumn{1}{l|}{\textbf{20}}  & \multicolumn{1}{l|}{0.86}         & \multicolumn{1}{l|}{0.85}          & \multicolumn{1}{l|}{0.85}         & \multicolumn{1}{l|}{0.84}          & \multicolumn{1}{l|}{0.83}         & \multicolumn{1}{l|}{0.83}          & \multicolumn{1}{l|}{0.82}       \\ \cline{3-10} 
\multicolumn{1}{|l|}{}                                                         & \multicolumn{1}{l|}{}                                     & \multicolumn{1}{l|}{\textbf{114}} & \multicolumn{1}{l|}{0.90}         & \multicolumn{1}{l|}{0.89}          & \multicolumn{1}{l|}{0.89}         & \multicolumn{1}{l|}{0.88}          & \multicolumn{1}{l|}{0.88}         & \multicolumn{1}{l|}{0.87}          & \multicolumn{1}{l|}{0.87}       \\ \hline
                                                                               &                                                           &                                   &                                   &                                    &                                   &                                    &                                   &                                    &                                 \\ \hline
\multicolumn{1}{|l|}{\multirow{2}{*}{\textbf{NCBS-1 $16\times 40$, $k = 1$}}}  & \multicolumn{2}{c|}{\textbf{q $\rightarrow$}}                                                               & \multicolumn{1}{l|}{\textbf{0.7}} & \multicolumn{1}{l|}{\textbf{0.75}} & \multicolumn{1}{l|}{\textbf{0.8}} & \multicolumn{1}{l|}{\textbf{0.85}} & \multicolumn{1}{l|}{\textbf{0.9}} & \multicolumn{1}{l|}{\textbf{0.95}} & \multicolumn{1}{l|}{\textbf{1}} \\ \cline{2-10} 
\multicolumn{1}{|l|}{}                                                         & \multicolumn{1}{l|}{\textbf{Sample ID}}                   & \multicolumn{1}{l|}{\textbf{14}}  & \multicolumn{1}{l|}{0.78}         & \multicolumn{1}{l|}{0.77}          & \multicolumn{1}{l|}{0.77}         & \multicolumn{1}{l|}{0.76}          & \multicolumn{1}{l|}{0.75}         & \multicolumn{1}{l|}{0.74}          & \multicolumn{1}{l|}{0.73}       \\ \hline
                                                                               &                                                           &                                   &                                   &                                    &                                   &                                    &                                   &                                    &                                 \\ \hline
\multicolumn{1}{|l|}{\multirow{3}{*}{\textbf{NCBS-2 $16\times 40$, $k = 2$}}}  & \multicolumn{2}{c|}{\textbf{q $\rightarrow$}}                                                               & \multicolumn{1}{l|}{\textbf{0.7}} & \multicolumn{1}{l|}{\textbf{0.75}} & \multicolumn{1}{l|}{\textbf{0.8}} & \multicolumn{1}{l|}{\textbf{0.85}} & \multicolumn{1}{l|}{\textbf{0.9}} & \multicolumn{1}{l|}{\textbf{0.95}} & \multicolumn{1}{l|}{\textbf{1}} \\ \cline{2-10} 
\multicolumn{1}{|l|}{}                                                         & \multicolumn{1}{l|}{\multirow{2}{*}{\textbf{Sample IDs}}} & \multicolumn{1}{l|}{\textbf{9}}   & \multicolumn{1}{l|}{0.59}         & \multicolumn{1}{l|}{0.58}          & \multicolumn{1}{l|}{0.58}         & \multicolumn{1}{l|}{0.57}          & \multicolumn{1}{l|}{0.57}         & \multicolumn{1}{l|}{0.56}          & \multicolumn{1}{l|}{0.56}       \\ \cline{3-10} 
\multicolumn{1}{|l|}{}                                                         & \multicolumn{1}{l|}{}                                     & \multicolumn{1}{l|}{\textbf{22}}  & \multicolumn{1}{l|}{0.60}         & \multicolumn{1}{l|}{0.59}          & \multicolumn{1}{l|}{0.59}         & \multicolumn{1}{l|}{0.59}          & \multicolumn{1}{l|}{0.58}         & \multicolumn{1}{l|}{0.58}          & \multicolumn{1}{l|}{0.57}       \\ \hline
                                                                               &                                                           &                                   &                                   &                                    &                                   &                                    &                                   &                                    &                                 \\ \hline
\multicolumn{1}{|l|}{\multirow{4}{*}{\textbf{NCBS-3 $16\times 40$, $k = 3$}}}  & \multicolumn{2}{c|}{\textbf{q $\rightarrow$}}                                                               & \multicolumn{1}{l|}{\textbf{0.7}} & \multicolumn{1}{l|}{\textbf{0.75}} & \multicolumn{1}{l|}{\textbf{0.8}} & \multicolumn{1}{l|}{\textbf{0.85}} & \multicolumn{1}{l|}{\textbf{0.9}} & \multicolumn{1}{l|}{\textbf{0.95}} & \multicolumn{1}{l|}{\textbf{1}} \\ \cline{2-10} 
\multicolumn{1}{|l|}{}                                                         & \multicolumn{1}{l|}{\multirow{3}{*}{\textbf{Sample IDs}}} & \multicolumn{1}{l|}{\textbf{4}}   & \multicolumn{1}{l|}{0.011}        & \multicolumn{1}{l|}{0.008}         & \multicolumn{1}{l|}{0.005}        & \multicolumn{1}{l|}{0.004}         & \multicolumn{1}{l|}{0.003}        & \multicolumn{1}{l|}{0.003}         & \multicolumn{1}{l|}{0.002}      \\ \cline{3-10} 
\multicolumn{1}{|l|}{}                                                         & \multicolumn{1}{l|}{}                                     & \multicolumn{1}{l|}{\textbf{6}}   & \multicolumn{1}{l|}{0.07}         & \multicolumn{1}{l|}{0.06}          & \multicolumn{1}{l|}{0.05}         & \multicolumn{1}{l|}{0.05}          & \multicolumn{1}{l|}{0.04}         & \multicolumn{1}{l|}{0.04}          & \multicolumn{1}{l|}{0.03}       \\ \cline{3-10} 
\multicolumn{1}{|l|}{}                                                         & \multicolumn{1}{l|}{}                                     & \multicolumn{1}{l|}{\textbf{23}}  & \multicolumn{1}{l|}{0.77}         & \multicolumn{1}{l|}{0.76}          & \multicolumn{1}{l|}{0.76}         & \multicolumn{1}{l|}{0.75}          & \multicolumn{1}{l|}{0.75}         & \multicolumn{1}{l|}{0.74}          & \multicolumn{1}{l|}{0.74}       \\ \hline
                                                                               &                                                           &                                   &                                   &                                    &                                   &                                    &                                   &                                    &                                 \\ \hline
\multicolumn{1}{|l|}{\multirow{6}{*}{\textbf{NCBS-4 $16\times 40$, $k = 4$}}}  & \multicolumn{2}{c|}{\textbf{q $\rightarrow$}}                                                               & \multicolumn{1}{l|}{\textbf{0.7}} & \multicolumn{1}{l|}{\textbf{0.75}} & \multicolumn{1}{l|}{\textbf{0.8}} & \multicolumn{1}{l|}{\textbf{0.85}} & \multicolumn{1}{l|}{\textbf{0.9}} & \multicolumn{1}{l|}{\textbf{0.95}} & \multicolumn{1}{l|}{\textbf{1}} \\ \cline{2-10} 
\multicolumn{1}{|l|}{}                                                         & \multicolumn{1}{l|}{\multirow{5}{*}{\textbf{Sample IDs}}} & \multicolumn{1}{l|}{\textbf{11}}  & \multicolumn{1}{l|}{0.014}        & \multicolumn{1}{l|}{0.009}         & \multicolumn{1}{l|}{0.006}        & \multicolumn{1}{l|}{0.004}         & \multicolumn{1}{l|}{0.003}        & \multicolumn{1}{l|}{0.003}         & \multicolumn{1}{l|}{0.003}      \\ \cline{3-10} 
\multicolumn{1}{|l|}{}                                                         & \multicolumn{1}{l|}{}                                     & \multicolumn{1}{l|}{\textbf{17}}  & \multicolumn{1}{l|}{0.04}         & \multicolumn{1}{l|}{0.04}          & \multicolumn{1}{l|}{0.04}         & \multicolumn{1}{l|}{0.04}          & \multicolumn{1}{l|}{0.03}         & \multicolumn{1}{l|}{0.03}          & \multicolumn{1}{l|}{0.03}       \\ \cline{3-10} 
\multicolumn{1}{|l|}{}                                                         & \multicolumn{1}{l|}{}                                     & \multicolumn{1}{l|}{\textbf{18}}  & \multicolumn{1}{l|}{0.001}        & \multicolumn{1}{l|}{0.002}         & \multicolumn{1}{l|}{0.002}        & \multicolumn{1}{l|}{0.003}         & \multicolumn{1}{l|}{0.003}        & \multicolumn{1}{l|}{0.003}         & \multicolumn{1}{l|}{0.004}      \\ \cline{3-10} 
\multicolumn{1}{|l|}{}                                                         & \multicolumn{1}{l|}{}                                     & \multicolumn{1}{l|}{\textbf{33}}  & \multicolumn{1}{l|}{0.94}         & \multicolumn{1}{l|}{0.94}          & \multicolumn{1}{l|}{0.94}         & \multicolumn{1}{l|}{0.94}          & \multicolumn{1}{l|}{0.94}         & \multicolumn{1}{l|}{0.93}          & \multicolumn{1}{l|}{0.93}       \\ \cline{3-10} 
\multicolumn{1}{|l|}{}                                                         & \multicolumn{1}{l|}{}                                     & \multicolumn{1}{l|}{\textbf{36}}  & \multicolumn{1}{l|}{0.11}         & \multicolumn{1}{l|}{0.10}          & \multicolumn{1}{l|}{0.09}         & \multicolumn{1}{l|}{0.08}          & \multicolumn{1}{l|}{0.07}         & \multicolumn{1}{l|}{0.06}          & \multicolumn{1}{l|}{0.05}       \\ \hline
\end{tabular}
\end{table*}

Typical RT-PCR output will not give us viral loads of the pools, rather, only the threshold cycle ($C_t$) values (see Sec. \ref{sec:rtpcr} of the main paper). In order to convert from the $C_t$ values to the relative viral load vector, we use Eqn. (8), as explained in Sec. \ref*{subsec:noisemodel} of the main paper. For this, we need the value of the parameter $q$, which we set to $0.95$, since we expect the viral amount in each pool to roughly double in each RT-PCR cycle.

The parameter $q$ may be estimated from raw RT-PCR data, which contains the fluorescence values found at the end of each RT-PCR thermal cycle for each pool.
If we take the natural logarithm on both Eqn. (5) (page 3 of the main paper), we can see that the logarithm of the fluorescence of a pool has a linear dependence on cycle time.
Hence, $q$ may be obtained by performing linear regression on log fluorescence values for any (positive) pool \cite{thermofisher}.

The value of $q$ thus obtained will have some dependence on the pool so chosen.
However, the relative viral loads of declared positives obtained by our algorithm show negligible variance in most cases, especially for samples with high viral load, over a large range of $q$ (see Table \ref{tab:viral_loads_q}). In some cases, especially for samples with low viral loads, we do observe some variance. However, it never happens that a sample with high relative viral load is reported to have low relative viral load, or vice-versa.
Moreover, the declared positives do not change for any experiment.

\section{Sensitivity of results to choice of threshold $\tau$}
\label{sup-sec:tau}

\begin{table*}
\centering
\caption{Sensitivity of CS algorithms to the choice of threshold $\tau$. Reported numbers are for $1000$ signals with $k=10$ on the $93\times 961$ Kirkman matrix.}
\label{tab:tau_sensitivity}
\begin{tabular}{|c|c|c|c|c|c|}
\hline
\multicolumn{6}{|c|}{\textbf{COMP-NNLAD}}                                                       \\ \hline
$\tau$ & \textbf{RMSE} & \textbf{\#FN} & \textbf{\#FP}  & \textbf{Sens.} & \textbf{Spec.} \\ \hline
\textbf{0}   & $0.1060\pm0.0527$ & $0.0000\pm0.0000$ & $15.4640\pm4.6299$ & $1.0000\pm0.0000$  & $0.9837\pm0.0049$  \\ \hline
\textbf{0.1} & $0.1060\pm0.0527$ & $0.2030\pm0.4242$ & $9.3480\pm3.1325 $ & $0.9797\pm0.0424$  & $0.9902\pm0.0033$  \\ \hline
\textbf{0.2} & $0.1060\pm0.0527$ & $0.2030\pm0.4242$ & $9.3450\pm3.1308 $ & $0.9797\pm0.0424$  & $0.9902\pm0.0033$  \\ \hline
\textbf{0.3} & $0.1060\pm0.0527$ & $0.2030\pm0.4242$ & $9.3440\pm3.1298 $ & $0.9797\pm0.0424$  & $0.9902\pm0.0033$  \\ \hline
\textbf{0.4} & $0.1060\pm0.0527$ & $0.2030\pm0.4242$ & $9.3420\pm3.1309 $ & $0.9797\pm0.0424$  & $0.9902\pm0.0033$  \\ \hline
\textbf{0.5} & $0.1060\pm0.0527$ & $0.2030\pm0.4242$ & $9.3420\pm3.1309 $ & $0.9797\pm0.0424$  & $0.9902\pm0.0033$  \\ \hline
\textbf{0.6} & $0.1060\pm0.0527$ & $0.2030\pm0.4242$ & $9.3400\pm3.1315 $ & $0.9797\pm0.0424$  & $0.9902\pm0.0033$  \\ \hline
\textbf{0.7} & $0.1060\pm0.0527$ & $0.2030\pm0.4242$ & $9.3380\pm3.1314 $ & $0.9797\pm0.0424$  & $0.9902\pm0.0033$  \\ \hline
\textbf{0.8} & $0.1060\pm0.0527$ & $0.2030\pm0.4242$ & $9.3360\pm3.1319 $ & $0.9797\pm0.0424$  & $0.9902\pm0.0033$  \\ \hline
\textbf{0.9} & $0.1060\pm0.0527$ & $0.2030\pm0.4242$ & $9.3340\pm3.1315 $ & $0.9797\pm0.0424$  & $0.9902\pm0.0033$  \\ \hline
\textbf{1}   & $0.1060\pm0.0527$ & $0.2030\pm0.4242$ & $9.3310\pm3.1317 $ & $0.9797\pm0.0424$  & $0.9902\pm0.0033$  \\ \hline
\multicolumn{6}{|c|}{\textbf{COMP-NNLS}}                                                        \\ \hline
$\tau$ & \textbf{RMSE} & \textbf{\#FN} & \textbf{\#FP}  & \textbf{Sens.} & \textbf{Spec.} \\ \hline
\textbf{0}   & $0.0981\pm0.0520$ & $0.0000\pm0.0000$ & $15.4250\pm4.4020$ & $1.0000\pm0.0000$  & $0.9838\pm0.0046$  \\ \hline
\textbf{0.1} & $0.0981\pm0.0520$ & $0.2540\pm0.4896$ & $7.9430\pm3.2712 $ & $0.9746\pm0.0490$  & $0.9916\pm0.0034$  \\ \hline
\textbf{0.2} & $0.0981\pm0.0520$ & $0.2540\pm0.4896$ & $7.9430\pm3.2712 $ & $0.9746\pm0.0490$  & $0.9916\pm0.0034$  \\ \hline
\textbf{0.3} & $0.0981\pm0.0520$ & $0.2540\pm0.4896$ & $7.9420\pm3.2701 $ & $0.9746\pm0.0490$  & $0.9916\pm0.0034$  \\ \hline
\textbf{0.4} & $0.0981\pm0.0520$ & $0.2540\pm0.4896$ & $7.9420\pm3.2701 $ & $0.9746\pm0.0490$  & $0.9916\pm0.0034$  \\ \hline
\textbf{0.5} & $0.0981\pm0.0520$ & $0.2540\pm0.4896$ & $7.9420\pm3.2701 $ & $0.9746\pm0.0490$  & $0.9916\pm0.0034$  \\ \hline
\textbf{0.6} & $0.0981\pm0.0520$ & $0.2540\pm0.4896$ & $7.9420\pm3.2701 $ & $0.9746\pm0.0490$  & $0.9916\pm0.0034$  \\ \hline
\textbf{0.7} & $0.0981\pm0.0520$ & $0.2540\pm0.4896$ & $7.9420\pm3.2701 $ & $0.9746\pm0.0490$  & $0.9916\pm0.0034$  \\ \hline
\textbf{0.8} & $0.0981\pm0.0520$ & $0.2540\pm0.4896$ & $7.9420\pm3.2701 $ & $0.9746\pm0.0490$  & $0.9916\pm0.0034$  \\ \hline
\textbf{0.9} & $0.0981\pm0.0520$ & $0.2540\pm0.4896$ & $7.9420\pm3.2701 $ & $0.9746\pm0.0490$  & $0.9916\pm0.0034$  \\ \hline
\textbf{1}   & $0.0981\pm0.0520$ & $0.2540\pm0.4896$ & $7.9420\pm3.2701 $ & $0.9746\pm0.0490$  & $0.9916\pm0.0034$  \\ \hline
\multicolumn{6}{|c|}{\textbf{COMP-NNLASSO}}                                                     \\ \hline
$\tau$ & \textbf{RMSE} & \textbf{\#FN} & \textbf{\#FP}  & \textbf{Sens.} & \textbf{Spec.} \\ \hline
\textbf{0}   & $0.0989\pm0.0538$ & $0.0000\pm0.0000$ & $15.1620\pm4.4452$ & $1.0000\pm0.0000$  & $0.9841\pm0.0047$  \\ \hline
\textbf{0.1} & $0.0989\pm0.0538$ & $0.2570\pm0.4870$ & $7.7850\pm3.3660 $ & $0.9743\pm0.0487$  & $0.9918\pm0.0035$  \\ \hline
\textbf{0.2} & $0.0989\pm0.0538$ & $0.2570\pm0.4870$ & $7.7760\pm3.3540 $ & $0.9743\pm0.0487$  & $0.9918\pm0.0035$  \\ \hline
\textbf{0.3} & $0.0989\pm0.0538$ & $0.2590\pm0.4880$ & $7.7730\pm3.3530 $ & $0.9741\pm0.0488$  & $0.9918\pm0.0035$  \\ \hline
\textbf{0.4} & $0.0989\pm0.0538$ & $0.2590\pm0.4880$ & $7.7670\pm3.3520 $ & $0.9741\pm0.0488$  & $0.9918\pm0.0035$  \\ \hline
\textbf{0.5} & $0.0989\pm0.0538$ & $0.2600\pm0.4885$ & $7.7660\pm3.3527 $ & $0.9740\pm0.0489$  & $0.9918\pm0.0035$  \\ \hline
\textbf{0.6} & $0.0989\pm0.0538$ & $0.2600\pm0.4885$ & $7.7640\pm3.3516 $ & $0.9740\pm0.0489$  & $0.9918\pm0.0035$  \\ \hline
\textbf{0.7} & $0.0989\pm0.0538$ & $0.2600\pm0.4885$ & $7.7640\pm3.3516 $ & $0.9740\pm0.0489$  & $0.9918\pm0.0035$  \\ \hline
\textbf{0.8} & $0.0989\pm0.0538$ & $0.2600\pm0.4885$ & $7.7620\pm3.3497 $ & $0.9740\pm0.0489$  & $0.9918\pm0.0035$  \\ \hline
\textbf{0.9} & $0.0989\pm0.0538$ & $0.2610\pm0.4890$ & $7.7620\pm3.3497 $ & $0.9739\pm0.0489$  & $0.9918\pm0.0035$  \\ \hline
\textbf{1}   & $0.0989\pm0.0538$ & $0.2610\pm0.4890$ & $7.7610\pm3.3513 $ & $0.9739\pm0.0489$  & $0.9918\pm0.0035$  \\ \hline
\end{tabular}
\end{table*}
As discussed in section \ref{subsec:comparison_criteria}, we use a threshold of $\tau = 0.2 \times x_{\min}$, below which entries of the estimated viral load vector $\hat{\boldsymbol x}$ are set to $0$. Here $x_{\min} = 1.0$ for our synthetic data. This is needed for the algorithms \textsc{Comp}-\textsc{Nnlasso}, \textsc{Comp}-\textsc{Nnlad} and \textsc{Comp}-\textsc{\textsc{Nnls}}.
Table \ref{tab:tau_sensitivity} shows the variance in performance of these algorithms for different choices of $\tau$, for the case when the number of infected samples $k = 10$. We see that $\tau = 0$ gives many false positives as compared to the other choices. For other values of $\tau$, only \textsc{Comp}-\textsc{Nnlasso} shows some variance. We chose $\tau = 0.2$ to provide good tradeoff between sensitivity and specificity.

\section{Optimal Expected Number of Dorfman Tests}
\label{sup-sec:optimal_dorfman}
A formula for the expected number of tests using Dorfman Testing is presented in \cite{Dorfman1943}. However, it is valid only for the case when the number of samples is a multiple of the pool size. We slightly modify their derivation to handle the case when it is not so. 
Let there be $n$ samples $\{1\dots n\}$, with disease prevalence rate $p = k/n$.
That is, any given sample is positive with probability $p = k/n$, independently of other samples.
Thus, out of $n$ samples, the expected number of samples which are infected is $k$.
Let the samples be divided into $\floor{n/g}$ pools of size $g \geq 2$, and one pool of size $r = n - \floor{n/g}g$.
Since $p$ is the probability that a given sample is infected, hence $1-p$ is the probability that a given sample is not infected.
Then $p'(l) = 1 - (1-p)^l$ is the probability of a pool of size $l$ being infected.
Let $T_g$ be the total number of first stage and second stage tests taken by the Dorfman testing method, given the pool size $g$.
Let the pools thus formed be numbered $\{1\dots \floor{n/g} + 1\}$, with a $0$-sized last pool if $r=0$ for the purposes of our computation. Also, let $t_j = 1$ if the $j^{\text{th}}$ pool tested positive, $0$ otherwise. The $0$-sized pool is considered to be tested as negative. Also note that if $r=1$, the last pool need not be retested even if it tested as positive. Let $\bar r$ denote the number of tests done for the last pool if it tested positive. Hence $\bar r = 0$ if $r \leq 1$, else $\bar r = r$. Then,

\begin{equation}
T_g = \sum_{j=1}^{\floor{\frac{n}{g}}} t_j g + \bar rt_{\floor{\frac{n}{g}} + 1} + \ceil{\frac{n}{g}}.
\end{equation}

By linearity of expectations, we get
\begin{equation}
E[T_g] = g\floor{\frac{n}{g}}p'(g) +  \bar rp'(r) + \ceil{\frac{n}{g}}.
\end{equation}
The optimal expected number of tests is $\underset{g}{\min} \text{ }E[T_g]$, and the optimal pool size is $g^* = \underset{g}{\arg\min} \text{ }E[T_g]$. Since there is no closed form for this, we implemented this numerically in practice -  see Table \ref{tab:dorfman} from the main paper. 
\section{Fraction of samples remaining after \textsc{Comp} using Kirkman matrices}
\label{sup-sec:kirkman_f}
\begin{proposition}
    Let $\A$ be an $m \times n$ full Kirkman matrix. Let $\y=\A\x$ be a measurement vector for some $\x \in \mathbb{R}^n$ such that $\|\x\|_0 = k$. Let $f \coloneqq \|\y\|_0/m \in (0, 1)$ be the fraction of pools that are tested positive. Then the fraction of samples declared positive by \textsc{Comp} is strictly less than $f^2$.
\end{proposition}

\begin{proof}
We will prove the proposition for the case when $\A$ is a Steiner Triple System matrix. Since every full Kirkman matrix is also a Steiner Triple System matrix, the same proof holds.

Let $\Y$ be the set of pools that tested positive and $\X$ be the set of samples declared positive by \textsc{Comp}.
Then, we have $|\Y| = fm$.
Recall from Sec. \ref{subsubsec:kirkman} of the main paper that since $\A$ is a Steiner Triple System matrix each column consists of $3$ entries with value $1$. 
Notice that any column which does not have all three of its $1$ entries in $\Y$ would have gotten eliminated by \textsc{Comp}.
Hence the reduced matrix $A_{\X, \Y}$ also has the property that each column consists of $3$ entries with value $1$.

Recall from Sec. \ref{subsubsec:kirkman} of the main paper that in a Steiner Triple System matrix, each column corresponds to a triplet of rows, and each (unordered) pair of rows occurs together in exactly $1$ such triplet.
Hence each column of $\A$ corresponds to ${3\choose 2} = 3$ unique pairs of rows of $\A$.
Consequently, in the reduced matrix $\redA$, each column corresponds to $3$ unique pairs of rows of $\redA$, since for each column in $\redA$, all the rows with $1$ entries are in $\Y$.

Hence, for $\redA$, total number of such unique pairs of rows as enumerated by its columns is $3|\X|$. This number must be less than or equal to the maximum number of unique pairs of rows of $\redA$, which is ${|\Y| \choose 2} = fm(fm-1)/2$.

Therefore, we have
\begin{equation}
|\X| \le \frac{fm(fm-1)}{6} < \frac{fm(fm-f)}{6} = f^2 \frac{m(m-1)}{6}.
\end{equation}

Recall from Sec. \ref{subsubsec:kirkman} of the main paper that the number of columns $n$ of a Steiner Triple System matrix is equal to ${m \choose 2}/3 = m(m-1)/6$. Hence,

\begin{equation}
\frac{|\X|}{n} < f^2.
\end{equation}

\end{proof}

This gives some intuition as to why \textsc{Comp} eliminates so many samples when using Kirkman matrices, as observed from Table \ref{tab:comp} of the main paper, especially in the regime $k \ll m$. Since each sample can only make $3$ tests positive, number of positive tests is at most $3k$. Hence $f \leq 3k/m$ for Kirkman matrices. For example, when using a full Kirkman matrix, if the fractions of positive tests $f$ is $0.1$, fraction of samples that remain after \textsc{Comp} is at most $f^2 = 0.01$. Similarly, if $f = 0.3$, then $f^2=0.09$, and so on.

\begin{table*}
\centering
\caption{Performance of \textsc{Comp} and \textsc{Dd} (on synthetic data) for $45 \times 105$ Kirkman triple matrix. For each criterion and each $k$ value, mean and standard deviation values are reported across 1000 signals.}
\label{tab:comp_45_105}
\begin{tabular}{|c|c|c|c|c|c|c|}
\hline
$k$ & \textbf{RMSE} & \textbf{\#FN} & \textbf{\#FP} & \textbf{Sens.} & \textbf{Spec.} & $\mathbf{\#\mathcal{HCP}}$\\ \hline
5          & 1.000 $\pm$ 0.000   & 0.0 $\pm$ 0.0       & 1.0 $\pm$ 1.0       & 1.0000 $\pm$ 0.0000  & 0.9899 $\pm$ 0.0099  & $4.8$\\ \hline
8          & 1.000 $\pm$ 0.000   & 0.0 $\pm$ 0.0       & 4.4 $\pm$ 2.2       & 1.0000 $\pm$ 0.0000  & 0.9541 $\pm$ 0.0223  & $5.2$\\ \hline
10         & 1.000 $\pm$ 0.000   & 0.0 $\pm$ 0.0       & 8.0 $\pm$ 3.2       & 1.0000 $\pm$ 0.0000  & 0.9163 $\pm$ 0.0338  & $4.0$\\ \hline
12         & 1.000 $\pm$ 0.000   & 0.0 $\pm$ 0.0       & 12.2 $\pm$ 4.1      & 1.0000 $\pm$ 0.0000  & 0.8689 $\pm$ 0.0446  & $2.5$\\ \hline
15         & 1.000 $\pm$ 0.000   & 0.0 $\pm$ 0.0       & 19.9 $\pm$ 5.8      & 1.0000 $\pm$ 0.0000  & 0.7791 $\pm$ 0.0647  & $0.9$\\ \hline
17         & 1.000 $\pm$ 0.000   & 0.0 $\pm$ 0.0       & 24.9 $\pm$ 6.6      & 1.0000 $\pm$ 0.0000  & 0.7174 $\pm$ 0.0747  & $0.5$\\ \hline
20         & 1.000 $\pm$ 0.000   & 0.0 $\pm$ 0.0       & 32.0 $\pm$ 8.1      & 1.0000 $\pm$ 0.0000  & 0.6233 $\pm$ 0.0955  & $0.1$\\ \hline
\end{tabular}
\end{table*}

\begin{table*}
\centering
\caption{Performance of COMP followed by NNLASSO (on synthetic data) for $45 \times 105$ Kirkman triple matrix. For each criterion and each $k$ value, mean and standard deviation values are reported across 1000 signals.}
\label{tab:comp_nnlasso_45_105}
\begin{tabular}{|c|c|c|c|c|c|}
\hline
k             & \textbf{RMSE}       & \textbf{\#FN}   & \textbf{\#FP}    & \textbf{Sens.}        & \textbf{Spec.}        \\ \hline
\textbf{$5$}  & $ 0.047 \pm 0.020 $ & $ 0.0 \pm 0.1 $ & $ 0.5 \pm 0.7 $  & $ 0.9994 \pm 0.0109 $ & $ 0.9949 \pm 0.0072 $ \\ \hline
\textbf{$8$}  & $ 0.064 \pm 0.024 $ & $ 0.0 \pm 0.2 $ & $ 2.3 \pm 1.6 $  & $ 0.9941 \pm 0.0265 $ & $ 0.9761 \pm 0.0165 $ \\ \hline
\textbf{$10$} & $ 0.079 \pm 0.031 $ & $ 0.1 \pm 0.3 $ & $ 3.9 \pm 2.2 $  & $ 0.9892 \pm 0.0317 $ & $ 0.9585 \pm 0.0230 $ \\ \hline
\textbf{$12$} & $ 0.100 \pm 0.041 $ & $ 0.3 \pm 0.5 $ & $ 6.1 \pm 2.9 $  & $ 0.9783 \pm 0.0409 $ & $ 0.9340 \pm 0.0313 $ \\ \hline
\textbf{$15$} & $ 0.143 \pm 0.076 $ & $ 0.5 \pm 0.7 $ & $ 9.4 \pm 3.9 $  & $ 0.9654 \pm 0.0465 $ & $ 0.8956 \pm 0.0432 $ \\ \hline
\textbf{$17$} & $ 0.178 \pm 0.097 $ & $ 0.9 \pm 0.9 $ & $ 12.0 \pm 5.3 $ & $ 0.9493 \pm 0.0521 $ & $ 0.8639 \pm 0.0605 $ \\ \hline
\textbf{$20$} & $ 0.256 \pm 0.131 $ & $ 1.2 \pm 1.1 $ & $ 17.1 \pm 9.5 $ & $ 0.9380 \pm 0.0557 $ & $ 0.7993 \pm 0.1123 $ \\ \hline
\end{tabular}
\end{table*}

\begin{table*}
\centering
\caption{Performance of COMP followed by SBL (on synthetic data) for $45 \times 105$ Kirkman triple matrix. For each criterion and each $k$ value, mean and standard deviation values are reported across 1000 signals.}
\label{tab:comp_sbl_45_105}
\begin{tabular}{|c|c|c|c|c|c|}
\hline
$k$ & \textbf{RMSE} & \textbf{\#FN} & \textbf{\#FP} & \textbf{Sens.} & \textbf{Spec.} \\ \hline
5          & 0.046 $\pm$ 0.019   & 0.0 $\pm$ 0.1       & 0.5 $\pm$ 0.7       & 0.9994 $\pm$ 0.0109  & 0.9945 $\pm$ 0.0075  \\ \hline
8          & 0.058 $\pm$ 0.020   & 0.0 $\pm$ 0.1       & 2.4 $\pm$ 1.6       & 0.9974 $\pm$ 0.0179  & 0.9749 $\pm$ 0.0169  \\ \hline
10         & 0.070 $\pm$ 0.023   & 0.1 $\pm$ 0.3       & 4.3 $\pm$ 2.3       & 0.9930 $\pm$ 0.0255  & 0.9550 $\pm$ 0.0241  \\ \hline
12         & 0.085 $\pm$ 0.035   & 0.1 $\pm$ 0.3       & 6.7 $\pm$ 2.9       & 0.9911 $\pm$ 0.0271  & 0.9281 $\pm$ 0.0312  \\ \hline
15         & 0.112 $\pm$ 0.041   & 0.2 $\pm$ 0.5       & 10.8 $\pm$ 4.0      & 0.9846 $\pm$ 0.0319  & 0.8799 $\pm$ 0.0445  \\ \hline
17         & 0.142 $\pm$ 0.081   & 0.4 $\pm$ 0.6       & 13.7 $\pm$ 4.5      & 0.9754 $\pm$ 0.0376  & 0.8445 $\pm$ 0.0517  \\ \hline
20         & 0.195 $\pm$ 0.145   & 0.7 $\pm$ 0.9       & 18.0 $\pm$ 5.8      & 0.9628 $\pm$ 0.0443  & 0.7885 $\pm$ 0.0685  \\ \hline
\end{tabular}
\end{table*}

\begin{table*}
\centering
\caption{Performance of COMP followed by NNOMP (on synthetic data) for $45 \times 105$ Kirkman triple matrix. For each criterion and each $k$ value, mean and standard deviation values are reported across 1000 signals.}
\label{tab:comp_nnomp_45_105}
\begin{tabular}{|c|c|c|c|c|c|}
\hline
$k$ & \textbf{RMSE} & \textbf{\#FN} & \textbf{\#FP} & \textbf{Sens.} & \textbf{Spec.} \\ \hline
5          & 0.046 $\pm$ 0.020   & 0.0 $\pm$ 0.1       & 0.2 $\pm$ 0.5       & 0.9984 $\pm$ 0.0178  & 0.9980 $\pm$ 0.0047  \\ \hline
8          & 0.060 $\pm$ 0.024   & 0.1 $\pm$ 0.3       & 1.0 $\pm$ 1.3       & 0.9882 $\pm$ 0.0377  & 0.9895 $\pm$ 0.0134  \\ \hline
10         & 0.073 $\pm$ 0.029   & 0.2 $\pm$ 0.5       & 2.2 $\pm$ 2.1       & 0.9774 $\pm$ 0.0478  & 0.9769 $\pm$ 0.0217  \\ \hline
12         & 0.090 $\pm$ 0.036   & 0.3 $\pm$ 0.6       & 3.9 $\pm$ 2.8       & 0.9713 $\pm$ 0.0503  & 0.9582 $\pm$ 0.0300  \\ \hline
15         & 0.135 $\pm$ 0.069   & 0.6 $\pm$ 0.8       & 7.7 $\pm$ 3.7       & 0.9573 $\pm$ 0.0528  & 0.9139 $\pm$ 0.0414  \\ \hline
17         & 0.175 $\pm$ 0.103   & 1.0 $\pm$ 1.0       & 10.2 $\pm$ 3.8      & 0.9406 $\pm$ 0.0590  & 0.8842 $\pm$ 0.0426  \\ \hline
20         & 0.266 $\pm$ 0.169   & 2.0 $\pm$ 1.6       & 12.7 $\pm$ 3.8      & 0.8998 $\pm$ 0.0789  & 0.8502 $\pm$ 0.0448  \\ \hline
\end{tabular}
\end{table*}

\begin{table*}
\centering
\caption{Performance of COMP followed by NNLAD (on synthetic data) for $45 \times 105$ Kirkman triple matrix. For each criterion and each $k$ value, mean and standard deviation values are reported across 1000 signals.}
\label{tab:comp_nnlad_45_105}
\begin{tabular}{|c|c|c|c|c|c|}
\hline
\textbf{k}    & \textbf{RMSE}       & \textbf{\#FN}   & \textbf{\#FP}     & \textbf{Sens.}        & \textbf{Spec.}        \\ \hline
\textbf{$5$}  & $ 0.050 \pm 0.022 $ & $ 0.0 \pm 0.0 $ & $ 0.6 \pm 0.8 $   & $ 0.9998 \pm 0.0063 $ & $ 0.9936 \pm 0.0079 $ \\ \hline
\textbf{$8$}  & $ 0.068 \pm 0.028 $ & $ 0.0 \pm 0.2 $ & $ 2.7 \pm 1.7 $   & $ 0.9968 \pm 0.0199 $ & $ 0.9721 \pm 0.0174 $ \\ \hline
\textbf{$10$} & $ 0.087 \pm 0.037 $ & $ 0.1 \pm 0.3 $ & $ 5.0 \pm 2.3 $   & $ 0.9903 \pm 0.0309 $ & $ 0.9479 \pm 0.0246 $ \\ \hline
\textbf{$12$} & $ 0.105 \pm 0.041 $ & $ 0.2 \pm 0.4 $ & $ 7.3 \pm 3.0 $   & $ 0.9841 \pm 0.0356 $ & $ 0.9215 \pm 0.0318 $ \\ \hline
\textbf{$15$} & $ 0.150 \pm 0.070 $ & $ 0.5 \pm 0.7 $ & $ 10.5 \pm 3.9 $  & $ 0.9693 \pm 0.0441 $ & $ 0.8837 \pm 0.0431 $ \\ \hline
\textbf{$17$} & $ 0.195 \pm 0.100 $ & $ 0.8 \pm 0.8 $ & $ 13.0 \pm 5.7 $  & $ 0.9554 \pm 0.0491 $ & $ 0.8527 \pm 0.0644 $ \\ \hline
\textbf{$20$} & $ 0.261 \pm 0.126 $ & $ 1.1 \pm 1.1 $ & $ 18.1 \pm 10.2 $ & $ 0.9438 \pm 0.0548 $ & $ 0.7870 \pm 0.1203 $ \\ \hline
\end{tabular}

\end{table*}

\begin{table*}
\centering
\caption{Performance of COMP followed by NNLS (on synthetic data) for $45 \times 105$ Kirkman triple matrix. For each criterion and each $k$ value, mean and standard deviation values are reported across 1000 signals.}
\label{tab:comp_nnls_45_105}
\begin{tabular}{|c|c|c|c|c|c|}
\hline
\textbf{k}    & \textbf{RMSE}       & \textbf{\#FN}   & \textbf{\#FP}     & \textbf{Sens.}        & \textbf{Spec.}        \\ \hline
\textbf{$5$}  & $ 0.046 \pm 0.019 $ & $ 0.0 \pm 0.0 $ & $ 0.5 \pm 0.7 $   & $ 0.9998 \pm 0.0063 $ & $ 0.9947 \pm 0.0071 $ \\ \hline
\textbf{$8$}  & $ 0.063 \pm 0.023 $ & $ 0.0 \pm 0.2 $ & $ 2.2 \pm 1.5 $   & $ 0.9944 \pm 0.0265 $ & $ 0.9772 \pm 0.0159 $ \\ \hline
\textbf{$10$} & $ 0.079 \pm 0.033 $ & $ 0.1 \pm 0.4 $ & $ 4.1 \pm 2.2 $   & $ 0.9863 \pm 0.0358 $ & $ 0.9565 \pm 0.0234 $ \\ \hline
\textbf{$12$} & $ 0.097 \pm 0.036 $ & $ 0.2 \pm 0.5 $ & $ 6.2 \pm 2.9 $   & $ 0.9810 \pm 0.0382 $ & $ 0.9337 \pm 0.0313 $ \\ \hline
\textbf{$15$} & $ 0.138 \pm 0.068 $ & $ 0.5 \pm 0.7 $ & $ 9.3 \pm 4.0 $   & $ 0.9660 \pm 0.0469 $ & $ 0.8966 \pm 0.0444 $ \\ \hline
\textbf{$17$} & $ 0.183 \pm 0.099 $ & $ 0.8 \pm 0.8 $ & $ 12.1 \pm 5.9 $  & $ 0.9525 \pm 0.0500 $ & $ 0.8629 \pm 0.0667 $ \\ \hline
\textbf{$20$} & $ 0.251 \pm 0.129 $ & $ 1.2 \pm 1.1 $ & $ 17.6 \pm 10.5 $ & $ 0.9425 \pm 0.0553 $ & $ 0.7935 \pm 0.1231 $ \\ \hline
\end{tabular}
\end{table*}


\begin{table*}
\centering
\caption{Performance of NNLASSO without COMP (on synthetic data) for $93 \times 961$ Kirkman triple matrix. For each criterion, mean and standard deviation are reported over 100 signals.}
\label{tab:only_nnlasso}
\begin{tabular}{|c|c|c|c|c|c|}
\hline
$k$ & \textbf{RMSE} & \textbf{\#FN} & \textbf{\#FP} & \textbf{Sens.} & \textbf{Spec.} \\ \hline
5  & 0.254 $\pm$ 0.368 & 0.1 $\pm$ 0.3 & 78.2 $\pm$ 28.2  & 0.9780 $\pm$ 0.0690 & 0.9182 $\pm$ 0.0295 \\ \hline
8  & 0.326 $\pm$ 0.360 & 0.2 $\pm$ 0.5 & 82.2 $\pm$ 33.3  & 0.9738 $\pm$ 0.0672 & 0.9137 $\pm$ 0.0349 \\ \hline
10 & 0.418 $\pm$ 0.346 & 0.5 $\pm$ 0.7 & 85.4 $\pm$ 36.3  & 0.9470 $\pm$ 0.0717 & 0.9102 $\pm$ 0.0382 \\ \hline
12 & 0.558 $\pm$ 0.295 & 0.9 $\pm$ 0.9 & 83.5 $\pm$ 30.9  & 0.9283 $\pm$ 0.0759 & 0.9120 $\pm$ 0.0326  \\ \hline
15 & 0.656 $\pm$ 0.196 & 1.9 $\pm$ 1.4 & 95.1 $\pm$ 34.2  & 0.8713 $\pm$ 0.0915 & 0.8995 $\pm$ 0.0362 \\ \hline
17 & 0.723 $\pm$ 0.161 & 2.3 $\pm$ 1.4 & 113.2 $\pm$ 43.2 & 0.8635 $\pm$ 0.0823 & 0.8801 $\pm$ 0.0457 \\ \hline
20 & 0.787 $\pm$ 0.099 & 2.8 $\pm$ 1.8 & 141.4 $\pm$ 43.2 & 0.8580 $\pm$ 0.0887 & 0.8497 $\pm$ 0.0459 \\ \hline
\end{tabular}
\end{table*}

\begin{table*}
\centering
\caption{Performance of SBL without COMP (on synthetic data) for $93 \times 961$ Kirkman triple matrix. For each criterion, mean and standard deviation are reported over 1000 signals.}
\label{tab:only_sbl}
\begin{tabular}{|c|c|c|c|c|c|}
\hline
$k$ & \textbf{RMSE} & \textbf{\#FN} & \textbf{\#FP} & \textbf{Sens.} & \textbf{Spec.} \\ \hline
5          & 0.105 $\pm$ 0.213   & 0.0 $\pm$ 0.0       & 388.4 $\pm$ 63.2    & 1.0000 $\pm$ 0.0000  & 0.5937 $\pm$ 0.0661  \\ \hline
8          & 0.113 $\pm$ 0.211   & 0.1 $\pm$ 0.2       & 420.2 $\pm$ 55.5    & 0.9925 $\pm$ 0.0297  & 0.5591 $\pm$ 0.0582  \\ \hline
10         & 0.065 $\pm$ 0.024   & 0.0 $\pm$ 0.1       & 443.9 $\pm$ 42.9    & 0.9990 $\pm$ 0.0099  & 0.5332 $\pm$ 0.0452  \\ \hline
12         & 0.070 $\pm$ 0.024   & 0.1 $\pm$ 0.4       & 449.9 $\pm$ 38.1    & 0.9942 $\pm$ 0.0295  & 0.5259 $\pm$ 0.0401  \\ \hline
15         & 0.108 $\pm$ 0.139   & 0.1 $\pm$ 0.4       & 450.0 $\pm$ 29.1    & 0.9913 $\pm$ 0.0277  & 0.5243 $\pm$ 0.0308  \\ \hline
17         & 0.179 $\pm$ 0.236   & 0.6 $\pm$ 1.2       & 456.5 $\pm$ 30.0    & 0.9659 $\pm$ 0.0702  & 0.5165 $\pm$ 0.0318  \\ \hline
20         & 0.317 $\pm$ 0.351   & 1.4 $\pm$ 2.2       & 467.4 $\pm$ 30.5    & 0.9290 $\pm$ 0.1098  & 0.5033 $\pm$ 0.0324  \\ \hline
\end{tabular}
\end{table*}

\begin{table*}
\centering
\caption{Performance of NNOMP without COMP (on synthetic data) for $93 \times 961$ Kirkman triple matrix. For each criterion, mean and standard deviation are reported over 1000 signals.}
\label{tab:only_nnomp}
\begin{tabular}{|c|c|c|c|c|c|}
\hline
$k$ & \textbf{RMSE} & \textbf{\#FN} & \textbf{\#FP} & \textbf{Sens.} & \textbf{Spec.} \\ \hline
5          & 0.048 $\pm$ 0.022   & 0.0 $\pm$ 0.2       & 2.6 $\pm$ 8.5       & 0.9940 $\pm$ 0.0341  & 0.9973 $\pm$ 0.0089  \\ \hline
8          & 0.062 $\pm$ 0.025   & 0.2 $\pm$ 0.4       & 9.5 $\pm$ 17.7      & 0.9750 $\pm$ 0.0559  & 0.9900 $\pm$ 0.0186  \\ \hline
10         & 0.168 $\pm$ 0.280   & 0.8 $\pm$ 1.7       & 16.7 $\pm$ 22.1     & 0.9180 $\pm$ 0.1740  & 0.9825 $\pm$ 0.0232  \\ \hline
12         & 0.242 $\pm$ 0.331   & 2.0 $\pm$ 3.3       & 19.4 $\pm$ 21.7     & 0.8292 $\pm$ 0.2745  & 0.9795 $\pm$ 0.0229  \\ \hline
15         & 0.526 $\pm$ 0.416   & 6.4 $\pm$ 5.3       & 11.9 $\pm$ 16.9     & 0.5747 $\pm$ 0.3565  & 0.9874 $\pm$ 0.0178  \\ \hline
17         & 0.734 $\pm$ 0.410   & 10.3 $\pm$ 6.2      & 9.8 $\pm$ 15.5      & 0.3918 $\pm$ 0.3632  & 0.9897 $\pm$ 0.0164  \\ \hline
20         & 0.902 $\pm$ 0.306   & 15.4 $\pm$ 5.4      & 6.2 $\pm$ 10.1      & 0.2290 $\pm$ 0.2697  & 0.9934 $\pm$ 0.0107  \\ \hline
\end{tabular}
\end{table*}

\begin{table*}
\centering
\caption{Performance of NNLAD without COMP (on synthetic data) for $93 \times 961$ Kirkman triple matrix. For each criterion, mean and standard deviation are reported over 1000 signals.}
\label{tab:only_nnlad}
\begin{tabular}{|c|c|c|c|c|c|}
\hline
$k$ & \textbf{RMSE} & \textbf{\#FN} & \textbf{\#FP} & \textbf{Sens.} & \textbf{Spec.} \\ \hline
5  & 0.080 $\pm$ 0.028 & 0.0 $\pm$ 0.1 & 19.9 $\pm$ 6.3  & 0.9972 $\pm$ 0.0235 & 0.9791 $\pm$ 0.0066 \\ \hline
8  & 0.103 $\pm$ 0.036 & 0.1 $\pm$ 0.3 & 43.8 $\pm$ 16.2 & 0.9914 $\pm$ 0.0322 & 0.9541 $\pm$ 0.0170 \\ \hline
10 & 0.130 $\pm$ 0.050 & 0.2 $\pm$ 0.5 & 45.9 $\pm$ 23.6 & 0.9750 $\pm$ 0.0479 & 0.9517 $\pm$ 0.0249 \\ \hline
12 & 0.178 $\pm$ 0.086 & 0.6 $\pm$ 0.7 & 38.3 $\pm$ 24.1 & 0.9539 $\pm$ 0.0596 & 0.9597 $\pm$ 0.0254 \\ \hline
15 & 0.293 $\pm$ 0.150 & 0.9 $\pm$ 1.0 & 37.7 $\pm$ 18.3 & 0.9390 $\pm$ 0.0672 & 0.9602 $\pm$ 0.0193 \\ \hline
17 & 0.404 $\pm$ 0.177 & 0.8 $\pm$ 1.2 & 53.9 $\pm$ 22.9 & 0.9558 $\pm$ 0.0721 & 0.9429 $\pm$ 0.0243 \\ \hline
20 & 0.542 $\pm$ 0.163 & 0.3 $\pm$ 1.0 & 88.1 $\pm$ 24.8 & 0.9855 $\pm$ 0.0489 & 0.9063 $\pm$ 0.0263 \\ \hline
\end{tabular}
\end{table*}

\begin{table*}
\centering
\caption{Performance of NNLS without COMP (on synthetic data) for $93 \times 961$ Kirkman triple matrix. For each criterion, mean and standard deviation are reported over 1000 signals.}
\label{tab:only_nnls}
\begin{tabular}{|c|c|c|c|c|c|}
\hline
$k$ & \textbf{RMSE} & \textbf{\#FN} & \textbf{\#FP} & \textbf{Sens.} & \textbf{Spec.} \\ \hline
5  & 0.072 $\pm$ 0.024 & 0.1 $\pm$ 0.2 & 64.0 $\pm$ 5.6  & 0.9880 $\pm$ 0.0492 & 0.9330 $\pm$ 0.0058 \\ \hline
8  & 0.091 $\pm$ 0.031 & 0.2 $\pm$ 0.5 & 66.7 $\pm$ 4.2  & 0.9739 $\pm$ 0.0564 & 0.9300 $\pm$ 0.0044 \\ \hline
10 & 0.120 $\pm$ 0.052 & 0.4 $\pm$ 0.6 & 69.0 $\pm$ 4.1  & 0.9634 $\pm$ 0.0564 & 0.9275 $\pm$ 0.0043 \\ \hline
12 & 0.164 $\pm$ 0.087 & 0.6 $\pm$ 0.8 & 70.1 $\pm$ 7.8  & 0.9472 $\pm$ 0.0634 & 0.9261 $\pm$ 0.0082 \\ \hline
15 & 0.301 $\pm$ 0.160 & 1.0 $\pm$ 1.1 & 67.2 $\pm$ 12.0 & 0.9359 $\pm$ 0.0706 & 0.9289 $\pm$ 0.0127 \\ \hline
17 & 0.412 $\pm$ 0.170 & 0.7 $\pm$ 1.2 & 69.3 $\pm$ 10.9 & 0.9561 $\pm$ 0.0713 & 0.9266 $\pm$ 0.0115 \\ \hline
20 & 0.538 $\pm$ 0.155 & 0.3 $\pm$ 1.0 & 92.1 $\pm$ 18.0 & 0.9855 $\pm$ 0.0499 & 0.9021 $\pm$ 0.0192 \\ \hline
\end{tabular}
\end{table*}


\begin{table*}
\centering
\caption{Performance of NNLASSO without COMP (on synthetic data) for $45 \times 105$ Kirkman triple matrix. For each criterion, mean and standard deviation values are reported across 1000 signals.}
\label{tab:only_nnlasso_45_105}
\begin{tabular}{|c|c|c|c|c|c|}
\hline
$k$ & \textbf{RMSE} & \textbf{\#FN} & \textbf{\#FP} & \textbf{Sens.} & \textbf{Spec.} \\ \hline
5  & 0.066 $\pm$ 0.022 & 0.0 $\pm$ 0.2 & 19.6 $\pm$ 2.8  & 0.9914 $\pm$ 0.0416 & 0.8037 $\pm$ 0.0283 \\ \hline
8  & 0.082 $\pm$ 0.024 & 0.1 $\pm$ 0.3 & 21.5 $\pm$ 2.7  & 0.9840 $\pm$ 0.0433 & 0.7788 $\pm$ 0.0277 \\ \hline
10 & 0.096 $\pm$ 0.030 & 0.2 $\pm$ 0.4 & 22.2 $\pm$ 2.7  & 0.9785 $\pm$ 0.0442 & 0.7666 $\pm$ 0.0283 \\ \hline
12 & 0.111 $\pm$ 0.040 & 0.4 $\pm$ 0.6 & 22.6 $\pm$ 2.7  & 0.9699 $\pm$ 0.0488 & 0.7573 $\pm$ 0.0289 \\ \hline
15 & 0.150 $\pm$ 0.075 & 0.7 $\pm$ 0.8 & 23.3 $\pm$ 5.1  & 0.9561 $\pm$ 0.0526 & 0.7414 $\pm$ 0.0562 \\ \hline
17 & 0.182 $\pm$ 0.089 & 0.9 $\pm$ 0.9 & 24.3 $\pm$ 7.4  & 0.9467 $\pm$ 0.0537 & 0.7239 $\pm$ 0.0845 \\ \hline
20 & 0.258 $\pm$ 0.136 & 1.2 $\pm$ 1.1 & 29.7 $\pm$ 16.9 & 0.9379 $\pm$ 0.0567 & 0.6504 $\pm$ 0.1983 \\ \hline
\end{tabular}
\end{table*}
\begin{table*}
\centering
\caption{Performance of SBL without COMP (on synthetic data) for $45 \times 105$ Kirkman triple matrix. For each criterion, mean and standard deviation values are reported across 1000 signals.}
\label{tab:only_sbl_45_105}
\begin{tabular}{|c|c|c|c|c|c|}
\hline
$k$ & \textbf{RMSE} & \textbf{\#FN} & \textbf{\#FP} & \textbf{Sens.} & \textbf{Spec.} \\ \hline
5  & 0.078 $\pm$ 0.026 & 0.0 $\pm$ 0.0 & 49.9 $\pm$ 3.6 & 0.9996 $\pm$ 0.0089 & 0.5012 $\pm$ 0.0364 \\ \hline
8  & 0.088 $\pm$ 0.029 & 0.0 $\pm$ 0.2 & 48.9 $\pm$ 3.8 & 0.9959 $\pm$ 0.0230 & 0.4962 $\pm$ 0.0395 \\ \hline
10 & 0.095 $\pm$ 0.029 & 0.1 $\pm$ 0.3 & 47.8 $\pm$ 4.0 & 0.9898 $\pm$ 0.0316 & 0.4968 $\pm$ 0.0418 \\ \hline
12 & 0.105 $\pm$ 0.034 & 0.2 $\pm$ 0.4 & 47.2 $\pm$ 3.9 & 0.9865 $\pm$ 0.0331 & 0.4924 $\pm$ 0.0420 \\ \hline
15 & 0.122 $\pm$ 0.051 & 0.3 $\pm$ 0.6 & 46.3 $\pm$ 4.1 & 0.9789 $\pm$ 0.0377 & 0.4855 $\pm$ 0.0457 \\ \hline
17 & 0.138 $\pm$ 0.081 & 0.4 $\pm$ 0.7 & 45.6 $\pm$ 4.2 & 0.9736 $\pm$ 0.0383 & 0.4815 $\pm$ 0.0474 \\ \hline
20 & 0.202 $\pm$ 0.154 & 0.8 $\pm$ 0.9 & 45.4 $\pm$ 4.6 & 0.9595 $\pm$ 0.0470 & 0.4663 $\pm$ 0.0542 \\ \hline
\end{tabular}
\end{table*}

\begin{table*}
\centering
\caption{Performance of NNOMP without COMP (on synthetic data) for $45 \times 105$ Kirkman triple matrix. For each criterion, mean and standard deviation values are reported across 1000 signals.}
\label{tab:only_nnomp_45_105}
\begin{tabular}{|c|c|c|c|c|c|}
\hline
$k$ & \textbf{RMSE} & \textbf{\#FN} & \textbf{\#FP} & \textbf{Sens.} & \textbf{Spec.} \\ \hline
5  & 0.050 $\pm$ 0.023 & 0.0 $\pm$ 0.1 & 1.6 $\pm$ 3.7  & 0.9986 $\pm$ 0.0167 & 0.9836 $\pm$ 0.0365 \\ \hline
8  & 0.070 $\pm$ 0.027 & 0.1 $\pm$ 0.4 & 5.7 $\pm$ 6.3  & 0.9819 $\pm$ 0.0515 & 0.9408 $\pm$ 0.0646 \\ \hline
10 & 0.085 $\pm$ 0.032 & 0.2 $\pm$ 0.5 & 9.9 $\pm$ 6.8  & 0.9759 $\pm$ 0.0520 & 0.8960 $\pm$ 0.0720 \\ \hline
12 & 0.103 $\pm$ 0.036 & 0.4 $\pm$ 0.6 & 14.4 $\pm$ 6.5 & 0.9667 $\pm$ 0.0509 & 0.8451 $\pm$ 0.0700 \\ \hline
15 & 0.141 $\pm$ 0.066 & 0.6 $\pm$ 0.8 & 18.8 $\pm$ 5.0 & 0.9575 $\pm$ 0.0509 & 0.7912 $\pm$ 0.0555 \\ \hline
17 & 0.182 $\pm$ 0.108 & 1.1 $\pm$ 1.1 & 20.0 $\pm$ 3.7 & 0.9368 $\pm$ 0.0639 & 0.7724 $\pm$ 0.0421 \\ \hline
20 & 0.278 $\pm$ 0.177 & 2.1 $\pm$ 1.7 & 20.1 $\pm$ 2.8 & 0.8944 $\pm$ 0.0828 & 0.7634 $\pm$ 0.0332 \\ \hline
\end{tabular}
\end{table*}


\begin{table*}
\centering
\caption{Performance of NNLAD without COMP (on synthetic data) for $45 \times 105$ Kirkman triple matrix. For each criterion, mean and standard deviation values are reported across 1000 signals.}
\label{tab:only_nnlad_45_105}
\begin{tabular}{|c|c|c|c|c|c|}
\hline
$k$ & \textbf{RMSE} & \textbf{\#FN} & \textbf{\#FP} & \textbf{Sens.} & \textbf{Spec.} \\ \hline
5          & 0.067 $\pm$ 0.026  & 0.0 $\pm$ 0.1      & 5.6 $\pm$ 2.4      & 0.9986 $\pm$ 0.0167 & 0.9441 $\pm$ 0.0242 \\ \hline
8          & 0.089 $\pm$ 0.031  & 0.1 $\pm$ 0.2      & 11.4 $\pm$ 4.1     & 0.9926 $\pm$ 0.0305 & 0.8822 $\pm$ 0.0419 \\ \hline
10         & 0.106 $\pm$ 0.040  & 0.1 $\pm$ 0.4      & 14.5 $\pm$ 5.0     & 0.9865 $\pm$ 0.0362 & 0.8475 $\pm$ 0.0529 \\ \hline
12         & 0.128 $\pm$ 0.047  & 0.3 $\pm$ 0.5      & 15.8 $\pm$ 5.3     & 0.9778 $\pm$ 0.0404 & 0.8301 $\pm$ 0.0565 \\ \hline
15         & 0.166 $\pm$ 0.076  & 0.6 $\pm$ 0.7      & 16.7 $\pm$ 5.0     & 0.9615 $\pm$ 0.0478 & 0.8141 $\pm$ 0.0560 \\ \hline
17         & 0.201 $\pm$ 0.094  & 0.9 $\pm$ 0.9      & 17.1 $\pm$ 5.5     & 0.9494 $\pm$ 0.0513 & 0.8052 $\pm$ 0.0629 \\ \hline
20         & 0.272 $\pm$ 0.125  & 1.2 $\pm$ 1.2      & 21.3 $\pm$ 9.9     & 0.9404 $\pm$ 0.0578 & 0.7500 $\pm$ 0.1164 \\ \hline
\end{tabular}
\end{table*}

\begin{table*}
\centering
\caption{Performance of NNLS without COMP (on synthetic data) for $45 \times 105$ Kirkman triple matrix. For each criterion, mean and standard deviation values are reported across 1000 signals.}
\label{tab:only_nnls_45_105}
\begin{tabular}{|c|c|c|c|c|c|}
\hline
$k$ & \textbf{RMSE} & \textbf{\#FN} & \textbf{\#FP} & \textbf{Sens.} & \textbf{Spec.} \\ \hline
5  & 0.066 $\pm$ 0.021 & 0.1 $\pm$ 0.2 & 19.3 $\pm$ 2.8 & 0.9882 $\pm$ 0.0471 & 0.8067 $\pm$ 0.0285 \\ \hline
8  & 0.082 $\pm$ 0.025 & 0.1 $\pm$ 0.4 & 21.4 $\pm$ 2.5 & 0.9831 $\pm$ 0.0452 & 0.7793 $\pm$ 0.0261 \\ \hline
10 & 0.094 $\pm$ 0.029 & 0.3 $\pm$ 0.5 & 22.0 $\pm$ 2.7 & 0.9745 $\pm$ 0.0492 & 0.7684 $\pm$ 0.0284 \\ \hline
12 & 0.110 $\pm$ 0.037 & 0.3 $\pm$ 0.5 & 22.4 $\pm$ 2.8 & 0.9722 $\pm$ 0.0446 & 0.7596 $\pm$ 0.0304 \\ \hline
15 & 0.147 $\pm$ 0.068 & 0.6 $\pm$ 0.7 & 23.1 $\pm$ 3.0 & 0.9571 $\pm$ 0.0493 & 0.7435 $\pm$ 0.0332 \\ \hline
17 & 0.183 $\pm$ 0.089 & 0.9 $\pm$ 0.9 & 23.6 $\pm$ 3.8 & 0.9474 $\pm$ 0.0537 & 0.7324 $\pm$ 0.0429 \\ \hline
20 & 0.262 $\pm$ 0.131 & 1.3 $\pm$ 1.2 & 25.3 $\pm$ 7.2 & 0.9368 $\pm$ 0.0599 & 0.7026 $\pm$ 0.0843 \\ \hline
\end{tabular}
\end{table*}
